\documentclass[10pt,aps,showpacs,nofootinbib,prd,aps,epsf,floats,
               amsmath,amssymb,amsfonts,axodraw,twocolumn]{revtex4}
\usepackage{amsmath, amssymb}
\newcommand{\mathsym}[1]{{}}

\usepackage{graphicx}% Include figure files
\usepackage{amsmath}
\usepackage{amssymb}
\usepackage{bm}% bold math
\newcommand{\be}{\begin{equation}}
\newcommand{\ee}{\end{equation}}
\newcommand{\bea}{\begin{eqnarray}}
\newcommand{\eea}{\end{eqnarray}}

\newcommand{\rem}[1]{}
\newsavebox{\PSLASH}
 \sbox{\PSLASH}{$p$\hspace{-1.8mm}/}
 
\renewcommand{\theequation}{\thesection.\arabic{equation}}
\newcounter{saveeqn}
\newcommand{\add}{\addtocounter{equation}{1}}
\newcommand{\alpheqn}{\setcounter{saveeqn}{\value{equation}}%
\setcounter{equation}{0}%
\renewcommand{\theequation}{\mbox{\thesection.\arabic{saveeqn}{\alph{equation}}}}}
\newcommand{\reseteqn}{\setcounter{equation}{\value{saveeqn}}%
\renewcommand{\theequation}{\thesection.\arabic{equation}}}

 \newsavebox{\notrightarrow}
 \sbox{\notrightarrow}{$\to$\hspace{-4mm}/}
 
 \newsavebox{\PARTIALSLASH}
 \sbox{\PARTIALSLASH}{$\partial$\hspace{-1.6mm}/}
 
 \newsavebox{\ASLASH}
 \sbox{\ASLASH}{$A$\hspace{-2.1mm}/}
 
 \newsavebox{\KSLASH}
 \sbox{\KSLASH}{$k$\hspace{-1.8mm}/}
 
 \newsavebox{\LSLASH}
 \sbox{\LSLASH}{$\ell$\hspace{-1.8mm}/}
 
 \newsavebox{\QSLASH}
 \sbox{\QSLASH}{$q$\hspace{-1.8mm}/}
 
 \newsavebox{\DSLASH}
 \sbox{\DSLASH}{$D$\hspace{-2.2mm}/}
 
 \newsavebox{\DbfSLASH}
 \sbox{\DbfSLASH}{${\mathbf D}$\hspace{-2.8mm}/}
 
 \newsavebox{\DELVECRIGHT}
 \sbox{\DELVECRIGHT}{$\stackrel{\rightarrow}{\partial}$}
 
 \newcommand{\blue}{\IfColor{\textCadetBlue}{}}
\newcommand{\black}{\IfColor{\textBlack}{}}
\newcommand{\red}{\IfColor{\textRed}{}}
\newcommand{\green}{\IfColor{\textOliveGreen}{}}
\newcommand{\lila}{\IfColor{\textRedViolet}{}}

\begin{document}
%\begin{flushright}
%[hep-th]
%\end{flushright}
\title{Properties of neutral mesons in a hot and magnetized quark matter}
\author{Sh. Fayazbakhsh$^{a}$}\email{shfayazbakhsh@ipm.ir}
\author{S. Sadeghian$^{b}$}\email{s_sadeghian@alzahra.ac.ir}
\author{N. Sadooghi$^{c}$}\email{sadooghi@physics.sharif.ir}
\affiliation{ $^{a}$Institute for Research in Fundamental Sciences
(IPM), School of Particles and Accelerators, P.O. Box 19395-5531,
Tehran-Iran\\
$^{b}$Department of Physics, Alzahra University, P. Code
19938-94336, Tehran-Iran
\\
$^{c}$Department of Physics, Sharif University of Technology, P.O.
Box 11155-9161, Tehran-Iran}
%%%%%%%%%%%%%%%%%%%%%%%%%%%%%%%%%%%%%%
\begin{abstract}
\noindent The properties of noninteracting $\sigma$ and $\pi^{0}$
mesons are studied at finite temperature, chemical potential and in
the presence of a constant magnetic field. To do this, the energy
dispersion relations of these particles, including nontrivial form
factors, are derived using a derivative expansion of the effective
action of a two-flavor, hot and magnetized Nambu--Jona-Lasinio (NJL)
model up to second order. The temperature dependence of the pole and
screening masses as well as the directional refraction indices of
magnetized \textit{neutral} mesons are explored for fixed magnetic
fields and chemical potentials. It is shown that, because of the
explicit breaking of the Lorentz invariance by the magnetic field,
the refraction index and the screening mass of neutral mesons
exhibit a certain anisotropy in the transverse and longitudinal
directions with respect to the direction of the external magnetic
field. In contrast to their longitudinal refraction indices, the
transverse indices of the neutral mesons are larger than unity.
\end{abstract}
 \maketitle
%%%%%%%%%%%%%%%%%%%%%%%%
\section{Introduction}\label{introduction}\label{sec1}
%%%%%%%%%%%%%%%%%%%%%%%%
The study of the states of quark matter under extreme conditions has
attracted much attention over the past few years. Extreme conditions
include high temperatures and finite baryonic chemical potentials as
well as strong magnetic fields. The latter is responsible for many
interesting effects on the properties of quark matter. Some of the
most important ones are magnetic catalysis of dynamical chiral
symmetry breaking \cite{klimenko1992, miransky1995, catalysis}, that
leads to a modification of the nature of electroweak
\cite{ayala2008}, chiral and color-superconducting phase transitions
\cite{superconducting, fayazbakhsh2010-1, fayazbakhsh2010,
skokov2011, pawlowski2012}, production of chiral density waves
\cite{klimenko2010}, chiral magnetic effect \cite{kharzeev2008}, and
last but not least inducing electromagnetic superconductivity and
superfluidity \cite{chernodub2011}. In this paper, we will focus on
the effect of constant magnetic fields on the properties of
\textit{neutral and noninteracting} mesons in a hot and dense quark
matter. In particular, the temperature dependence of meson masses as
well as their direction-dependent refraction indices\footnote{The
term ``refraction index'' is used in \cite{shuryak1990} for pions
modified by the matter (quasipions). Although, the same terminology
is also used in \cite{ayala2002}, the definitions of refraction
index in \cite{shuryak1990} and \cite{ayala2002} are slightly
different, as will be explained later.} and screening masses will be
explored in the presence of various fixed magnetic fields. The
largest observed magnetic field in nature is about $10^{12}-10^{13}$
Gau\ss~in pulsars and up to $10^{14}-10^{15}$ Gau\ss~on the surface
of some magnetars, where the inner field is estimated to be of order
$10^{18}-10^{20}$ Gau\ss~\cite{incera2010}. There are also evidences
for the creation of very strong and short-living magnetic fields in
the early stages of non-central heavy ion collisions at RHIC
\cite{kharzeev51-STAR, mclerran2007}. Depending on the collision
energies and impact parameters, the magnetic fields produced at RHIC
and LHC are estimated to be in the order $eB\sim 1.5~m_{\pi}^{2}$,
corresponding to $0.03$ GeV$^{2}$ for $m_{\pi}=138$ MeV, and $eB\sim
15~m_{\pi}^{2}$, corresponding to $0.3$ GeV$^{2}$, respectively
\cite{skokov2010}.\footnote{Note that $eB=1$ GeV$^{2}$ corresponds
to $B\sim 1.7\times 10^{20}$ Gau\ss.} On the other hand, it is known
that the quark-gluon plasma, produced in high-energy heavy-ion
collisions, passes over many stages during its evolution. The last
of which consists of a large amount of hadrons, including pions,
until a final freeze-out \cite{ayala2002}. Thus, the presence of a
background magnetic field created in heavy ion experiments may
affect the properties of ``charged quarks'' in the earliest stage of
the collision and although the created strong magnetic field is
extremely short living and decays very fast \cite{mclerran2007,
skokov2010}, it may affect the properties of the hadrons made of
these ``magnetized'' quarks. Even the properties \textit{neutral
mesons} may be affected by the external magnetic field produced in
the earliest phase of heavy-ion collisions. In the present paper, we
do not intend to go through the phenomenology of heavy-ion
collisions. Our computation is only a theoretical attempt to study
the effect of external magnetic fields on ``magnetized''
\textit{neutral mesons}, that, because of the lack of electric
charge, do not interact directly with the external magnetic field.
Our study is indeed in contrast with the recent studies in
\cite{andersen2011-pions, anderson2012-2}, where chiral perturbation
theory is used to study the effect of external magnetic fields on
the pole and screening masses as well as the decay rates of
\textit{charged pions} interacting directly with the external
magnetic field.
\par
There are several attempts to study the effect of temperature and
chemical potential on the properties pions in a hot and dense
medium, in the absence of external magnetic fields
\cite{shuryak1990,  pisarski1996, ayala2002, chiral-perturbation,
son2000}. In \cite{shuryak1990}, the energy dispersion relation of
the so-called ``quasipions'' (or pions modified by the matter) is
introduced by
\begin{eqnarray}\label{int1a}
\omega^{2}(p)=u^{2}\mathbf{p}^{2}+m_{\pi}^{2}.
\end{eqnarray}
Here, $u(T)$ is the temperature-dependent refraction index (also
called ``mean quasipion velocity'' \cite{shuryak1990}), and
$m_{\pi}$ is the pole mass of the pions. To determine $m_{\pi}$, one
can either start from the Lagrangian density of a linear
$\sigma$-model including four-pion interaction or use the chiral
perturbation theory Lagrangian within certain approximation.
Considering the pion (one-loop) self-energy of the model, and
computing, in particular, its pole, it is possible to determine the
pion pole mass (at one-loop level). As concerns the screening mass
of pions, $m^{s}_{\pi}$, it is related to  $m_{\pi}$ through the
relation $m^{s}_{\pi}=m_{\pi}/v_{\pi}$, where $v_{\pi}$ is the pion
velocity \cite{pisarski1996}. As it is shown in \cite{pisarski1996},
the velocity $v_{\pi}$ of \textit{massless} pions is in general
given by
\begin{eqnarray}\label{int1}
\omega^{2}=v_{\pi}^{2}p^{2}\equiv \frac{\mbox{Re}
f_{\pi}^{s}}{\mbox{Re}f_{\pi}^{t}}p^{2},
\end{eqnarray}
where $\omega\equiv p_{0}$ is the energy, $p\equiv |\mathbf{p}|$ is
the absolute value of pion three momentum, and  $f_{\pi}^{t}$ and
$f_{\pi}^{s}$ are temporal and spatial pion decay constants,
respectively. As it turns out, at zero temperature, because of
relativistic invariance, $f_{\pi}^{t}=f_{\pi}^{s}$, and therefore
$v_{\pi}=1$. At finite temperature, however, since a privileged rest
frame is provided by the medium, relativistic invariance does not
apply anymore, and, as it is shown in \cite{pisarski1996}, ``cool''
pions propagate at a velocity $v_{\pi}<1$. Moreover, it is shown in
\cite{pisarski1996}, that for approximate chiral symmetry, the
Gell-Mann, Ookes and Renner (GOR) relation between the pion mass
$m_{\pi}$ and the pion decay constant $f_{\pi}$ still holds at
finite temperature, except that instead of $f_{\pi}$, the real part
of $f_{\pi}^{t}$ enters the GOR relation, i.e.
$m_{\pi}^{2}=\frac{2m_{0}\langle
\bar{\psi}\psi\rangle}{(\mbox{Re}f_{\pi}^{t})^{2}}$. Let us also
notice that at finite temperature and in the absence of external
magnetic fields, no distinction is to be made between neutral and
charged pion masses.
\par
Nontrivial energy dispersion relation of mesons is also introduced
in \cite{ayala2002} and \cite{son2000}. In \cite{ayala2002}, using
the definition of the group velocity, a momentum dependent
``refraction index'' $\tilde{n}(p)$ is defined for pions by the
ratio of the group velocity in matter and in vacuum,
$\tilde{n}(p)\equiv v_{gr}^{\mbox{\tiny{vac}}}/v_{gr}$. Here, the
matter pion group velocity is defined by $v_{gr}\equiv
\frac{dp_{0}}{dp}$ with
$p_{0}=[n^{-1}(T,\mu)p^{2}+M^{2}(T,\mu)]^{1/2}$, and the vacuum pion
group velocity is defined by $v_{gr}^{\mbox{\tiny{vac}}}\equiv
\frac{p}{p_{0}^{\mbox{\tiny{vac}}}}$. The momentum dependent
refraction index is therefore given by
$n(p)=\left(\frac{p_{0}}{p_{0}^{\mbox{\tiny{vac}}}}\right)n$. It is
argued that since for finite temperature $T$ and chemical potential
$\mu$, we always have both $n>1$ and
$\frac{p_{0}}{p_{0}^{\mbox{\tiny{vac}}}}>1$ for all values of $p$,
the index of refraction developed by the pion medium at finite $T$
and $\mu$ is always larger than unity \cite{ayala2002}. Let us
notice that the definition of the refraction index $n$ in
\cite{ayala2002} is slightly different from what is used in
\cite{shuryak1990}: In \cite{ayala2002}, $n^{-1}$ appearing in the
dispersion relation $p_{0}=[n^{-1}(T,\mu)p^{2}+M^{2}(T,\mu)]^{1/2}$
is the same as $u^{2}$ appearing in the dispersion relation
(\ref{int1a}) from \cite{shuryak1990}. In the latter, $u=n^{-1/2}$
is called refraction index.\footnote{In the present paper, we have
adopted the terminology used in \cite{shuryak1990}.} Having this in
mind, it turns out that the results presented in \cite{ayala2002},
coincides with those obtained in \cite{son2000}. Here, the quantity
$u$ appears as in \cite{shuryak1990}, in the pion energy dispersion
relation, $\omega^{2}=u^{2}(\mathbf{p}^{2}+m^{2})$, and is termed
``velocity'', although the authors mention that $u$ is the pion
velocity only when $m=0$. Here, $m$ is the screening mass. The pion
pole mass is then defined by $m_{p}=um$. Using scaling and
universality arguments, the authors predict that ``when critical
temperature is approached from below, the pole mass of the pion
drops despite the growth of the pion screening mass. This fact is
attributed to the decrease of the pion velocity near the phase
transition'' \cite{son2000}.
\par
As concerns the effect of external magnetic fields on the low energy
properties of QCD, in \cite{agasian2001}, the GOR relation between
the neutral pion mass $m_{\pi^{0}}$ and its decay constant
$f_{\pi^{0}}$, is shown to be valid in the first order a chiral
perturbation theory in the presence of constant and weak magnetic
fields, whose Lagrangian includes, in particular,
$(\vec{\pi}^{2})^{2}$ self-interaction terms. This method is also
used recently in \cite{andersen2011-pions, anderson2012-2} to
determine the pion thermal mass and the pion decay constants in the
presence of a constant magnetic field and at finite temperature. It
is shown, that the magnetic field gives rise to a splitting between
$m_{\pi^{0}}$ and $m_{\pi^{\pm}}$ as well as $f_{\pi^{0}}$ and
$f_{\pi^{\pm}}$. The pion decay constants $f_{\pi^{0}}$ and
$f_{\pi^{\pm}}$ are computed by evaluating the matrix elements
$\langle0|A_{\mu}^{0}|\pi^{0}\rangle$ and
$\langle0|A_{\mu}^{\pm}|\pi^{\mp}\rangle$, respectively. However, no
distinction between the temporal ($\mu=0$) and spatial ($\mu=1,2,3$)
directions is made.
\par
In the present paper, we will mainly focus on nontrivial energy
dispersion relations of \textit{noninteracting} $\sigma$ and
$\vec{\pi}$ mesons, arising from an appropriate evaluation of the
one-loop effective action of a two-flavor NJL model in a derivative
expansion up to second order. Our method is therefore different from
the method used in \cite{andersen2011-pions, anderson2012-2}, and
involves, in contrast to \cite{andersen2011-pions, anderson2012-2},
the effect of external magnetic fields on \textit{charged quarks}
from which the mesons are built. This will give us the possibility
to explore the effect of external magnetic fields on \textit{neutral
mesons} at finite temperature and chemical potential. Using the
method originally introduced in \cite{miranskybook, miransky1995}
for a single flavor NJL model, we will arrive at the effective
action of $\sigma$ and $\vec{\pi}=(\pi_{1},\pi_{2},\pi_{3})$ mesons,
\begin{eqnarray}\label{int2}
\lefteqn{\hspace{-0.6cm}\Gamma_{\mbox{\tiny{eff}}}[\sigma,\vec{\pi}]=\Gamma_{\mbox{\tiny{eff}}}[\sigma_{0}]}\nonumber\\
&&\hspace{-0.8cm}-\frac{1}{2}\int
d^{d}x~{\sigma}(x)\left(M_{\sigma}^{2}+{\cal{G}}^{\mu\mu}\partial_{\mu}^{2}\right){\sigma}(x)\nonumber\\
&&\hspace{-0.8cm}-\frac{1}{2}\sum\limits_{\ell=1}^{3}\int
d^{d}x~{\pi}_{\ell}(x)\left(M_{\vec{\pi}}^{2}+
{\cal{F}}^{\mu\mu}\partial_{\mu}^{2}\right)_{\ell\ell}{\pi}_{\ell}(x),
\end{eqnarray}
including nontrivial meson squared mass matrices $(M_{\sigma}^{2},
M_{\vec{\pi}}^{2})$ and form factors $({\cal{G}}^{\mu\nu},
{\cal{F}}^{\mu\nu})$, and leading to the energy dispersion relations
of $\sigma$ and $\vec{\pi}$ mesons
\begin{eqnarray}\label{int3}
E_{\sigma}^{2}&=&\sum_{i}(u_{\sigma}^{(i)}p_{i})^{2}+m_{\sigma}^{2},
\nonumber\\
E_{\vec{\pi}}^{2}&=&\sum_{i}(u_{\vec{\pi}}^{(i)}p_{i})^{2}+m_{\vec{\pi}}^{2}.
\end{eqnarray}
Here, the pole masses $(m_{\sigma}^{2},m_{\vec{\pi}}^{2})$ and
refraction indices $(\mathbf{u}_{\sigma}, \mathbf{u}_{\vec{\pi}})$
of the mesons are defined by
\begin{eqnarray*}
m_{\sigma}^{2}=\frac{\mbox{Re}[
M_{\sigma}^{2}]}{\mbox{Re}[{\cal{G}}^{00}]},\qquad
m_{\pi_{\ell}}^{2}=\frac{\mbox{Re}[M_{\pi_{\ell}}^{2}]}{\mbox{Re}[({\cal{F}}^{00})_{\ell\ell}]},
\end{eqnarray*}
and
\begin{eqnarray*}
u_{\sigma}^{(i)}=\left(\frac{\mbox{Re}[{\cal{G}}^{ii}]}{\mbox{Re}[{\cal{G}}^{00}]}\right)^{1/2},\qquad
u_{\pi_{\ell}}^{(i)}=\left(\frac{\mbox{Re}[({\cal{F}}^{ii})_{\ell\ell}]}{\mbox{Re}[({\cal{F}}^{00})_{\ell\ell}]}\right)^{1/2},
\end{eqnarray*}
for all space directions $i=1,2,3$ and isospin indices $\ell=1,2,3$.
These quantities can be computed using the one-loop effective action
of a two-flavor NJL model at finite temperature $T$, chemical
potential $\mu$ and for a constant magnetic field $B$, according to
the formalism presented in \cite{miranskybook, miransky1995}. Using
the definition of the screening mass from \cite{son2000}, the
screening masses of $\sigma$ and $\vec{\pi}$ mesons,
$m_{\sigma}^{(i)}$ and $m_{\vec{\pi}}^{(i)}$ are given by
\begin{eqnarray*}
m_{\sigma}^{(i)}=\frac{m_{\sigma}}{u_{\sigma}^{(i)}},\qquad\mbox{and}\qquad
m_{\vec{\pi}}^{(i)}=\frac{m_{\vec{\pi}}}{u_{\vec{\pi}}^{(i)}},~~~\forall
i=1,2,3,
\end{eqnarray*}
respectively. Later, we will, in particular, show that in the
presence of a uniform magnetic field, directed in a specific
direction, the refraction indices and screening masses in the
transverse and longitudinal directions with respect to the direction
of the background magnetic field will be different.
\par
The organization of this paper is as follows. In Sec. \ref{sec2}, we
will generalize the method introduced in \cite{miransky1995} to a
multi-flavor system, and will derive the effective action
(\ref{int2}), using an appropriate derivative expansion up to second
order. In Sec. \ref{sec3}, we will determine the one-loop effective
potential of a two-flavor NJL model including $(\sigma,\vec{\pi})$
mesons. In Sec. \ref{sec4}, the squared mass matrices
$(M_{\sigma}^{2}, M_{\pi^{0}}^{2})$ and kinetic coefficients
$({\cal{G}}^{\mu\nu}, {\cal{F}}^{\mu\nu})$ corresponding to neutral
mesons $\sigma$ and $\pi^{0}$ will be analytically computed at
finite $(T,\mu,eB)$ and up to an integration over $p_{3}$-momentum
as well as a summation over Landau levels. In Sec. \ref{sec5p1}, we
will first use the one-loop effective potential, evaluated in Sec.
\ref{sec3}, to explore the phase portrait of the model. Here, the
effect of magnetic catalysis \cite{klimenko1992, miransky1995} and
inverse magnetic catalysis \cite{fayazbakhsh2010, rebhan2011} on the
critical $(T,\mu,eB)$ will be scrutinized. Performing numerically
the remaining $p_{3}$-integration and the summation over Landau
levels from Sec. \ref{sec4}, we will present, in Sec. \ref{sec5p2},
the $T$-dependence of $(M_{\sigma}^{2}, M_{\pi^{0}}^{2})$ and
$({\cal{G}}^{\mu\nu}, {\cal{F}}^{\mu\nu})$ for various fixed
magnetic fields and at $\mu=0$. Using these results, the
$T$-dependence of pole masses of neutral mesons as well as their
directional refraction indices and screening masses will be
determined in Sec. \ref{sec5p3} for various fixed $eB=0.03, 0.2,
0.3$ GeV$^{2}$. We will in particular show that, for non-vanishing
magnetic fields, the refraction index of noninteracting mesons in
the longitudinal direction is equal to unity, while their transverse
refraction index is \textit{larger} than unity. Let us notice that
since the mesons are massive, this does not mean that magnetized
mesons propagate with speed larger than the speed of
light.\footnote{The effect of constant magnetic fields on the
propagation of massless particles is recently discussed in
\cite{alexandre2012}.} The observed anisotropy in the meson
refraction indices is because of the explicit breaking of Lorentz
invariance by uniform magnetic fields. The same anisotropy is also
reflected in the screening masses of neutral mesons in the
longitudinal and transverse directions with respect to the direction
of the background magnetic field. We will plot the $T$-dependence of
mesons screening masses for various fixed $eB$ and $\mu$, and will
show that, in the transverse directions, they are always smaller
than the screening masses in the longitudinal direction. Motivated
by recent experimental activities at RHIC and LHC, we will only
consider the effects of relatively weak and intermediate magnetic
field strength ($eB=0.03, 0.2, 0.3$ GeV$^{2}$). As concerns the
effect of stronger magnetic fields, we will show that they lead to
certain instabilities at low temperature. Our results for $eB=0.5,
0.7$ GeV$^{2}$ are consistent with the main conclusions presented
recently in \cite{gorbar2012}, where a single flavor NJL model is
studied in $2+1$ dimensions in the presence of a strong magnetic
field and at finite temperature. A summary of our results will be
presented in Section \ref{sec6}.
%%%%%%%%%%%%%%%%%%%%%%%%%%%%%%%%%%%%%%%%%%%%%%%%%%%%%%%
\section{Mathematical Tool: Derivative Expansion of the Quantum Effective Action}\label{sec2}
%%%%%%%%%%%%%%%%%%%%%%%%%%%%%%%%%%%%%%%%%%%%%%%%%%%%%%%%%%%
\setcounter{equation}{0}\par\noindent Let us consider a theory
containing $N$ real scalar fields $(\varphi_{0},\varphi_{1}\cdots,
\varphi_{N-1})\equiv\Phi$, whose dynamics are described by the
effective action $\Gamma_{\mbox{\tiny{eff}}}[\Phi]$. Using an
appropriate derivative expansion, and, in particular, generalizing
the method introduced in \cite{miranskybook, miransky1995} to a
multi-flavor system, we will derive, in this section, the energy
dispersion relations of $\varphi_{\ell},~ \ell=0,\cdots, N-1$. Using
the energy dispersion relation, the pole and screening mass as well
as the \textit{directional} refraction index corresponding to
$\varphi_{\ell}, \ell=0,\cdots,N-1$ will be defined.
\par
Let us start by expanding $\Phi(x)$ around an $x$-independent
configuration $\Phi_{0}$,
\begin{eqnarray}\label{NN1}
\Phi(x)=\Phi_{0}+\bar{\Phi}(x).
\end{eqnarray}
Plugging (\ref{NN1}) in the effective action, we arrive first at
\begin{eqnarray}\label{NN2}
\lefteqn{\Gamma_{\mbox{\tiny{eff}}}[\Phi]=\Gamma_{\mbox{\tiny{eff}}}[\Phi_{0}]+\int
d^{d}x\frac{\delta\Gamma_{\mbox{\tiny{eff}}}}{\delta\varphi_{i}(x)}\bigg|_{\Phi_{0}}\bar{\varphi}_{i}(x)
}\nonumber\\
&&+\frac{1}{2}\int d^{d}x
d^{d}y\frac{\delta^{2}\Gamma_{\mbox{\tiny{eff}}}}{\delta\varphi_{i}(x)\delta\varphi_{j}(y)}
\bigg|_{\Phi_{0}}\bar{\varphi}_{i}(x)\bar{\varphi}_{j}(y)+\cdots.\nonumber\\
\end{eqnarray}
Assuming that $\Phi_{0}$ describes a configuration that minimizes
the effective action, the second term in (\ref{NN2}) vanishes. Using
then the Taylor expansion
\begin{eqnarray}\label{NN3}
\bar{\Phi}(y)=\bar{\Phi}(x)+z^{\mu}\partial_{\mu}\bar{\Phi}(x)+\frac{1}{2}z^{\mu}z^{\nu}
\partial_{\mu}\partial_{\nu}\bar{\Phi}(x)+\cdots,\nonumber\\
\end{eqnarray}
with $z\equiv y-x$, and neglecting the terms linear in $z$, we get
\begin{eqnarray}\label{NN4}
\lefteqn{\hspace{-0.3cm}\Gamma_{\mbox{\tiny{eff}}}[\Phi]=\Gamma_{\mbox{\tiny{eff}}}[\Phi_{0}]-\frac{1}{2}\int
d^{d}x\
{\cal{M}}^{2}_{ij}[\Phi_{0}]\bar{\varphi}_{i}(x)\bar{\varphi}_{j}(x)}\nonumber\\
&&\hspace{-0.3cm}+\frac{1}{2}\int d^{d}x\
\chi_{ij}^{\mu\nu}[\Phi_{0}]\partial_{\mu}\bar{\varphi}_{i}(x)\partial_{\nu}\bar{\varphi}_{j}(x)+\cdots,
\end{eqnarray}
where the summation over $i,j=0,\cdots N-1$ is skipped. In
(\ref{NN4}), the ``squared mass matrix'' ${\cal{M}}_{ij}^{2}$ and
the ``kinetic matrix'' $\chi_{ij}^{\mu\nu}$ are given by
\begin{eqnarray}
\hspace{-0.3cm}{\cal{M}}_{ij}^{2}[\Phi_{0}]&\equiv&-\int
d^{d}z\frac{\delta^{2}\Gamma_{\mbox{\tiny{eff}}}}{\delta\varphi_{i}(0)\delta\varphi_{j}(z)}\bigg|_{\Phi_{0}},\label{NN5}\\
\hspace{-0.3cm}\chi_{ij}^{\mu\nu}[\Phi_{0}]&\equiv& -\frac{1}{2}\int
d^{d}z
z^{\mu}z^{\nu}\frac{\delta^{2}\Gamma_{\mbox{\tiny{eff}}}}{\delta\varphi_{i}(0)\delta\varphi_{j}(z)}\bigg|_{\Phi_{0}}.\label{NN6}
\end{eqnarray}
The above derivative expansion of $\Gamma_{\mbox{\tiny{eff}}}[\Phi]$
from (\ref{NN4}) can alternatively be given as
\begin{eqnarray}\label{NN7}
\lefteqn{\hspace{-0.5cm}\Gamma_{\mbox{\tiny{eff}}}[\Phi]=\int
d^{d}x\left(-V[\Phi]\right.}\nonumber\\
&&\left.+\frac{1}{2}\chi_{ij}^{\mu\nu}[\Phi]
\partial_{\mu}\varphi_{i}(x)\partial_{\nu}\varphi_{j}(x)+\cdots\right),
\end{eqnarray}
where, all non-derivative terms in (\ref{NN4}) are summed up into
the potential part of the effective action $V[\Phi]$, and the terms
with two derivatives yield the kinetic part of the effective action,
proportional to $\chi_{ij}^{\mu\nu}$. To have a connection to the
example that will be worked out in the subsequent sections, let us
assume a fixed configuration for
$\Phi_{0}=\left(\varphi_{0(0)},0,0,\cdots,0\right)$, with
$\varphi_{0(0)}$= const., that spontaneously breaks the $O(N)$
symmetry of the original action. Using (\ref{NN5}), or equivalently
\begin{eqnarray}\label{NN8}
\hspace{-0cm}{\cal{M}}_{00}^{2}[\Phi_{0}]&=&-\int
d^{d}z\frac{\delta^{2}\Gamma_{\mbox{\tiny{eff}}}}{\delta\varphi_{0}(z)\delta\varphi_{0}(0)}\bigg|_{\Phi_{0}},\nonumber\\
\hspace{-0cm}{\cal{M}}_{\ell m}^{2}[\Phi_{0}]&=&-\int
d^{d}z\frac{\delta^{2}\Gamma_{\mbox{\tiny{eff}}}}{\delta\varphi_{\ell}(z)\delta\varphi_{m}(0)}\bigg|_{\Phi_{0}},
\end{eqnarray}
$\forall~\ell,m\geq 1$, it is possible to determine the squared mass
matrices corresponding to the collective modes
$\varphi_{0},\varphi_{1},\cdots,\varphi_{N-1}$. To determine the
kinetic part of the effective action, we use, as in
\cite{miransky1995}, the Ansatz
\begin{eqnarray}\label{NN9}
\tilde{\chi}_{ij}^{\mu\nu}[\Phi]=(F_{1}^{\mu\nu})_{ij}+2F_{2}^{\mu\nu}\frac{\varphi_{i}\varphi_{j}}{\Phi^{2}},
\end{eqnarray}
$\forall~i,j=0,1,\cdots N-1$. Here,
$\Phi^{2}=\sum_{i=0}^{N-1}\varphi_{i}^{2}$ and
$\tilde{\chi}^{\mu\nu}_{ij}[\Phi_{0}]=\chi_{ij}^{\mu\nu}[\Phi_{0}]$,
appearing in (\ref{NN4}). Plugging (\ref{NN9}) in (\ref{NN7}), the
kinetic part of the effective Lagrangian density including two
derivatives is given by
\begin{eqnarray}\label{NN10}
{\cal{L}}_{k}=\frac{1}{2}(F_{1}^{\mu\nu})_{ij}\partial_{\mu}\varphi_{i}\partial_{\nu}\varphi_{j}+\frac{F_{2}^{\mu\nu}}{\Phi^{2}}\left(\varphi_{i}\partial_{\mu}\varphi_{i}\right)
\left(\varphi_{j}\partial_{\nu}\varphi_{j}\right).\nonumber\\
\end{eqnarray}
To determine the form factors $F_{1}^{\mu\nu}$ and $F_{2}^{\mu\nu}$,
or at least a combination of these two form factors, we will use the
definition of $\Gamma_{\mbox{\tiny{eff}}}^{k}\equiv \int d^{d}x
{\cal{L}}_{k}$, as a part of the effective action including only two
derivatives \cite{miransky1995}. We get
\begin{eqnarray}\label{NN11}
\hspace{-0.3cm}\frac{\delta^{2}\Gamma_{\mbox{\tiny{eff}}}^{k}}{\delta\varphi_{0}(x)\delta\varphi_{0}(0)}\bigg|_{\Phi_{0}}
=-{\cal{G}}^{\mu\nu}\bigg|_{\Phi_{0}}\partial_{\mu}\partial_{\nu}\delta^{d}(x),
\end{eqnarray}
with
${\cal{G}}^{\mu\nu}\equiv\big[(F_{1}^{\mu\nu})_{00}+2F_{2}^{\mu\nu}\big]$,
and
\begin{eqnarray}\label{NN12}
\hspace{-0.3cm}\frac{\delta^{2}\Gamma_{\mbox{\tiny{eff}}}^{k}}{\delta\varphi_{\ell}(x)\delta\varphi_{m}(0)}\bigg|_{\Phi_{0}}=-({\cal{F}}^{\mu\nu})_{\ell
m}\bigg|_{\Phi_{0}}\partial_{\mu}\partial_{\nu}\delta^{d}(x),
\end{eqnarray}
$\forall~\ell,m\geq 1$, where $({\cal{F}}^{\mu\nu})_{\ell
m}\equiv\frac{1}{2}\big[(F_{1}^{\mu\nu})_{\ell
m}+(F_{1}^{\mu\nu})_{m\ell}\big]$. From (\ref{NN11}) and
(\ref{NN12}) we have
\begin{eqnarray}\label{NN13}
{\cal{G}}^{\mu\nu}[\Phi_{0}]&=&-\frac{1}{2}\int d^{d}z
z^{\mu}z^{\nu}\frac{\delta^{2}\Gamma_{\mbox{\tiny{eff}}}^{k}}{\delta\varphi_{0}(z)\delta\varphi_{0}(0)}\bigg|_{\Phi_{0}},\nonumber\\
%%%%%%
({\cal{F}}^{\mu\nu})_{\ell m}[\Phi_{0}]&=&-\frac{1}{2}\int d^{d}z
z^{\mu}z^{\nu}\frac{\delta^{2}\Gamma_{\mbox{\tiny{eff}}}^{k}}{\delta\varphi_{\ell}(z)\delta\varphi_{m}(0)}\bigg|_{\Phi_{0}},\nonumber\\
\end{eqnarray}
$\forall~\ell,m\geq 1$. Comparing the above relations with
$\chi_{ij}^{\mu\nu}$ from (\ref{NN6}), it turns out that
$\chi_{00}^{\mu\nu}={\cal{G}}^{\mu\nu}$ and $\chi_{\ell
m}^{\mu\nu}=({\cal{F}}^{\mu\nu})_{\ell m},\forall~\ell,m\geq 1$.
Assuming then $ ({\cal{M}}^{2})_{\ell m}=-({\cal{M}}^{2})_{m\ell}$,
and $({\cal{F}}^{\mu\nu})_{\ell m}=-({\cal{F}}^{\mu\nu})_{m\ell}$,
$\forall~\ell\neq m$ and $\ell,m\geq 1$,\footnote{This will be shown
in our specific example in the subsequent sections.} and denoting
${\cal{M}}^{2}_{00}$ by $M_{0}^{2}$, as well as
${\cal{M}}^{2}_{\ell\ell}$ by $M_{\ell}^{2}$ for
$\ell=1,\cdots,N-1$, the effective action (\ref{NN4}) simplifies as
\begin{eqnarray}\label{NN14}
\lefteqn{\hspace{-1cm}\Gamma_{\mbox{\tiny{eff}}}[\Phi]=\Gamma_{\mbox{\tiny{eff}}}[\Phi_{0}]-\frac{1}{2}\int
d^{d}x\bar{\varphi_{0}}\left(M_{0}^{2}+{\cal{G}}^{\mu\mu}\partial_{\mu}^{2}\right)\bar{\varphi}_{0}
}\nonumber\\
&&\hspace{-.5cm}-\frac{1}{2}\sum\limits_{\ell=1}^{N-1}\int
d^{d}x\bar{\varphi_{\ell}}\big[M_{\ell}^{2}+({\cal{F}}^{\mu\mu})_{\ell\ell}\partial_{\mu}^{2}\big]\bar{\varphi}_{\ell}.
\end{eqnarray}
Here, we have used the fact that ${\cal{G}}^{\mu\nu}$ and
${\cal{F}}^{\mu\nu}$ are diagonal, i.e.
${\cal{G}}^{\mu\nu}={\cal{G}}^{\mu\mu}g^{\mu\nu}$ as well as
${\cal{F}}^{\mu\nu}={\cal{F}}^{\mu\mu}g^{\mu\nu}$. The same
relations are shown to be valid in a single-flavor case
\cite{miransky1995}. From (\ref{NN14}), the general expressions for
the energy dispersion relation of noninteracting $\varphi_{\ell},
\ell=0,\cdots,N-1$ fields can be determined. For $\ell=0$, we have
\begin{eqnarray}\label{NN15}
\hspace{-1cm}E_{\varphi_{0}}^{2}\equiv\frac{1}{{\cal{G}}^{00}}\left({\cal{G}}^{11}p_{1}^{2}+{\cal{G}}^{22}p_{2}^{2}+{\cal{G}}^{33}
p_{3}^{2}+M_{0}^{2}\right),
\end{eqnarray}
and for $\forall \ell\geq 1$, we have
\begin{eqnarray}\label{NN15b}
\lefteqn{E_{\varphi_{\ell}}^{2}\equiv}\nonumber\\
&&\hspace{-0.5cm}\frac{1}{({\cal{F}}^{00})_{\ell\ell}}\big[({\cal{F}}^{11})_{\ell\ell}~p_{1}^{2}+({\cal{F}}^{22})_{\ell\ell}~p_{2}^{2}+
({\cal{F}}^{33})_{\ell\ell}~p_{3}^{2}+M_{\ell}^{2}\big].\nonumber\\
\end{eqnarray}
Using the above energy dispersion relations, the pole masses of free
$\varphi_{0}$ and $\varphi_{\ell}, \ell\geq 1$ are given by
\begin{eqnarray}\label{NN16}
m_{0}^{2}=\frac{M_{0}^{2}}{{\cal{G}}^{00}}, \qquad\mbox{and}\qquad
m_{\ell}^{2}=\frac{M_{\ell}^{2}}{({\cal{F}}^{00})_{\ell\ell}},
\end{eqnarray}
respectively. The screening masses $m _{\ell}^{(i)}$, and
``directional'' refraction indices $u_{\ell}^{(i)}$ of
noninteracting $\varphi_{\ell}, \ell=0,1,\cdots,N-1$ fields in the
$i$-th directions ($i=1,2,3$) are defined by
\begin{eqnarray}\label{NN17}
\hspace{-0.8cm}m_{0}^{(i)}=\frac{m_{0}}{u_{0}^{(i)}},\qquad\mbox{where}\qquad
(u_{0}^{(i)})^{2}=\frac{{\cal{G}}^{ii}}{{\cal{G}}^{00}},
\end{eqnarray}
for $\ell=0$, as well as
\begin{eqnarray}\label{NN18}
\hspace{-0.4cm}m_{\ell}^{(i)}=\frac{m_{\ell}}{u_{\ell}^{(i)}},\qquad\mbox{where}\qquad
(u_{\ell}^{(i)})^{2}=\frac{({\cal{F}}^{ii})_{\ell\ell}}{({\cal{F}}^{00})_{\ell\ell}},
\end{eqnarray}
for $\ell\geq 1$ [see Sec. \ref{sec5} for more details on the
definition of screening masses and refraction indices].
\par
In the present paper, we will use the above dispersion relations, to
describe the properties of \textit{noninteracting} $\sigma$ and
$\vec{\pi}$ mesons in a hot and magnetized medium. We will focus, in
particular, on $\sigma$ and $\pi_{3}$ mesons. The latter will be
identified with the neutral pion, $\pi_{3}\equiv \pi^{0}$. To do
this, we will first consider, in the next section, a two-flavor NJL
model including appropriate four-fermion interactions. Defining the
meson fields $\sigma$ and $\vec{\pi}$ in terms of fermionic fields,
and eventually integrating the fermions in the presence of a
constant magnetic field, we arrive at the one-loop effective action
$\Gamma_{\mbox{\tiny{eff}}}[\sigma,\vec{\pi}]$, describing the
dynamics of magnetized meson fields. We will spontaneously break the
chiral symmetry of the original theory, by choosing a fixed
configuration
$(\sigma_{0},\vec{\pi}_{0})=(\mbox{const.},{\mathbf{0}})$, that
minimizes $\Gamma_{\mbox{\tiny{eff}}}[\sigma,\vec{\pi}]$. Using then
the formalism described in the present section for the specific case
of $N=4$, and identifying $\varphi_{0}$ with the $\sigma$-meson and
$\varphi_{\ell}, \ell=1,2,3$ with the pions $\pi_{\ell},
\ell=1,2,3$, we will determine the temperature dependence of the
pole and screening mass, as well as the directional refraction
indices of noninteracting neutral $\sigma$ and $\pi^{0}$ mesons at
finite temperature and in the presence of a constant magnetic field.
We will postpone the discussion on the properties of charged and
magnetized pions to a future publication \cite{sadooghi2012-3}.
%%%%%%%%%%%%%%%%%%%%%%%%%%%%%%%%%%%%%%%%%%%%%%%%%%%%%%%
\section{One-loop effective potential of a two-flavor NJL model at finite $(T,\mu,eB)$}\label{sec3}
%%%%%%%%%%%%%%%%%%%%%%%%%%%%%%%%%%%%%%%%%%%%%%%%%%%%%%%%%%%
\setcounter{equation}{0}\par\noindent In this section, we will
determine the one-loop effective potential corresponding to a
two-flavor magnetized NJL model at finite temperature and density.
The minima of this effective potential will then be used in the
subsequent sections to determine the kinetic coefficients and mass
matrices corresponding to neutral $\sigma$ and $\pi^{0}$ mesons.
\par
Let us start by introducing the Lagrangian density of a two-flavor
gauged NJL model in the presence of a constant magnetic field
\begin{eqnarray}\label{NE1b}
{\cal{L}}&=&\bar{\psi}(x)\left(i\gamma^{\mu}D_{\mu}-m_{0}\right)\psi(x)+G~\{[\bar{\psi}(x)\psi(x)]^2\nonumber\\
&&+
[\bar{\psi}(x)i\gamma_5\vec{\tau}\psi(x)]^2\}-\frac{1}{4}F^{\mu\nu}F_{\mu\nu}.
\end{eqnarray}
Here, the fermionic fields $\psi^{c}_{f}$ carry apart from the Dirac
index, a flavor index $f\in(1,2)=(u,d)$ and a color index
$c\in(1,2,3)=(r,g,b)$. In the chiral limit $m_{0}\to 0$, this
implies the $SU_{L}(2)\times SU_{R}(2)$ chiral and $SU(3)$ color
symmetry of the theory. The isospin symmetry of the theory is
guaranteed by setting $m_{u}=m_{d}\equiv m_{0}$. The covariant
derivative $D_{\mu}$ in (\ref{NE1b}) is defined by $D_{\mu}\equiv
\partial_{\mu}+ieQA_{\mu}^{ext.}$, where
$Q=\mbox{diag}\left(2/3,-1/3\right)$ is the fermionic charge matrix
coupled to the $U(1)$ gauge field $A_{\mu}^{ext.}$, and
$\vec{\tau}=(\tau_{1},\tau_{2},\tau_{3})$ are the Pauli matrices.
Choosing, the vector potential $A_{\mu}^{ext.}$ in the Landau gauge
$A_{\mu}^{ext.}=(0,0,Bx_{1},0)$, (\ref{NE1b}) describes a two-flavor
NJL model in the presence of a uniform magnetic field
$\mathbf{B}=B\mathbf{e}_{3}$, aligned in the third direction. The
field strength tensor $F_{\mu\nu}$ is defined as usual by
$F_{\mu\nu}=\partial_{[\mu}A_{\nu]}^{ext.}$, with $A_{\mu}^{ext.}$
fixed as above. As it turns out, the above Lagrangian is equivalent
with the semi-bosonized Lagrangian
\begin{eqnarray}\label{NE2b}
{\cal{L}}_{sb}&=&\bar{\psi}(x)\left(i\gamma^{\mu}D_{\mu}-m_{0}\right)\psi(x)-\bar{\psi}\left(\sigma+i\gamma_5\vec{\tau}\cdot\vec{\pi}\right)\psi\nonumber\\
&&-\frac{(\sigma^2+\vec{\pi}^2)}{4G}-\frac{B^2}{2},
\end{eqnarray}
where the Euler-Lagrange equations of motion for the auxiliary
fields lead to the constraints
\begin{eqnarray}\label{NE3b}
\sigma(x)&=&-2G\bar{\psi}(x)\psi(x),\nonumber\\
\vec{\pi}(x)&=&-2G\bar{\psi}(x)i\gamma_5\vec{\tau}\psi(x).
\end{eqnarray}
To determine the one-loop effective action corresponding to
(\ref{NE1b}) as a functional of $\sigma$ and $\vec{\pi}$, the
fermionic fields $\psi$ and $\bar{\psi}$ in (\ref{NE2b}) are to be
integrated out. Using
\begin{eqnarray}\label{NE4b}
e^{i\Gamma_{\mbox{\tiny{eff}}}[\sigma,\vec{\pi}]}=\int{\cal{D}}\psi{\cal{D}}\bar{\psi}\exp\left(i\int
d^{4}x~{\cal{L}}_{sb}\right),
\end{eqnarray}
the one-loop effective action $\Gamma_{\mbox{\tiny{eff}}}$ is then
given by
\begin{eqnarray}\label{NE5b}
\Gamma_{\mbox{\tiny{eff}}}[\sigma,\vec{\pi}]=\Gamma_{\mbox{\tiny{eff}}}^{(0)}[\sigma,\vec{\pi}]
+\Gamma_{\mbox{\tiny{eff}}}^{(1)}[\sigma,\vec{\pi}],
\end{eqnarray}
where the tree level part, $\Gamma_{\mbox{\tiny{eff}}}^{(0)}$, and
the one-loop part, $\Gamma_{\mbox{\tiny{eff}}}^{(1)}$, are given by
\begin{eqnarray}\label{NE6b}
\Gamma_{\mbox{\tiny{eff}}}^{(0)}[\sigma,\vec{\pi}]=-\int
d^{4}x\left(\frac{\sigma^{2}+\vec{\pi}^{2}}{4G}+\frac{B^{2}}{2}\right),
\end{eqnarray}
and
\begin{eqnarray}\label{NE7b}
\Gamma_{\mbox{\tiny{eff}}}^{(1)}[\sigma,\vec{\pi}]=-i
{\mbox{Tr}}_{\{cfsx\}}\ln[i{S_{Q}^{-1}(\sigma,\vec{\pi})}].
\end{eqnarray}
Here, $m\equiv m_{0}+\sigma(x)$ and
\begin{eqnarray}\label{NE8b}
iS^{-1}_{Q}(\sigma,\vec{\pi})\equiv
i\gamma^{\mu}D_{\mu}-\left(m+i\gamma^{5}\vec{\tau}\cdot\vec{\pi}\right),
\end{eqnarray}
is the inverse fermion propagator. To determine
$\Gamma_{\mbox{\tiny{eff}}}^{(1)}[\sigma,\vec{\pi}]$, let us assume
a constant and fixed configuration
$(\sigma_{0},\vec{\pi}_{0})=(\mbox{const.}, {\mathbf{0}})$ for the
collective modes $(\sigma,\vec{\pi})$, that breaks the
$SU_{L}(2)\times SU_{R}(2)$ chiral symmetry of the original action
in the chiral limit. Only in this case, $m$ can be replaced by the
constant constituent quark mass $m=m_{0}+\sigma_{0}$,  where
$\sigma_{0}=$const. The one-loop effective potential is given by
evaluating the trace operation in (\ref{NE7b}), that includes a
trace over color $c$, flavor $f$, and spinor $s$ degrees of freedom,
as well as a trace over a four-dimensional space-time coordinate
$x$. Following the standard method introduced e.g. in
\cite{fayazbakhsh2010}, and after a straightforward computation, the
one-loop part of the effective action
$\Gamma_{\mbox{\tiny{eff}}}^{(1)}[\sigma_{0}]$ reads
\begin{eqnarray}\label{NE9b}
\Gamma^{(1)}_{\mbox{\tiny{eff}}}[\sigma_{0}]=-6i\sum\limits_{q\in\{\frac{2}{3},-\frac{1}{3}\}}
\ln{\det}_{x}[E_{q}^2-p_{0}^2],
\end{eqnarray}
where the energy of a charged fermion in a constant magnetic field
is given by
\begin{eqnarray}\label{NE10b}
E_{q}\equiv\sqrt{\bar{\mathbf{p}}_{q}^{2}+m^{2}}=\sqrt{2|qeB|p+p_{3}^{2}+m^{2}}.
\end{eqnarray}
Here, the Ritus four-momentum
\begin{eqnarray}\label{NE11b}
\bar{p}_{q}=(p_{0},0,-\mbox{sgn}(q eB)\sqrt{2|q eB|p}, p_{3}),
\end{eqnarray}
arises from the solutions of Dirac equation in the presence of a
constant magnetic field (see \cite{ritus, sadooghi2012} for more
details on the Ritus Eingenfunction method). In (\ref{NE10b}), $p$
labels the corresponding Landau levels appearing in the presence of
a uniform magnetic field. Performing the remaining determinant over
the coordinate space in (\ref{NE9b}) leads to the effective
(thermodynamic) potential $\Omega_{\mbox{\tiny{eff}}}^{(1)}$ defined
by $\Omega_{\mbox{\tiny{eff}}}^{(1)}\equiv
-{\cal{V}}^{-1}\Gamma_{\mbox{\tiny{eff}}}^{(1)}$, where the factor
${\cal{V}}$ denotes the four-dimensional space-time volume. The
final form of $\Omega_{\mbox{\tiny{eff}}}^{(1)}$ is then determined
in the momentum space, where the effect of finite temperature and
chemical potential is introduced by replacing $p_{0}$ in
(\ref{NE9b}) with $p_{0}=i\omega_{n}-\mu$. Here, the Matsubara
frequencies $\omega_{n}$ are defined by $\omega_{n}=(2n+1)\pi T$.
Using the standard replacement
\begin{eqnarray}\label{NE12b}
\lefteqn{\int\frac{d^4
p}{(2\pi)^4}f\left(p_{0},\bar{\mathbf{p}}\right)}\nonumber\\
&=&\frac{|qeB|}{\beta}\sum_{n=-\infty}^{+\infty}\sum\limits_{p=0}^{+\infty}\alpha_{p}\int
_{-\infty}^{+\infty}\frac{dp_{3}}{8\pi^{2}}~f(i\omega_{n}-\mu,p,p_{3}),\nonumber\\
\end{eqnarray}
with $p$ labeling the Landau levels and $\beta\equiv T^{-1}$, and
after summing over the Matsubara frequencies $n$, the (one-loop)
effective potential of the model reads
\begin{eqnarray}\label{NE13b}
\lefteqn{\Omega_{\mbox{\tiny{eff}}}^{(1)}
=-3\sum\limits_{q\in\{\frac{2}{3},-\frac{1}{3}\}}\frac{|q
eB|}{\beta}
}\nonumber\\
&&\times
\sum^{+\infty}_{p=0}\alpha_{p}\int_{-\infty}^{+\infty}\frac{dp_{3}}{4\pi^{2}}\left\{\beta
E_{q}+\ln\left(1+e^{-\beta(E_{q}+\mu)}\right)\right.\nonumber\\
&&\left.+\ln\left(1+e^{-\beta(E_{q}-\mu)}\right)\right\}.
\end{eqnarray}
Here, $\alpha_{p}=2-\delta_{p0}$ is the spin degeneracy factor. As
it turns out, the above expression for
$\Omega_{\mbox{\tiny{eff}}}^{(1)}$ consists of a
$(T,\mu)$-independent and a $(T,\mu)$-dependent term. The
$(T,\mu)$-independent part of $\Omega_{\mbox{\tiny{eff}}}^{(1)}$ is
divergent and is to be appropriately regulated. In the Appendix, we
have followed the method presented in \cite{providencia2008}, and
shown that the $(T,\mu)$-independent part of
$\Omega_{\mbox{\tiny{eff}}}^{(1)}$ is given by (\ref{appB13}).
Adding this part to the tree level part of the effective potential,
(\ref{NE6b}), as well as to the $(T,\mu)$-dependent part of
$\Omega_{\mbox{\tiny{eff}}}^{(1)}$, we arrive at the final
expression for the one-loop effective potential of a two-flavor NJL
model at finite $(T,\mu)$ and in the presence of a uniform magnetic
field aligned in the third direction
\begin{widetext}
\begin{eqnarray}\label{NE14b}
\lefteqn{\hspace{-0.8cm}\Omega_{\mbox{\tiny{eff}}}(m;T,\mu,eB)=\frac{\sigma^{2}}{4G}+\frac{B^{2}}{2}-\frac{3}{2\pi^{2}}\sum\limits_{q\in\{\frac{2}{3},-\frac{1}{3}\}}|qeB|^{2}\left\{\zeta'\left(-1,x_{q}\right)+\frac{x_{q}^{2}}{4}+\frac{x_{q}}{2}(1-x_{q})\ln
x_{q}\right\}
}\nonumber\\
&&
+\frac{3}{4\pi^{2}}\left\{m^{4}\ln\left(\frac{\Lambda+\sqrt{\Lambda^{2}+m^{2}}}{m}\right)-\Lambda(2\Lambda^{2}+m^{2})\sqrt{\Lambda^{2}+m^{2}}\right\}\nonumber\\
&&-3\sum\limits_{q\in\{\frac{2}{3},-\frac{1}{3}\}}\frac{|q
eB|}{\beta}\sum^{+\infty}_{p=0}\alpha_{p}\int_{-\infty}^{+\infty}\frac{dp_{3}}{4\pi^{2}}\left
\{\ln\left(1+e^{-\beta(E_{q}+\mu)}\right)+\ln\left(1+e^{-\beta(E_{q}-\mu)}\right)\right\}.
\end{eqnarray}
\end{widetext}
Here, $x_{q}\equiv \frac{m^{2}}{2|qeB|}$, $\Lambda$ is an
appropriate ultraviolet (UV) momentum cutoff,
$\zeta'(-1,x_{q})\equiv \frac{d\zeta(s,x_{q})}{ds}\big|_{s=1}$ and
$E_{q}$ is given in (\ref{NE10b}). In Sec. \ref{sec5}, after fixing
a number of free parameters, such as the coupling $G$ and the UV
cutoff $\Lambda$, the global minima of
$\Omega_{\mbox{\tiny{eff}}}(m;T,\mu,eB)$ will be determined
numerically. They will be then used to determine the squared mass
matrices $M_{\sigma}^{2}$ and $M_{{\pi}^{0}}^{2}$ and the form
factors (kinetic coefficients) ${\cal{G}}^{\mu\nu}$ and
$({\cal{F}}^{\mu\nu})_{33}$, corresponding to the neutral mesons
$\sigma$ and $\pi^{0}$, at finite $(T,\mu)$ and $eB$.
%%%%%%%%%%%%%%%%%%%%%%%%%%%%%%%%%%%%%%%%%%%
\section{Effective kinetic part of the one-loop effective action of a two-flavor NJL model at finite $(T,\mu,eB)$}\label{sec4}
%%%%%%%%%%%%%%%%%%%%%%%%%%%%%%%%%%%%%%%%%%%
\setcounter{equation}{0}
\par\noindent
In the previous section, the one-loop effective potential of a
magnetized two-flavor NJL model at finite $(T,\mu)$ is computed by
evaluating the trace operation in (\ref{NE7b}) for a fixed field
configuration $
\Phi_{0}=(\sigma_{0},\vec{\pi}_{0})=(\mbox{const.},{\mathbf{0}})$,
which is supposed to minimize the one-loop effective potential
(\ref{NE14b}) of the model. In the next two sections, we will
compute the squared meson mass matrices and form factors of the
effective kinetic part of the one-loop effective action
corresponding to neutral mesons $\sigma$ and $\pi^{0}$. This
computation includes an analytical and a numerical part. In this
section, after reformulating the general derivation presented in
Sec. \ref{sec2}, and making it compatible with our case of
magnetized two-flavor NJL model, we will present the analytical
results of the squared mass matrices $(M_{\sigma}^{2},
M_{\pi^{0}}^{2})$ and form factors $({\cal{G}}^{\mu\nu},
{\cal{F}}^{\mu\nu})$  for neutral mesons up to a one-dimensional
integration over $p_{3}$-momentum and a summation over Landau levels
$p$. They shall be performed numerically. The results of the
numerical computation will be presented in Sec. \ref{sec5}, where we
explore the $(T,\mu,eB)$ dependence of $(M_{\sigma}^{2},
M_{\pi^{0}}^{2})$ and $({\cal{G}}^{\mu\nu}, {\cal{F}}^{\mu\nu})$.
Using these quantities the pole and screening masses of free neutral
mesons and their directional refraction indices will be determined
for various $(T,\mu,eB)$.
\par
As we have described in Sec. \ref{sec2}, our goal is to bring the
effective action of a two-flavor NJL model including
$(\sigma,\vec{\pi})$ mesons, in the form
\begin{eqnarray}\label{ND1b}
\lefteqn{\hspace{-0.6cm}\Gamma_{\mbox{\tiny{eff}}}[\sigma,\vec{\pi}]=\Gamma_{\mbox{\tiny{eff}}}[\sigma_{0}]}\nonumber\\
&&\hspace{-0.8cm}-\frac{1}{2}\int
d^{d}x~\bar{\sigma}(x)\left(M_{\sigma}^{2}+{\cal{G}}^{\mu\mu}\partial_{\mu}^{2}\right)\bar{\sigma}(x)\nonumber\\
&&\hspace{-0.8cm}-\frac{1}{2}\sum\limits_{\ell=1}^{3}\int
d^{d}x~\bar{\pi}_{\ell}(x)\left(M_{\vec{\pi}}^{2}+
{\cal{F}}^{\mu\mu}\partial_{\mu}^{2}\right)_{\ell\ell}\bar{\pi}_{\ell}(x),
\end{eqnarray}
which is valid in a truncation of the derivative expansion of the
full effective action $\Gamma_{\mbox{\tiny{eff}}}[\sigma,\vec{\pi}]$
up to two derivatives. According to (\ref{NN8}), the squared mass
matrices of neutral mesons, $\sigma$ and $\pi^{0}$, are given
by\footnote{Here, the third component of $\vec{\pi}$ is identified
with $\pi^{0}$, i.e. $\pi_{3}=\pi^{0}$.}
\begin{eqnarray}\label{ND2b}
M_{\sigma}^{2}&\equiv&-\int
d^{4}z\frac{\delta^{2}\Gamma_{\mbox{\tiny{eff}}}}{\delta\sigma(0)\delta\sigma(z)}\bigg|_{(\sigma_{0},{\mathbf{0}})},\nonumber\\
\hspace{-0.5cm}(M_{\vec{\pi}}^{2})_{33}&\equiv&-\int
d^{4}z\frac{\delta^{2}\Gamma_{\mbox{\tiny{eff}}}}{\delta\pi_{3}(0)\delta\pi_{3}(z)}\bigg|_{(\sigma_{0},{\mathbf{0}})},
\end{eqnarray}
and, according to (\ref{NN13}), the form factors of the effective
kinetic part of the effective action, corresponding to $\sigma$ and
$\pi^{0}$, read
\begin{eqnarray}\label{ND3b}
{\cal{G}}^{\mu\nu}&\equiv&-\frac{1}{2}\int
d^{4}z z^{\mu}z^{\nu}\frac{\delta^{2}\Gamma_{\mbox{\tiny{eff}}}^{k}}{\delta\sigma(0)\delta\sigma(z)}\bigg|_{(\sigma_{0},{\mathbf{0}})},\nonumber\\
({\cal{F}}^{\mu\nu})_{33}&\equiv&-\frac{1}{2}\int d^{4}z
z^{\mu}z^{\nu}\frac{\delta^{2}\Gamma_{\mbox{\tiny{eff}}}^{k}}{\delta\pi_{3}(0)\delta\pi_{3}(z)}\bigg|_{
(\sigma_{0},{\mathbf{0}})}\hspace{-0.7cm}.
\end{eqnarray}
To simplify our notations, we will denote in the rest of this paper,
the mass squared matrix $(M_{\vec{\pi}}^{2})_{33}$ from (\ref{ND2b})
corresponding to $\pi^{0}$ by $M_{\pi^{0}}^{2}$. Similarly,
$({\cal{F}}^{\mu\nu})_{33}$ will be denoted by ${\cal{F}}^{\mu\nu}$.
Whereas the mesons squared mass matrices at zero temperature and
chemical potential are given by plugging the effective action
(\ref{NE5b})-(\ref{NE8b}) in (\ref{ND2b}) and read
\begin{eqnarray}
M_{\sigma}^{2}&=&\frac{1}{2G}-i\int
d^{4}z\mbox{tr}_{sfc}\big[S_{Q}(z,0)S_{Q}(0,z)\big],\label{ND4b}\\
%%%%%%%%%%%%%%%%%%%%%%%%%%%
\hspace{-0.5cm}M_{\pi^{0}}^{2}&=&\frac{1}{2G}+i\int
d^{4}z\mbox{tr}_{sfc}\big[S_{Q}(z,0)\tau_{3}\gamma^{5}S_{Q}(0,z)\gamma^{5}\tau_{3}\big],\nonumber\\
\label{ND5b}
\end{eqnarray}
the form factors (\ref{ND3b}) arise by replacing
$\Gamma_{\mbox{\tiny{eff}}}^{k}$ with the one-loop effective
potential $\Gamma_{\mbox{\tiny{eff}}}^{(1)}$ from
(\ref{NE6b})-(\ref{NE8b}),
\begin{eqnarray}
{\cal{G}}^{\mu\nu}&=&-\frac{i}{2}\int
d^{4}z z^{\mu}z^{\nu}\mbox{tr}_{sfc}\big[S_{Q}(z,0)S_{Q}(0,z)\big],\label{ND6b}\\
{\cal{F}}^{\mu\nu}&=&\frac{i}{2}\int
d^{4}z z^{\mu}z^{\nu}\mbox{tr}_{sfc}\big[S_{Q}(z,0)\tau_{3}\gamma^{5}S_{Q}(0,z)\gamma^{5}\tau_{3}\big].\nonumber\\
\label{ND7b}
\end{eqnarray}
Similar expressions for ${\cal{G}}^{\mu\nu}$ and
${\cal{F}}^{\mu\nu}$ are also presented in
\cite{miransky1995,sadooghi2009} for a single-flavor NJL model. To
study the effect of very strong magnetic fields, the authors of
\cite{miransky1995, sadooghi2009} use the fermion propagator,
arising from Schwinger proper-time method \cite{schwinger1960}, in
the LLL approximation. In the present paper, however, we are
interested on the full $eB$ dependence of these coefficients for the
whole range of $eB\in [0,1]$ GeV$^{2}$, and have to consider, in
contrast to \cite{miransky1995, sadooghi2009}, the contributions of
higher Landau levels too. To do this, we use the Ritus fermion
propagator
\begin{eqnarray}\label{ND8b}
S_{Q}(x,y)&=&i\sum_{p=0}^{\infty}\int{\cal{D}}\tilde{p}~e^{-i\tilde{p}\cdot
(x-y)}\nonumber\\
&&\times P_{p}(x_{1})D_{Q}^{-1}(\bar{p})~P_{p}(y_{1}),
\end{eqnarray}
arising from the solution of Dirac equation in the presence of
uniform magnetic field using Ritus eigenfunction method. The same
expression for $S_{Q}(x,y)$ appears also in \cite{fukushima2009}. In
(\ref{ND8b}), $\tilde{p}\equiv (p_{0},0,p_{2},p_{3})$,
${\cal{D}}\tilde{p}\equiv \frac{dp_{0}dp_{2}dp_{3}}{(2\pi)^{3}}$,
and $P_{p}(x_{1})$ is given by
\begin{eqnarray}\label{ND9b}
\hspace{-0.2cm}P_{p}(x_{1})&=&\frac{1}{2}[f_{p}^{+s}(x_{1})+\Pi_{p}f_{p}^{-s}(x_{1})]
\nonumber\\
&&\hspace{-0.2cm}+\frac{is}{2}[f_{p}^{+s}(x_{1})-\Pi_{p}f_{p}^{-s}(x_{1})]
\gamma^{1}\gamma^{2},
\end{eqnarray}
where, $s\equiv \mbox{sgn}(QeB)$, and $\Pi_{p}\equiv 1-\delta_{p0}$
considers the spin degeneracy in the LLL.  The functions $f_{p}^{\pm
s}(x_{1})$ are defined by
\begin{eqnarray}\label{ND10b}
\begin{array}{rclcrcl}
f_{p}^{+s}(x_{1})&=&\phi_{p}\left(x_{1}-sp_{2}\ell_{B}^{2}\right),&&
p&=&0,1,2,\cdots,\nonumber\\
f_{p}^{-s}(x_{1})&=&\phi_{p-1}\left(x_{1}-sp_{2}\ell_{B}^{2}\right),&&
p&=&1,2,3,\cdots,
\end{array}
\hspace{-0.2cm}\nonumber\\
\end{eqnarray}
where $\phi_{p}(x)$ is a function of Hermite polynomials $H_{p}(x)$
in the form
\begin{eqnarray}\label{ND11b}
\phi_{p}(x)=a_{p}\exp\left(-\frac{x^{2}}{2\ell_{B}^{2}}\right)H_{p}\left(\frac{x}{\ell_{B}}\right).
\end{eqnarray}
Here, $a_{p}\equiv (2^{p}p!\sqrt{\pi}\ell_{B})^{-1/2}$ is the
normalization factor and $\ell_{B}\equiv |QeB|^{-1/2}$ is the
magnetic length. In (\ref{ND8b}),
$D_{Q}(\bar{p})\equiv\gamma\cdot\bar{p}_{Q}-m$, with the Ritus
four-momentum from (\ref{NE11b}). Note that since $Q$ is a $2\times
2$ matrix in the flavor space,  $f_{p}^{\pm s}$ and therefore
$P_{p}(x_{1})$ are matrices in the flavor space. In what follows, we
will first determine $(M_{\sigma}^{2}, M^{2}_{\pi^{0}})$ and
$({\cal{G}}^{\mu\nu}, {\cal{F}}^{\mu\nu})$ at zero $(T,\mu)$ and in
the presence of a constant magnetic field. We then introduce $T$ and
$\mu$ using standard replacements
\begin{eqnarray}\label{ND12b}
p_{0}=i(2n+1)\pi T-\mu,~~\mbox{and}~~\int\frac{dp_{0}}{2\pi}\to
iT\sum_{n},\nonumber\\
\end{eqnarray}
and present the result for $(M_{\sigma}^{2}, M^{2}_{\pi^{0}})$ and
$({\cal{G}}^{\mu\nu}, {\cal{F}}^{\mu\nu})$ at finite $(T,\mu,eB)$ up
to an integration over $p_{3}$-momentum and a summation over Landau
levels $p$.
%%%%%%%%%%%%%%%%%%%%%%%%%%%%%%%%%%%%%%%%%%%%
\subsection{$(M_{\sigma}^{2}, M_{\pi^{0}}^{2})$ at finite $(T,\mu,eB)$}\label{sec4p1}
%%%%%%%%%%%%%%%%%%%%%%%%%%%%%%%%%%%%%%%%%%%%
\subsubsection{$M_{\sigma}^{2}$ at finite $(T,\mu,eB)$}
%%%%%%%%%%%%%%%%%%%%%%%%%%%%%%%%%%%%%%%%%%%%
\par\noindent
To compute $M_{\sigma}^{2}$ from (\ref{ND4b}), we use the definition
of the Ritus fermion propagator (\ref{ND8b}), and arrive first at
\begin{eqnarray}\label{ND13b}
\lefteqn{M_{\sigma}^{2}=\frac{1}{2G}+i\sum_{q}\int
d^{4}z\sum\limits_{p,k=0}^{\infty}{\cal{D}}\tilde{p}~{\cal{D}}\tilde{k}~e^{-iz\cdot(\tilde{p}-\tilde{k})}
}\nonumber\\
&&
\times\mbox{tr}_{sc}\left(D_{q}^{-1}(\bar{p})P_{p}(0)K_{k}(0)D^{-1}_{q}(\bar{k})K_{k}(z_{1})P_{p}(z_{1})\right).
\nonumber\\
\end{eqnarray}
Here, the summation over $q\in\{\frac{2}{3},-\frac{1}{3}\}$ replaces
the trace in the flavor space, and in $D_{q}$, $q$ are the
eigenvalues of the charge matrix $Q=\mbox{diag}(2/3,-1/3)$. After
performing the integration over $z_{i}, i=0,2,3$, and using the
definition of $D_{q}^{-1}$ as well as the Ritus-momentum
(\ref{NE11b}), with $Q$ replaced by $q$, we get
\begin{eqnarray}\label{ND14b}
\lefteqn{M_{\sigma}^{2}=\frac{1}{2G}+3i\sum_{q}\sum\limits_{p,k=0}^{\infty}\int
\frac{dp_{0}dp_{3}}{(2\pi)^{3}}
}\nonumber\\
&&\times\int
dp_{2}\mbox{tr}_{s}\bigg[\frac{1}{\gamma\cdot\bar{p}_{q}-m}I_{pk}(p_{2},k_{2})\frac{1}{\gamma\cdot\bar{k}_{q}-m}
\nonumber\\
&&\qquad\qquad\times J^{(0)}_{kp}(k_{2},p_{2})
\bigg]\bigg|_{\tilde{k}=\tilde{p}},
\end{eqnarray}
where the factor $3$ behind the integral arises from the trace in
the color space using $\mbox{tr}_{c}(\mathbb{I}_{N_{c}\times
N_{c}})=3$, and two functions $I_{pk}$ and $J_{kp}^{(0)}$ in
(\ref{ND14b}) are given by
\begin{eqnarray}\label{ND15b}
I_{pk}(p_{2},k_{2})&\equiv& P_{p}(0)K_{k}(0),\nonumber\\
J_{kp}^{(0)}(k_{2},p_{2})&\equiv& \int
dz_{1}K_{k}(z_{1})P_{p}(z_{1}).
\end{eqnarray}
Here, $K_{k}(x_{1})$ is defined similar to $P_{p}(x_{1})$ from
(\ref{ND9b})
\begin{eqnarray}\label{ND16b}
\hspace{-0.5cm}K_{k}(x_{1})&=&\frac{1}{2}[g_{k}^{+s}(x_{1})+\Pi_{k}g_{k}^{-s}(x_{1})]\nonumber\\
&&\hspace{-0.3cm}+\frac{is}{2}[g_{k}^{+s}(x_{1})-\Pi_{k}g_{k}^{-s}(x_{1})]
\gamma^{1}\gamma^{2},
\end{eqnarray}
with $g^{\pm s}_{k}(x_{1})$ defined as in (\ref{ND10b}), with
$p_{2}$ replaced by $k_{2}$. Note that for $k_{2}=p_{2}$, which is
included in the condition $\tilde{k}=\tilde{p}$ in (\ref{ND14b}), we
have $g^{\pm s}_{k}|_{k_{2}=p_{2}}=f^{\pm s}_{k}$. In what follows,
we will first evaluate the integration over $z_{1}$ in
(\ref{ND15b}). Using then the orthonormality of the Hermite
polynomials appearing in $P_{p}$ from (\ref{ND9b}), the
$p_{2}$-integration can also be performed. We will eventually end
with an expression for $M_{\sigma}^{2}$, that includes only two
integrations over $p_{0}$ and $p_{3}$ momenta. To start, let us
first rewrite $I_{pk}$ and $J_{kp}^{(0)}$ from (\ref{ND15b}) using
the definition of $P_{p}(x_{1})$ from (\ref{ND9b}) and
(\ref{ND16b}). We get
\begin{eqnarray}\label{ND17b}
I_{pk}(p_{2},k_{2})&\equiv&
\alpha^{+}_{pk}(p_{2},k_{2})+is\gamma^{1}\gamma^{2}\alpha^{-}_{pk}(p_{2},k_{2}),\nonumber\\
J_{kp}^{(0)}(k_{2},p_{2})&\equiv&
A^{+(0)}_{kp}(k_{2},p_{2})+is\gamma^{1}\gamma^{2}A^{-(0)}_{kp}(k_{2},p_{2}),\nonumber\\
\end{eqnarray}
where
\begin{eqnarray}\label{ND18b}
\alpha_{pk}^{\pm}(p_{2},k_{2})&\equiv&\frac{1}{2}[f_{p}^{+s}(0)g_{k}^{+s}(0)\pm\Pi_{p}\Pi_{k}
f^{-s}_{p}(0)g^{-s}_{k}(0)], \nonumber\\
\lefteqn{\hspace{-1.6cm}A^{\pm(0)}_{kp}(k_{2},p_{2})
}\nonumber\\
&&\hspace{-2cm}\equiv\frac{1}{2}\int
dz_{1}[f_{p}^{+s}(z_{1})g_{k}^{+s}(z_{1})\pm\Pi_{p}\Pi_{k}
f^{-s}_{p}(z_{1})g^{-s}_{k}(z_{1})].\nonumber\\
\end{eqnarray}
In this way, the integration over $z_{1}$ in (\ref{ND15b}) reduces
to an integration over $z_{1}$ in $A^{\pm(0)}_{kp}(k_{2},p_{2})$.
The latter can be performed using
\begin{eqnarray}\label{ND19b}
\lefteqn{\hspace{-0.6cm}\int
dz_{1}f_{p}^{+s}(z_{1})g^{+s}_{k}(z_{1})
}\nonumber\\
&&\hspace{-1cm}=\frac{(-1)^{p}2^{k}a^{k-p}e^{-a^{2}}}{\sqrt{2^{k+p}
k! p!}}U\left(-p,1+k-p,2a^{2}\right),
\end{eqnarray}
where $a\equiv \frac{\ell_{B}(p_{2}-k_{2})}{2}$ and
$\ell_{B}=|qeB|^{-1/2}$, and $U(m,n,z)$ is the confluent
hypergeometric function of the second kind \cite{gradshteyn}. This
can, however, be simplified by implementing the condition
$k_{2}=p_{2}$, which is required in (\ref{ND14b}). In this case $a$
vanishes, and (\ref{ND19b}) therefore reduces to
\begin{eqnarray}\label{ND20b}
\int
dz_{1}f_{p}^{+s}(z_{1})g^{+s}_{k}(z_{1})\bigg|_{k_{2}=p_{2}}=\delta_{pk}.
\end{eqnarray}
Plugging this result in (\ref{ND18b}) and using
$\Pi_{p}^{2}=\Pi_{p}$, we arrive at
\begin{eqnarray}\label{ND21b}
A_{kp}^{\pm(0)}(p_{2},k_{2}=p_{2})=\frac{1}{2}\left(1\pm\Pi_{p}\right)\delta_{pk}.
\end{eqnarray}
Plugging further (\ref{ND17b}) in (\ref{ND14b}), and performing the
traces over the $\gamma$-matrices, using
$\mbox{tr}_{s}(\gamma_{\mu}\gamma_{\nu})=4g_{\mu\nu}$ and
$\mbox{tr}_{s}\left(\gamma_{\mu}\gamma_{\nu}\gamma_{\rho}\gamma_{\sigma}\right)=
4\left(g_{\mu\nu}g_{\rho\sigma}-g_{\mu\rho}g_{\nu\sigma}+g_{\mu\sigma}g_{\nu\rho}\right)$,
the $\sigma$-meson squared mass matrix is given by
\begin{eqnarray}\label{ND22b}
\lefteqn{M_{\sigma}^{2}=\frac{1}{2G}+12i\sum_{q}\sum\limits_{p,k=0}^{\infty}\int
\frac{dp_{0}dp_{3}}{(2\pi)^{3}}
}\nonumber\\
&&\times\int
dp_{2}\left\{\frac{\left(\alpha_{pk}^{+}A_{kp}^{+(0)}+\alpha_{pk}^{-}A_{kp}^{-(0)}\right)\left(m^{2}+\bar{p}_{q}\cdot
\bar{k}_{q}\right)}{(\bar{p}^{2}_{q}-m^{2})(\bar{k}^{2}_{q}-m^{2})}\right.\nonumber\\
&&~~~~\left.+\frac{2\bar{p}_{2}\bar{k}_{2}\alpha_{pk}^{-}A_{kp}^{-(0)}}{(\bar{p}^{2}_{q}-m^{2})(\bar{k}^{2}_{q}-m^{2})}\right\}
\bigg|_{\tilde{p}=\tilde{k}}.
\end{eqnarray}
Here, $\bar{p}_{q}^{2}=p_{0}^{2}-2|qeB|p-p_{3}^{2}$ and for
$\tilde{p}=\tilde{k}$,
$\bar{k}_{q}^{2}=p_{0}^{2}-2|qeB|k-p_{3}^{2}$. To perform the
integration over $p_{2}$, we first compute
\begin{eqnarray}\label{ND23b}
W_{pk}^{(0)}\equiv\int dp_2 f_p^{+s}(0)f_k^{+s}(0).
\end{eqnarray}
This can be done using the definition of $f^{+s}_{p}(0)$ in terms of
Hermite polynomials [see (\ref{ND10b}) and (\ref{ND11b})], and their
orthonormality relation
\begin{eqnarray}\label{ND24b}
\int_{-\infty}^{+\infty}d\ell~
e^{-\ell^{2}}H_{p}(\ell)H_{k}(\ell)=\frac{\delta_{pk}}{\ell_{B}a_{k}^{2}},
\end{eqnarray}
leading to
\begin{eqnarray}\label{ND25b}
W_{pk}^{(0)}&=&\frac{a_p a_k}{\ell_B}(-1)^{p+k}\int dp'_2 e^{-
p'^2_2}
H_p(p'_2)H_k(p'_2)\nonumber\\
&=&\frac{\delta_{pk}}{\ell_B^2},
\end{eqnarray}
with $p'_{2}\equiv \ell_{B}p_{2}$. Moreover, we arrive at the useful
relation
\begin{eqnarray}\label{ND26b}
\hspace{-3.5cm}\int
dp_{2}\alpha^{\pm}_{pk}(p_{2},k_{2})A_{kp}^{\pm(0)}(k_{2},p_{2})\big|_{k_{2}=p_{2}}
\nonumber\\
\hspace{2.5cm}=\frac{\delta_{pk}}{4\ell_{B}^{2}}\left(1\pm\Pi_{p}\right)^{2},
\end{eqnarray}
arising from (\ref{ND23b}). Plugging these results in (\ref{ND22b})
and summing over $k$, the $\sigma$-meson squared mass matrix at zero
temperature, chemical potential and non-vanishing magnetic field is
given by
\begin{eqnarray}\label{ND27b}
\lefteqn{M_{\sigma}^{2}=\frac{1}{2G}
}\nonumber\\
&&+6i\sum\limits_{q\in\{\frac{2}{3},-\frac{1}{3}\}}|qeB|\sum\limits_{p=0}^{\infty}\alpha_{p}\int
\frac{dp_{0}dp_{3}}{(2\pi)^{3}}~\frac{(\bar{p}_{q}^{2}+m^{2})}{(\bar{p}_{q}^{2}-m^{2})^{2}},\nonumber\\
\end{eqnarray}
where $\alpha_{p}\equiv 1+\Pi_{p}$ is the same spin degeneracy
factor that appears in (\ref{NE14b}). To introduce the temperature
$T$ and the chemical potential $\mu$, we use the method described at
the beginning of this section [see (\ref{ND12b})]. The mass squared
matrix corresponding to $\sigma$-meson at finite $(T,\mu,eB)$ is
therefore given by
\begin{eqnarray}\label{ND28b}
\lefteqn{\hspace{-1cm}M_{\sigma}^{2}=\frac{1}{2G}-6\sum\limits_{q\in\{\frac{2}{3},-\frac{1}{3}\}}|qeB|
}\nonumber\\
&&\hspace{-1.2cm}\times
\sum_{p=0}^{\infty}\alpha_{p}\int\frac{dp_{3}}{(2\pi)^{2}}
\big[{\cal{S}}_{1}^{(0)}(\omega_{p})+2m^{2}{\cal{S}}_{2}^{(0)}(\omega_{p})\big],
\end{eqnarray}
where $\omega_{p}^{2}\equiv p_{3}^{2}+2|qeB|p+m^{2}$, and
${\cal{S}}_{\ell}^{(0)}(\omega_{p}), \ell=1,2$ are defined by
\begin{eqnarray}\label{ND29b}
{\cal{S}}_{\ell}^{(m)}(\omega_{p})\equiv
T\sum\limits_{n=-\infty}^{+\infty}\frac{(p_{0}^{2})^{m}}{(p_{0}^{2}-\omega_{p}^{2})^{\ell}},
\end{eqnarray}
with $\ell\geq 1, m\geq 0$. Using
\begin{eqnarray}\label{ND30b}
{\cal{S}}_{1}^{(0)}(\omega_{p})=
\frac{1}{2\omega_{p}}[1-N_{f}(\omega_{p})],
\end{eqnarray}
and assuming that ${\cal{S}}_{0}^{(m)}=0, \forall m\geq 0$,
following recursion relations can be used to evaluate
${\cal{S}}_{\ell}^{(m)}(\omega_{p})$ from (\ref{ND29b}) for all
$\ell\geq 1$ and $m\geq 0$,
\begin{eqnarray}\label{ND31b}
{\cal{S}}_{\ell}^{(0)}(\omega_{p})&=&\frac{1}{2(\ell-1)\omega_{p}}\frac{d{\cal{S}}_{\ell-1}^{(0)}(\omega_{p})}
{d\omega_{p}},~~~\forall\ell\geq 2,\nonumber\\
{\cal{S}}_{\ell}^{(m)}(\omega_{p})&=&{\cal{S}}_{\ell-1}^{(m-1)}(\omega_{p})+\omega_{p}^{2}{\cal{S}}_{\ell}^{(m-1)}(\omega_{p}).
\end{eqnarray}
In (\ref{ND30b}), $N_{f}(\omega_{p})\equiv
n_{f}^{+}(\omega_{p})+n_{f}^{-}(\omega_{p})$ and
$n_{f}^{\pm}(\omega_{p})$ are fermionic distribution functions
\begin{eqnarray}\label{ND32b}
n_{f}^{\pm}(\omega_{p})\equiv
\frac{1}{e^{\beta(\omega_{p}\mp\mu)}+1}.
\end{eqnarray}
In the following paragraph, the same method will be used to
determine $M_{\pi^{0}}^{2}$ at zero and nonzero $(T,\mu)$ and for
non-vanishing $eB$.
%%%%%%%%%%%%%%%%%%%%%%%%%%%%%%%%%%%%%%%%%%%%
\subsubsection{$M_{\pi^{0}}^{2}$ at finite $(T,\mu,eB)$}
%%%%%%%%%%%%%%%%%%%%%%%%%%%%%%%%%%%%%%%%%%%%
\par\noindent
To determine the squared mass matrix $M_{\pi^{0}}^{2}$ from
(\ref{ND5b}), corresponding to $\pi^{0}$, we use the definition of
the fermion propagator (\ref{ND8b})-(\ref{ND9b}), and arrive first
at
\begin{eqnarray}\label{ND33b}
\lefteqn{M_{\pi^{0}}^{2}=\frac{1}{2G} -i\int
d^{4}z\sum\limits_{p,k=0}^{\infty}\int{\cal{D}}\tilde{p}~{\cal{D}}\tilde{k}~e^{-iz\cdot(\tilde{p}-\tilde{k})}
}\nonumber\\
&&\times
~\mbox{tr}_{sfc}\bigg[D_{Q}^{-1}(\bar{p})P_{p}(0)\tau_{3}\gamma_{5}K_{k}(0)D^{-1}_{Q}(\bar{k})\nonumber\\
&&\hspace{1.5cm}\times K_{k}(z_{1})
\gamma_{5}\tau_{3}P_{p}(z_{1})\bigg].
\end{eqnarray}
Using the anticommutation relation $\{\gamma_{5},\gamma_{\mu}\}=0$
leading to $[\gamma_{5},K_{k}]=0$, we simplify first the combination
$\gamma_{5}K_{k}(0)D^{-1}_{Q}(\bar{k})K_{k}(z_{1})\gamma_{5}$ in
(\ref{ND33b}), and arrive at
\begin{eqnarray}\label{ND34b}
\gamma_{5}K_{k}(0)D^{-1}_{Q}(\bar{k})K_{k}(z_{1})\gamma_{5}
=-K_{k}(0)\frac{1}{\gamma\cdot \bar{k}+m}K_{k}(z_{1}). \hspace{-0.5cm}\nonumber\\
\end{eqnarray}
Plugging this relation in (\ref{ND33b}) and performing the
integration over $z_{i}, i=0,2,3$, we arrive at
\begin{eqnarray}\label{ND35b}
\lefteqn{M_{\pi^{0}}^{2}=\frac{1}{2G}+3i
}\nonumber\\
&&\times\sum\limits_{p,k=0}^{\infty}\int
\frac{dp_{0}dp_{3}}{(2\pi)^{3}}\int
dp_{2}~\mbox{tr}_{fs}\left\{\frac{1}{\gamma\cdot\bar{p}-m}I_{pk}(p_{2},k_{2})\right.\nonumber\\
&&\left.\times \tau_{3}
\frac{1}{\gamma\cdot\bar{k}+m}\tau_{3}J^{(0)}_{kp}(k_{2},p_{2})
\right\}\bigg|_{\tilde{k}=\tilde{p}},
\end{eqnarray}
where $I_{pk}(p_{2},k_{2})$ and $J_{kp}^{(0)}(k_{2},p_{2})$ are
given in (\ref{ND15b}). We follow the same method leading from
(\ref{ND14b}) to (\ref{ND27b}) to evaluate the traces over the
$\gamma$-matrices and to perform the integrations over $z_{1}$ and
$p_{2}$ in (\ref{ND35b}). We arrive after a lengthy but
straightforward computation at
\begin{eqnarray}\label{ND36b}
M_{\pi^{0}}^{2}&=&\frac{1}{2G}+6i\sum\limits_{q\in\{\frac{2}{3},-\frac{1}{3}\}}
|qeB|\nonumber\\
&&\times\sum\limits_{p=0}^{\infty}\alpha_{p}\int\frac{dp_{0}dp_{3}}{(2\pi)^{3}}\frac{1}{(\bar{p}_{q}^{2}-m^{2})}.
\end{eqnarray}
Thus, the mass squared matrix corresponding to $\pi^{0}$ at finite
$(T,\mu,eB)$ is given by
\begin{eqnarray}\label{ND37b}
M_{\pi^{0}}^{2}&=&\frac{1}{2G}-6\sum_{q\in\{\frac{2}{3},-\frac{1}{3}\}}|qeB|\nonumber\\
&&\times\sum\limits_{p=0}^{\infty}\alpha_{p}
\int\frac{dp_{3}}{(2\pi)^{2}}{\cal{S}}_{1}^{(0)}(\omega_{p}),
\end{eqnarray}
where ${\cal{S}}_{1}^{(0)}(\omega_{p})$ is given in (\ref{ND30b}).
In Sec. \ref{sec5}, the integration over $p_{3}$ and the summation
over Landau level $p$, appearing in (\ref{ND37b}), will be performed
numerically.
%%%%%%%%%%%%%%%%%%%%%%%%%%%%%%%%%%%%%%%%%%%%%%%%%%%%%%%%%%%%%%%%%%%%%%%%%%%%
\subsection{$({\cal{G}}^{\mu\nu}, {\cal{F}}^{\mu\nu})$ at finite $(T,\mu,eB)$}\label{sec4p2}
%%%%%%%%%%%%%%%%%%%%%%%%%%%%%%%%%%%%%%%%%%%%%%%%%%%%%%%%%%%%%%%%%%%%%%%%%%%%
\subsubsection{${\cal{G}}^{\mu\nu}$ at finite $(T,\mu,eB)$}
%%%%%%%%%%%%%%%%%%%%%%%%%%%%%%%%%%%%%%%%%%%%%%%%%%%%%%%%%%%%%%%%%%%%%%%%%%%%
\par\noindent
We start by computing ${\cal{G}}^{\mu\nu}$ from (\ref{ND6b}) at zero
$(T,\mu)$ but non-vanishing $eB$. To do this, we use the definition
of the Ritus propagator (\ref{ND8b}), and arrive first at
\begin{eqnarray}\label{ND38b}
\lefteqn{{\cal{G}}^{\mu\nu}=\frac{i}{2}\sum_{q}\int
d^{4}z~z^{\mu}z^{\nu}\sum\limits_{p,k=0}^{\infty}\int{\cal{D}}\tilde{p}~{\cal{D}}\tilde{k}~e^{-iz\cdot(\tilde{p}-\tilde{k})}
}\nonumber\\
&&\times\mbox{tr}_{sc}\bigg[D_{q}^{-1}(\bar{p})P_{p}(0)K_{k}(0)D^{-1}_{q}(\bar{k})K_{k}(z_{1})P_{p}(z_{1})\bigg].\nonumber\\
\end{eqnarray}
After performing the integration over $z_{i}, i=0,2,3$, and using
the definition of $D^{-1}_{q}$, the diagonal elements of
${\cal{G}}^{\mu\nu}$ are given by
\begin{eqnarray}\label{ND39b}
\lefteqn{{\cal{G}}^{jj}=-\frac{i}{2}\sum_{q}\sum_{k,r=0}^{\infty}\int
{\cal{D}}\tilde{k}~\frac{\partial^{2}}{\partial\ell_{j}^{2}}
\mbox{tr}_{sc}\bigg[\frac{1}{(\gamma\cdot
\bar{r}_{q}-m)}
}\nonumber\\
&&\times
I_{rk}(r_{2},k_{2})\frac{1}{(\gamma\cdot\bar{k}_{q}-m)}J_{kr}^{(0)}
(k_{2},r_{2})\bigg]\Bigg|_{ \tilde{\ell}=0
},\nonumber\\
\lefteqn{{\cal{G}}^{11}=+\frac{i}{2}\sum_{q}\sum_{k,p=0}^{\infty}\int
{\cal{D}}\tilde{p}~\mbox{tr}_{sc}\bigg[\frac{1}{(\gamma\cdot
\bar{p}_{q}-m)}
}\nonumber\\
&&\times I_{pk}(p_{2},k_{2})\frac{1}{(\gamma\cdot\bar{k}_{q}-m)}J_{kp}^{(2)}(k_{2},p_{2})\bigg]\Bigg|_{\tilde{k}=\tilde{p}},\nonumber\\
\lefteqn{
{\cal{G}}^{22}=-\frac{i}{2}\sum_{q}\sum_{k,p=0}^{\infty}\int
{\cal{D}}\tilde{p}~\mbox{tr}_{sc}\bigg[\frac{1}{(\gamma\cdot
\bar{p}_{q}-m)}
}\nonumber\\
&&\times\frac{\partial^{2}}{\partial
p_{2}^{2}}\bigg[I_{pk}(p_{2},k_{2})\frac{1}{(\gamma\cdot\bar{k}_{q}-m)}J_{kp}^{(0)}(k_{2},p_{2})\bigg]\bigg]
\Bigg|_{\tilde{k}=\tilde{p}}.\nonumber\\
\end{eqnarray}
In ${\cal{G}}^{jj}$, $j=0,3$, $\bar{r}_{q}\equiv
\bar{k}_{q}+\bar{\ell}_{q}$ and $r\equiv k+\ell$. Moreover, two
functions $I_{pk}(p_{2},k_{2})$ and $J^{(0)}_{kp}(k_{2},p_{2})$ are
defined in (\ref{ND15b}), and $J^{(2)}_{kp}(k_{2},p_{2})$ is defined
by
\begin{eqnarray}\label{ND40b}
J_{kp}^{(2)}(k_{2},p_{2})\equiv \int dz_{1} z_{1}^{2}
K_{k}(z_{1})P_{p}(z_{1}),
\end{eqnarray}
with $P_{p}$ and $K_{k}$ given in (\ref{ND9b}) and (\ref{ND16b}),
respectively. Following the method presented in the first part of
this section, leading from (\ref{ND14b}) to (\ref{ND27b}), all
non-diagonal elements of ${\cal{G}}^{\mu\nu}$ turn out to vanish,
and therefore, as it is claimed in Sec. \ref{sec2},
${\cal{G}}^{\mu\nu}={\cal{G}}^{\mu\mu}g^{\mu\nu}$ (no summation over
$\mu$). This is similar to what also happens in the single-flavor
NJL model \cite{miransky1995}. We therefore focus on
${\cal{G}}^{\mu\mu}, \mu=0,\cdots,3$ from (\ref{ND39b}), which shall
be evaluated using the same method as before. Evaluating the
$\bar{k}$-integration in ${\cal{G}}^{jj}, j=0,3$ from (\ref{ND39b}),
using an additional Feynman parametrization, we arrive first at
\begin{eqnarray}\label{ND41b}
\lefteqn{
{\cal{G}}^{00}=-{\cal{G}}^{33}=3i\sum\limits_{q\in\{\frac{2}{3},
-\frac{1}{3}\}}|qeB|
}\nonumber\\
&&\times\sum\limits_{p=0}\alpha_{p}\int
\frac{dp_{0}dp_{3}}{(2\pi)^{3}}\left\{\frac{1}{(\bar{p}_{q}^{2}-m^{2})^{2}}+
\frac{4}{3}\frac{m^{2}}{(\bar{p}_{q}^{2}-m^{2})^{3}}
\right\}.\nonumber\\
\end{eqnarray}
At finite $(T,\mu, eB)$, we therefore have
\begin{eqnarray}\label{ND42b}
\lefteqn{
{\cal{G}}^{00}=-{\cal{G}}^{33}=-3\sum\limits_{q\in\{\frac{2}{3},-\frac{1}{3}\}}|qeB|}\nonumber\\
&&\times\sum_{p=0}^{\infty}\alpha_{p}\int\frac{dp_{3}}{(2\pi)^{2}}
\left\{{\cal{S}}_{2}^{(0)}(\omega_{p})
+\frac{4}{3}m^{2}{\cal{S}}_{3}^{(0)}(\omega_{p})\right\}.\nonumber\\
\end{eqnarray}
To determine ${\cal{G}}^{11}$ from (\ref{ND39b}), we shall first
evaluate the $z_{1}$ integration in $J_{kp}^{(2)}(k_{2},p_{2})$ from
(\ref{ND40b}) at $k_{2}=p_{2}$, as it is required from
(\ref{ND39b}). To do this, we first define
\begin{eqnarray}\label{ND43b}
J_{kp}^{(2)}(k_{2},p_{2})\equiv
A_{kp}^{+(2)}(k_{2},p_{2})+i\gamma^{1}\gamma^{2}sA_{kp}^{-(2)}(k_{2},p_{2}),\nonumber\\
\end{eqnarray}
with
\begin{eqnarray}\label{ND44b}
\lefteqn{\hspace{-1.3cm}A_{kp}^{\pm(2)}(k_{2},p_{2}) }\nonumber\\
&&\hspace{-1.5cm}\equiv
\frac{1}{2}\bigg[{\cal{L}}_{kp}(k_{2},p_{2})\pm
\Pi_{p}\Pi_{k}{\cal{L}}_{k-1,p-1}(k_{2},p_{2})\bigg].
\end{eqnarray}
Here, ${\cal{L}}_{kp}(k_{2},p_{2})$ is defined by
\begin{eqnarray}\label{ND45b}
\hspace{-.2cm}{\cal{L}}_{kp}(k_{2},p_{2})\equiv\int
dz_{1}~[z_{1}g_{k}^{+s}(z_{1})][z_{1}f_{p}^{+s}(z_{1})].
\end{eqnarray}
To determine ${\cal{L}}_{kp}$ for $k_{2}=p_{2}$, we use the
definition of $f^{+s}_{p}$ from (\ref{ND10b}) in terms of the
Hermite polynomials $H_{p}$ and their standard recursion relations $
\frac{d H_{k}(x)}{dx}=2k H_{k-1}(x)$ and $H_{k+1}(x)=2x H_{k}(x)-2
kH_{k-1}(x)$, to arrive first at
\begin{eqnarray}\label{ND46b}
\lefteqn{z_{1}f_{p}^{+s}(z_{1})}\nonumber\\
&&=\ell_{B}\left(C_{p+1}f_{p+1}^{+s}(z_{1})+C_{p}f_{p-1}^{+s}(z_{1})+p'_{2}f_{p}^{+s}(z_{1})\right),\nonumber\\
\end{eqnarray}
where $C_{p}\equiv \sqrt{\frac{p}{2}}$ and
$p'_{2}\equiv\ell_{B}p_{2}$. Replacing (\ref{ND46b}) in
(\ref{ND45b}), setting $k_{2}=p_{2}$, and integrating over $z_{1}$,
we get
\begin{eqnarray}\label{ND47b}
\lefteqn{{\cal{L}}_{kp}(k_{2}=p_{2},p_{2})=\ell_{B}^{2}\bigg[\left(C_{2p+1}^{2}+p_{2}^{'2}\right)\delta_{kp}
}\nonumber\\
&&+
C_{p}C_{p-1}\delta_{k,p-2}+C_{p+1}C_{p+2}\delta_{k,p+2}+2p'_{2}\left(C_{p}\delta_{k,p-1}\right.\nonumber\\
&&\left.+C_{p+1}\delta_{k,p+1}\right)\bigg].
\end{eqnarray}
Thus, $A_{kp}^{\pm(2)}(p_{2},p_{2})$ in (\ref{ND43b}) are given by
\begin{eqnarray}\label{ND48b}
\lefteqn{\hspace{-1cm}A_{kp}^{\pm(2)}(k_{2}=p_{2},p_{2})=\frac{\ell_{B}^{2}}{2}\bigg[C^{\pm}\delta_{kp}}\nonumber\\
&&\qquad+C_{p-1}\left(C_{p}\pm \Pi_{p}\Pi_{k}
C_{p-2}\right)\delta_{k,p-2}\nonumber\\
&&\qquad +C_{p+1}\left(C_{p+2}\pm
\Pi_{p}\Pi_{k}C_{p}\right)\delta_{k,p+2}
\nonumber\\
&&\qquad+2p'_{2}\left( C_{p}\pm \Pi_{p}\Pi_{k}
C_{p-1}\right)\delta_{k,p-1}\nonumber\\
&&\qquad+2p'_{2}\left(C_{p+1}\pm \Pi_{p}\Pi_{k}
C_{p}\right)\delta_{k,p+1}\bigg],
\end{eqnarray}
where the coefficients $C^{\pm}\equiv
D^{\pm}+p_{2}^{'2}(1\pm\Pi_{p})$ with $D^{\pm}\equiv C_{2p+1}^{2}\pm
C_{2p-1}^{2}\Pi_{p}$. Plugging (\ref{ND48b}) in (\ref{ND43b}) and
the resulting expression in ${\cal{G}}^{11}$ from (\ref{ND39b}), and
performing the trace over $\gamma$-matrices, we arrive at
\begin{eqnarray}\label{ND49b}
\lefteqn{{\cal{G}}^{11}=6i\sum_{q}\sum\limits_{p,k=0}^{\infty}\int
\frac{dp_{0}dp_{3}}{(2\pi)^{3}} }\nonumber\\
&&\times \int
dp_{2}\left\{\frac{\left(\alpha_{pk}^{+}A_{kp}^{+(2)}+\alpha_{pk}^{-}A_{kp}^{-(2)}\right)\left(\bar{p}_{q}\cdot
\bar{k}_{q}+m^{2}\right)}{(\bar{p}_{q}^{2}-m^{2})(\bar{k}_{q}^{2}-m^{2})}
\right.\nonumber\\
&&~~~~+\left.\frac{2\bar{p}_{2}\bar{k}_{2}\alpha_{pk}^{-}A_{kp}^{-(2)}}{(\bar{p}_{q}^{2}-m^{2})(\bar{k}_{q}^{2}-m^{2})}
\right\}\bigg|_{\tilde{k}=\tilde{p}},
\end{eqnarray}
where $\alpha^{\pm}_{pk}$ are defined in (\ref{ND18b}). The
integration over $p_{2}$ is then performed using
\begin{eqnarray}\label{ND50b}
W^{(1)}_{pk}&\equiv&\int
dp_{2}p'_{2}f_{p}^{+s}(0)f_{k}^{+s}(0)\nonumber\\
&=&-\frac{1}{\ell_{B}^{2}}\left(C_{p+1}\delta_{k,p+1}+C_{p}\delta_{k,p-1}\right),\nonumber\\
W^{(2)}_{pk}&\equiv&\int
dp_{2}p^{'2}_{2}f_{p}^{+s}(0)f_{k}^{+s}(0)\nonumber\\
&=&+\frac{1}{\ell_{B}^{2}}\left(C_{2p+1}^{2}\delta_{kp}+C_{p+2}C_{p+1}
\delta_{k,p+2}\right.\nonumber\\
&&\left.+C_{p}C_{p-1}\delta_{k,p-2}\right).
\end{eqnarray}
These results arise from the orthonormality relations of the Hermite
polynomials (\ref{ND24b}), in the same way that $W^{(0)}_{pk}$ from
(\ref{ND23b}) is derived. Using $W_{pk}^{(1)}$ and $W_{pk}^{(2)}$
from (\ref{ND50b}), we get
\begin{eqnarray}\label{ND51b}
\lefteqn{\hspace{-0.5cm}\int dp_{2}\
\alpha^{\pm}_{pk}A^{\pm(2)}_{kp}|_{\tilde{k}=\tilde{p}}=\frac{1}{4}\bigg[
\left(D^{\pm}+C_{2p+1}^{2}\pm\Pi_{p}C_{2p-1}^{2}\right)}\nonumber\\
&&\times(1\pm\Pi_{p})\delta_{kp}-2\left(C_{p}\pm\Pi_{p}\Pi_{k}C_{p-1}\right)^{2}\delta_{k,p-1}\nonumber\\
&&-2\left(C_{p+1}\pm\Pi_{p}\Pi_{k}C_{p}\right)^{2}
\delta_{k,p+1}\bigg].
\end{eqnarray}
Plugging these relations in (\ref{ND49b}), we finally arrive at
\begin{eqnarray}\label{ND52b}
\lefteqn{{\cal{G}}^{11}=3i
}\nonumber\\
&&\times
\sum\limits_{q}\sum_{p,k=0}^{\infty}\int\frac{dp_{0}dp_{3}}{(2\pi)^{3}}
\bigg\{\frac{\big(\bar{p}_{q}\cdot\bar{k}_{q}+m^{2}\big)}{(\bar{p}_{q}^{2}-m^{2})(\bar{k}_{q}^{2}-m^{2})}C^{(1)}_{pk}\nonumber\\
&&-\frac{\bar{p}_{2}\bar{k}_{2}}
{(\bar{p}_{q}^{2}-m^{2})(\bar{k}_{q}^{2}-m^{2})}C^{(2)}_{pk}
\bigg\}\bigg|_{\tilde{k}=\tilde{p}},
\end{eqnarray}
where
\begin{eqnarray}\label{ND53b}
\lefteqn{C^{(1)}_{pk}\equiv[(2p+1)+\Pi_{p}(2p-1)]\delta_{kp}}\nonumber\\
&&\hspace{-0.3cm}-[p+\Pi_{p}\Pi_{k}(p-1)]\delta_{k,p-1}-[(p+1)+\Pi_{p}\Pi_{k}p]\delta_{k,p+1},\nonumber\\
\lefteqn{C^{(2)}_{pk}\equiv-[(2p+1)-(2p-1)\Pi_{p}](1-\Pi_{p})\delta_{pk}}\nonumber\\
&&\hspace{-0.3cm}+[p+\Pi_{p}\Pi_{k}(p-1-2\sqrt{p(p-1)})]\delta_{k,p-1}\nonumber\\
&&\hspace{-0.3cm}+[(p+1)+\Pi_{p}\Pi_{k}(p-2\sqrt{p(p+1)})]\delta_{k,p+1}.
\end{eqnarray}
To determine ${\cal{G}}^{22}$ from (\ref{ND39b}), we perform the
traces over the $\gamma$-matrices and arrive first at
\begin{eqnarray}\label{ND54b}
\lefteqn{{\cal{G}}^{22}=-6i\sum\limits_{q\in\{\frac{2}{3},-\frac{1}{3}\}}\sum_{p,k=0}^{\infty}
\int \frac{dp_0 dp_3}{(2\pi)^3}
}\nonumber\\
&&\times\int
dp_2\left\{\frac{\big(\bar{p}_{q}\cdot\bar{k}_{q}+m^{2}\big)N^{(1)}_{pk}+2\bar{p}_{2}\bar{k}_{2}N^{(2)}_{pk}}
{(\bar{p}_{q}^{2}-m^{2})(\bar{k}_{q}^{2}-m^{2})}\right\}
\bigg|_{\tilde{k}=\tilde{p}},\nonumber\\
\end{eqnarray}
where
\begin{eqnarray*}
N^{(1)}_{pk}(p_{2},k_{2})&\equiv&\frac{d^{2}}{dp_{2}^{2}}\left(\alpha_{pk}^{+}A_{kp}^{+(0)}+\alpha_{pk}^{-}A_{kp}^{-(0)}\right)
,\nonumber\\
N^{(2)}_{pk}(p_{2},k_{2})&\equiv&\frac{d^{2}}{dp_{2}^{2}}\left(\alpha_{pk}^{-}A_{kp}^{-(0)}\right).
\end{eqnarray*}
Plugging the definitions of  $\alpha_{pk}^{\pm}(p_{2},k_{2})$ and
$A_{kp}^{\pm(0)}(p_{2},k_{2})$ from (\ref{ND18b}) in (\ref{ND54b}),
and performing the integration over $p_{2}$ in (\ref{ND54b}) by
making use of $W^{(0)}_{pk}$ from (\ref{ND25b}), we arrive after a
lengthy but straightforward computation at
\begin{eqnarray}\label{ND55b}
\int
dp_{2}N^{(1)}_{pk}(p_{2},k_{2}=p_{2})&=&-\frac{1}{2}C_{pk}^{(1)},\nonumber\\
\int
dp_{2}N^{(2)}_{pk}(p_{2},k_{2}=p_{2})&=&\frac{1}{4}C_{pk}^{(2)},
\end{eqnarray}
where $C^{(1)}_{pk}$ and $C^{(2)}_{pk}$ are given in (\ref{ND53b}).
This leads eventually to
\begin{eqnarray}\label{ND56b}
{\cal{G}}^{22}={\cal{G}}^{11},
\end{eqnarray}
with ${\cal{G}}^{11}$ given in (\ref{ND52b}). Note that the equality
${\cal{G}}^{11}={\cal{G}}^{22}$ arises also in a single-flavor NJL
model in \cite{miransky1995}, where the form factors of the
effective kinetic term are computed at zero temperature and chemical
potential and in the regime of LLL dominance. In (\ref{ND56b}), this
regime is characterized by $k=p=0$, where $k$ and $p$ label the
Landau levels. At finite $(T,\mu)$, ${\cal{G}}^{11}={\cal{G}}^{22}$
is therefore given by
\begin{eqnarray}\label{ND57b}
\lefteqn{\hspace{-0.8cm}{\cal{G}}^{11}={\cal{G}}^{22}=-3\sum\limits_{q\in\{\frac{2}{3},-\frac{1}{3}\}}\sum_{p=0}^{\infty}
\int\frac{dp_3}{(2\pi)^{2}}
\left\{8m^{2}p{\cal{S}}_{2}^{(0)}(\omega_{p}) \right.}\nonumber\\
&&\left.
+2[(2p+1)\ell_{B}^{2}m^{2}+p]~{\cal{S}}_{1}^{(0)}(\omega_{p})\right.\nonumber\\
&&\left.-2[(2p+1)\ell_{B}^{2}m^{2}+(p+1)]~{\cal{S}}_{1}^{(0)}(\omega_{p+1})\right.\nonumber\\
&&\left.+\delta_{p0}[{\cal{S}}_{1}^{(0)}(\omega_{p})
+2m^{2}{\cal{S}}_{2}^{(0)}(\omega_{p})]\right\},
\end{eqnarray}
where ${\cal{S}}_{1}^{(0)}(\omega_{p})$ is given in (\ref{ND30b})
and ${\cal{S}}_{2}^{(0)}(\omega_{p})$ can be evaluated using the
recursion relations (\ref{ND31b}).
%%%%%%%%%%%%%%%%%%%%%%%%%%%%%%%%%%%%%%%%%%%%%%%%%%%%%%%%%%%%
\subsubsection{${\cal{F}}^{\mu\nu}$ at finite $(T,\mu,eB)$}
%%%%%%%%%%%%%%%%%%%%%%%%%%%%%%%%%%%%%%%%%%%%%%%%%%%%%%%%%%%%
\par\noindent
We start the computation of the elements of the matrix
${\cal{F}}^{\mu\nu}$ by considering its definition from
(\ref{ND7b}), and arrive after plugging the Ritus propagator
(\ref{ND8b}) in (\ref{ND7b}) at
\begin{eqnarray}\label{ND58b}
\lefteqn{\hspace{-0.8cm}{\cal{F}}^{\mu\nu}=-\frac{i}{2}\int
d^{4}z~z^{\mu}z^{\nu}\sum\limits_{p,k=0}^{\infty}\int{\cal{D}}\tilde{p}~{\cal{D}}\tilde{k}~e^{-iz\cdot(\tilde{p}-\tilde{k})}
}\nonumber\\
&&\times
\mbox{tr}_{sfc}\bigg[D_{Q}^{-1}(\bar{p})P_{p}(0)\tau_{3}\gamma_{5}K_{k}(0)D^{-1}_{Q}(\bar{k})\nonumber\\
&&\qquad~~~~\times K_{k}(z_{1})
\gamma_{5}\tau_{3}P_{p}(z_{1})\bigg].
\end{eqnarray}
Using (\ref{ND34b}) and following the same method as is used to
determine ${\cal{G}}^{\mu\nu}$ in the previous section, we arrive
after some work at
\begin{eqnarray}\label{ND59b}
\lefteqn{\hspace{-1cm}{\cal{F}}^{00}=-{\cal{F}}^{33}=3i\sum\limits_{q\in\{\frac{2}{3},-\frac{1}{3}\}}|qeB|}\nonumber\\
&&\times
\sum\limits_{p=0}^{\infty}\alpha_{p}\int\frac{dp_{0}dp_{3}}{(2\pi)^{3}}
\frac{1}{(\bar{p}_{q}^{2}-m^{2})^{2}},
\end{eqnarray}
and
\begin{eqnarray}\label{ND60b}
\lefteqn{{\cal{F}}^{11}={\cal{F}}^{22}}\nonumber\\
&&=3i\sum\limits_{q}\sum\limits_{p,k=0}^{\infty}\int\frac{dp_{0}dp_{3}}{(2\pi)^{3}}
\bigg[\frac{(\bar{p}_{q}\cdot\bar{k}_{q}-m^{2})}{(\bar{p}_{q}^{2}-m^{2})(\bar{k}_{q}^{2}-m^{2})}C_{pk}^{(1)}\nonumber\\
&&~~~-
\frac{2|qeB|\sqrt{pk}}{(\bar{p}_{q}^{2}-m^{2})(\bar{k}_{q}^{2}-m^{2})}C_{pk}^{(2)}\bigg]\bigg|_{\tilde{k}=\tilde{p}}.
\end{eqnarray}
At finite $(T,\mu)$, ${\cal{F}}^{\mu\mu}, \mu=0,\cdots,3$ are
therefore given by
\begin{eqnarray}\label{ND61b}
\lefteqn{\hspace{0cm} {\cal{F}}^{00}=-{\cal{F}}^{33}
}\nonumber\\
&&=-3\sum\limits_{q\in\{\frac{2}{3},-\frac{1}{3}\}}|qeB|\sum_{p=0}^{\infty}\alpha_{p}\int\frac{dp_{3}}{(2\pi)^{2}}~
{\cal{S}}_{2}^{(0)}(\omega_{p}),\nonumber\\
\end{eqnarray}
as well as
\begin{eqnarray}\label{ND62b}
\lefteqn{{\cal{F}}^{11}={\cal{F}}^{22}}\nonumber\\
&&\hspace{-0.3cm}=-3\sum\limits_{q\in\{\frac{2}{3},-\frac{1}{3}\}}\sum_{p=0}^{\infty}
\int\frac{dp_{3}}{(2\pi)^{2}}\bigg\{2\big[p~{\cal{S}}_{1}^{(0)}(\omega_{p})\nonumber\\
&&-
(p+1)~{\cal{S}}_{1}^{(0)}(\omega_{p+1})\big]+\delta_{p0}{\cal{S}}_{1}^{(0)}(\omega_{p})\bigg\}.
\end{eqnarray}
All non-diagonal elements of ${\cal{F}}^{\mu\nu}$ turn out to
vanish. As we have described before, the remaining
$p_{3}$-integration and the summation over Landau levels appearing
in the final results of Secs. \ref{sec4p1} and \ref{sec4p2} for the
squared mass matrices $(M_{\sigma}^{2}, M_{\pi^{0}}^{2})$ as well as
form factors (kinetic coefficients) $({\cal{G}}^{\mu\nu},
{\cal{F}}^{\mu\nu})$, will be evaluated numerically in the next
section. Using these results, the $(T,\mu,eB)$ dependence of pole
and screening masses as well as the refraction indices of neutral
mesons will be explored.
%%%%%%%%%%%%%%%%%%%%%%%%%%%%%%%%%%%%%%%%%
\section{Numerical Results}\label{sec5}
%%%%%%%%%%%%%%%%%%%%%%%%%%%%%%%%%%%%%%%%%
\par\noindent
In Sec. \ref{sec3}, we have introduced the one-loop effective action
$\Gamma_{\mbox{\tiny{eff}}}[\sigma,\vec{\pi}]$ of a two-flavor NJL
model describing the dynamics of non-interacting $\sigma$ and
$\vec{\pi}$ mesons in a hot and magnetized medium. We have then
determined the corresponding one-loop effective potential of this
model $\Omega_{\mbox{\tiny{eff}}}(m;T,\mu,eB)$, up to an integration
over $p_{3}$-momentum and a summation over Landau levels, labeled by
$p$. According to our description in Sec. \ref{sec2}, the global
minima of $\Omega_{\mbox{\tiny{eff}}}(m;T,\mu,eB)$ can be used to
determine the squared mass matrices $(M_{\sigma}^{2},
M_{\pi^{0}}^{2})$ and the coefficients of the form factors (kinetic
coefficients) $({\cal{G}}^{\mu\nu}, {\cal{F}}^{\mu\nu})$,
corresponding to neutral mesons and appearing in the effective
action (\ref{ND1b}). In Sec. \ref{sec4}, we have described the
analytical method leading to $(M_{\sigma}^{2}, M_{\pi^{0}}^{2})$ and
$({\cal{G}}^{\mu\nu}, {\cal{F}}^{\mu\nu})$ at finite $(T,\mu,eB)$.
The squared mass matrices are given in (\ref{ND28b}) as well as
(\ref{ND37b}) and the form factors in (\ref{ND57b}), (\ref{ND61b})
as well as (\ref{ND62b}). All these results are presented up to an
integration over $p_{3}$-momentum and a summation over Landau levels
$p$. In this section, we will first use the one-loop effective
potential (\ref{NE14b}), to determine numerically the
$(T,\mu,eB)$-dependence of the constituent quark mass
$m=m_{0}+\sigma_{0}$ for non-vanishing bare quark mass $m_{0}$. This
will be done in Sec. \ref{sec5p1} by keeping one of these three
parameters fixed and varying two other parameters. We then continue
to explore the complete phase portrait of our magnetized two-flavor
NJL model in the chiral limit $m_{0}\to 0$. Our results are
comparable with the results previously presented in \cite{sato1997,
inagaki2003}. Similar results are also obtained in
\cite{fayazbakhsh2010}, where the two-flavor NJL model, used in the
present paper, is considered with additional diquark degrees of
freedom to study the chiral and color-superconductivity phases in a
hot and magnetized quark matter. In Sec. \ref{sec5p2}, we will then
evaluate the above mentioned $p_{3}$-integration and the summation
over Landau levels numerically.  This gives us the possibility to
study, in particular, the $T$-dependence of $(M_{\sigma}^{2},
M_{\pi^{0}}^{2})$ as well as $({\cal{G}}^{\mu\nu},
{\cal{F}}^{\mu\nu})$ for $\mu=0$ and various $eB=0,0.03, 0.2, 0.3$
GeV$^{2}$ (or equivalently $eB\simeq 0, 1.5m_{\pi}^{2},
10.5m_{\pi}^{2}, 15.7 m_{\pi}^{2}$ for $m_{\pi}=138$ MeV). As we
have described in Sec. \ref{sec1}, the magnetic fields produced in
the con-central heavy ion collisions at RHIC and LHC are estimated
to be in the order of $eB\sim 1.5 m_{\pi}^{2}$ and $eB\sim 15
m_{\pi}^{2}$ (or equivalently, $eB\sim 0.03$ GeV$^{2}$ and $eB\sim
0.3$ GeV$^{2}$, respectively) \cite{mclerran2007, skokov2010}.
Hence, our results for small values of magnetic fields (here,
$eB=0.03$ GeV$^{2}$) may be relevant for the physics of heavy ion
collisions at RHIC, while our results in the intermediate magnetic
fields (here, $eB= 0.2, 0.3$ GeV$^{2}$) seem to be relevant for the
heavy ion collision at LHC. In Sec. \ref{sec5p3}, we will finally
present a number of applications of the results presented in the
second part of this section. In particular, we will determine the
$T$-dependence of the pole mass as well as the refraction index and
screening mass of neutral mesons for $\mu=0$ and $eB=0,0.03, 0.2,
0.3$ GeV$^{2}$. To do this, we will use the corresponding dispersion
relations of $\sigma$- and $\pi^{0}$ mesons. The goal is to study
the effect of uniform magnetic fields on meson masses and refraction
indices and explore the interplay between the effects of temperature
and the external magnetic fields on these quantities. We will, in
particular, show that uniform magnetic fields induce a certain
anisotropy in the mesons refraction indices and the screening masses
in the longitudinal and transverse directions with respect to the
external magnetic field. Detailed studies on $eB$ and $\mu$
dependence of all the above physical quantities, together with other
possible applications of $(M_{\sigma}^{2}, M_{\pi^{0}}^{2})$ and
$({\cal{G}}^{\mu\nu}, {\cal{F}}^{\mu\nu})$, e.g. in studying the
mass splitting between charged pion masses will be presented
elsewhere \cite{sadooghi2012-3}.\footnote{The mass splitting between
$\pi^{+}$ and $\pi^{-}$ is recently discussed in
\cite{andersen2011-pions, anderson2012-2}, using chiral perturbation
theory in the presence of constant magnetic field.}
%%%%%%%%%%%%%%%%%%%%%%%%%%%%%%%%%%%%%%%%%
\subsection{Chiral condensate and complete phase portrait of a magnetized and hot two-flavor NJL model in the chiral limit}\label{sec5p1}
%%%%%%%%%%%%%%%%%%%%%%%%%%%%%%%%%%%%%%%%%
\setcounter{equation}{0}
\begin{figure*}[hbt]
\includegraphics[width=5.5cm,height=4cm]{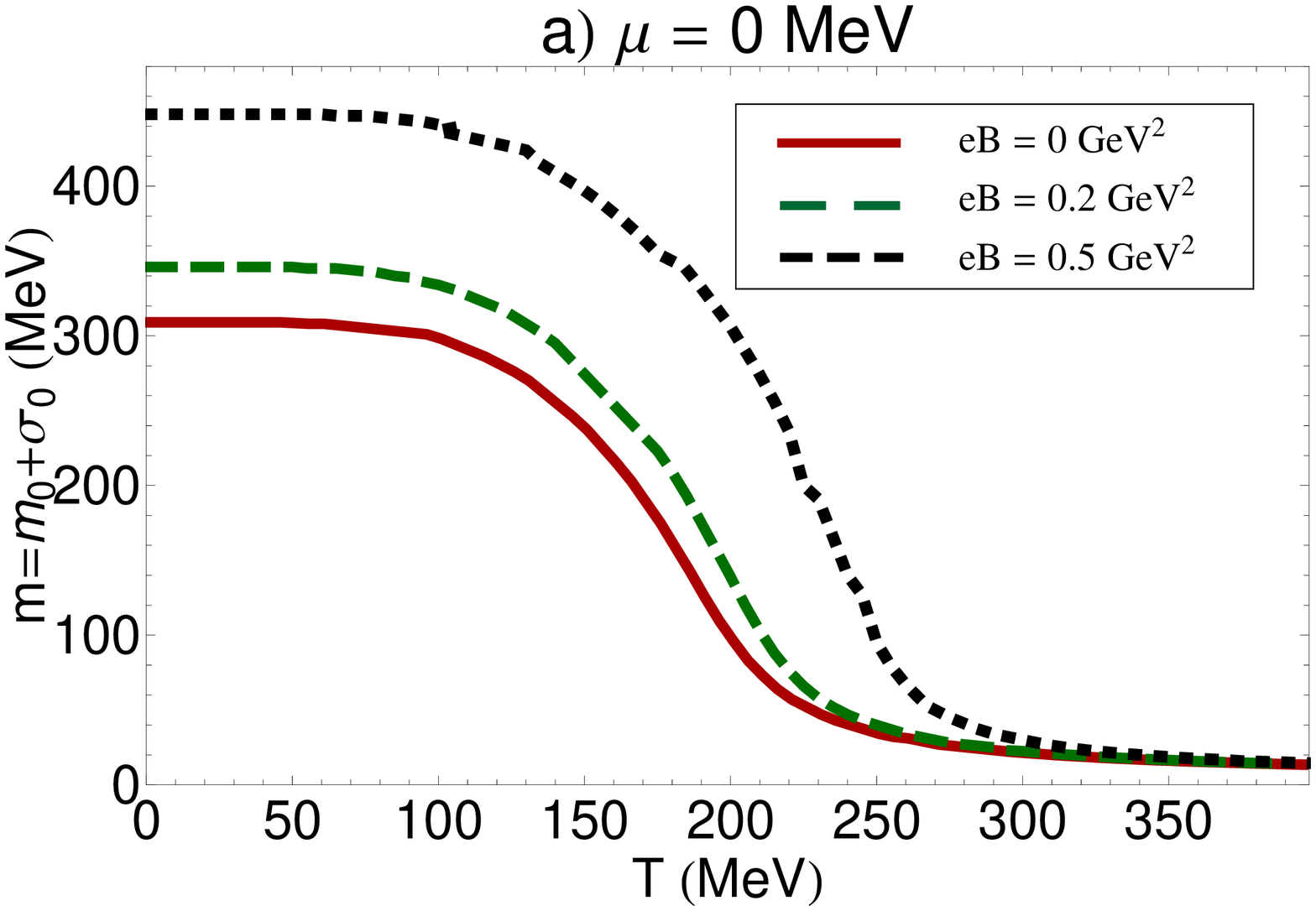}
\includegraphics[width=5.5cm,height=4cm]{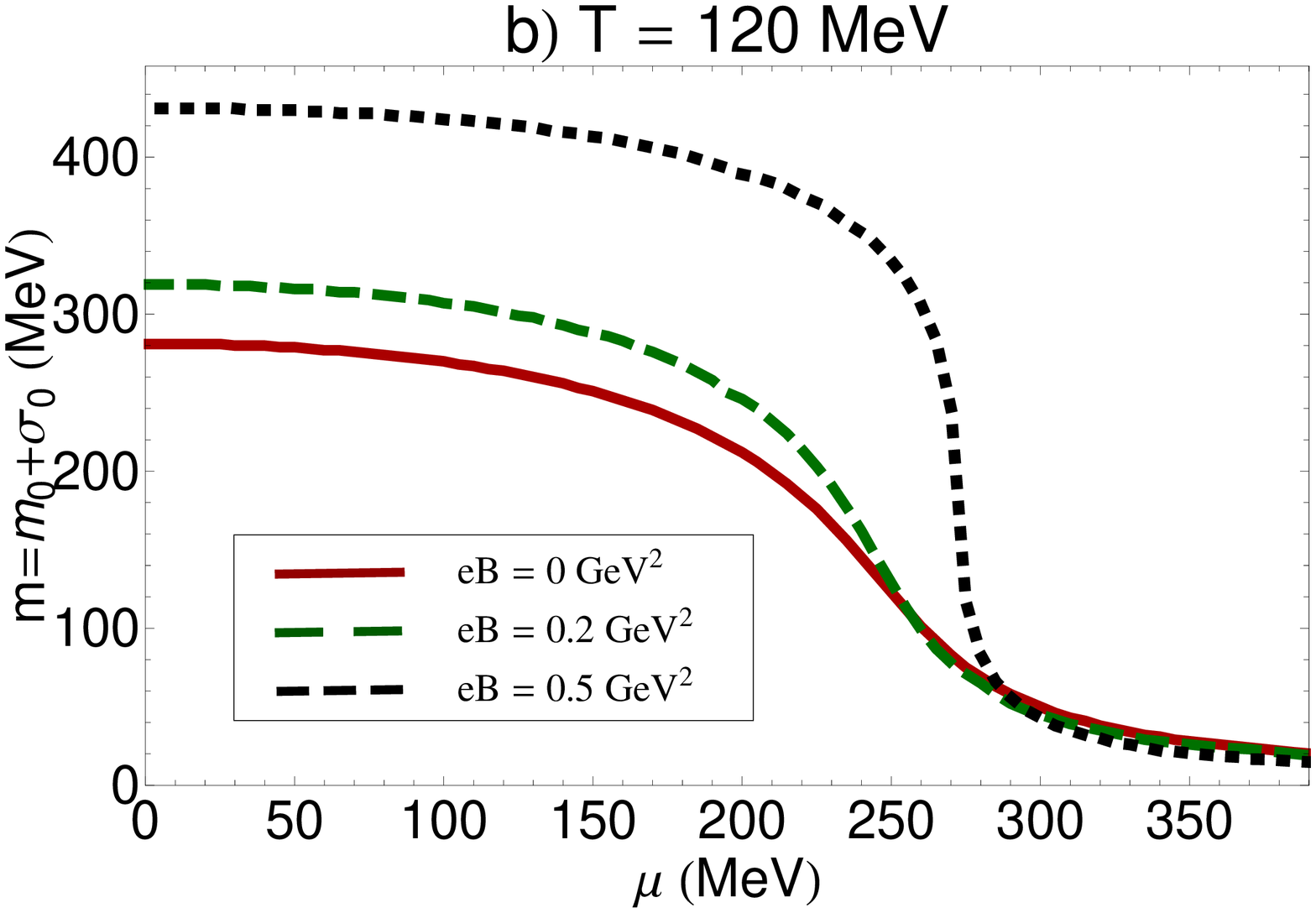}
\includegraphics[width=5.5cm,height=4cm]{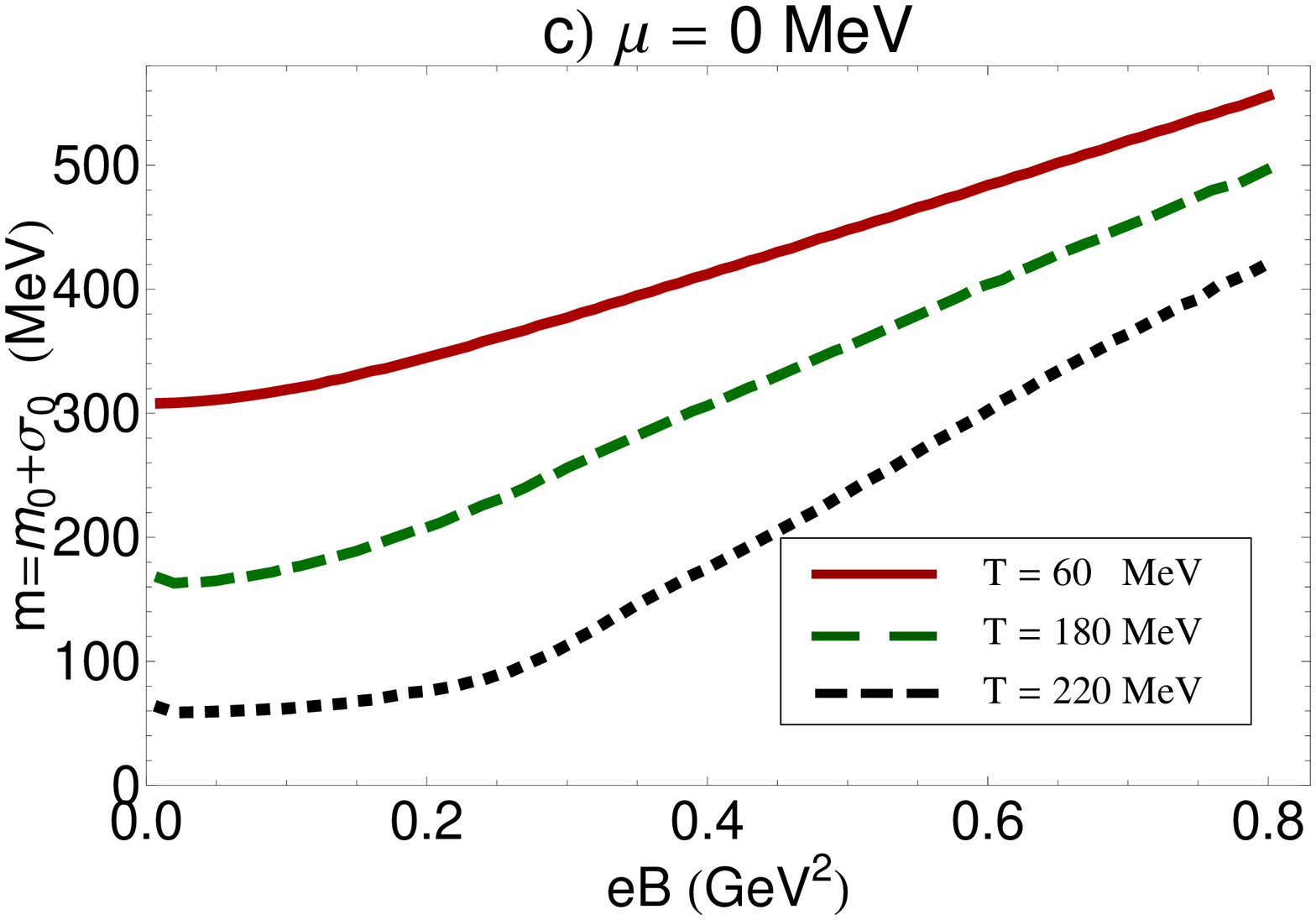}
\caption{(a) and (b) The $T$ and $\mu$ dependence of the constituent
quark mass $m=m_{0}+\sigma_{0}$ for fixed $eB=0,0.2, 0.5$ GeV$^{2}$
and for fixed $\mu=0$ and $T=120$, respectively. Here, $m_{0}\simeq
5$ MeV is the bare quark mass and $\sigma_{0}$ is the chiral
condensate. (c) The $eB$ dependence of $m$ is demonstrated for fixed
$\mu=0$ and various $T=60, 180$ and $220$ MeV.}\label{fig1}
\end{figure*}
\par\noindent
Using the thermodynamic potential from (\ref{NE14b}), we will
determine, in what follows, the chiral condensate and the complete
phase portrait of the two-flavor NJL model at finite $T,\mu$ and
$eB$. The notations and mathematical method used in this paragraph
are similar to what was previously used in \cite{fayazbakhsh2010}.
To determine the chiral condensate, we have to solve the gap
equation numerically
\begin{eqnarray}\label{NA1}
\frac{\partial\Omega_{\mbox{\tiny{eff}}}(\bar{m};T,\mu,eB)}{\partial
\bar{m}}\bigg|_{\bar{m}=m}=0.
\end{eqnarray}
Here, $\bar{m}=m_{0}+\sigma$ and $m=m_{0}+\sigma_{0}$, as are
introduced in Sec. \ref{sec3}. Our specific choice of parameters is
\cite{buballa2004}
\begin{eqnarray}\label{NA2}
\Lambda=0.6643~\mbox{GeV},~G=4.668~\mbox{GeV}^{-2},
m_{0}=5~\mbox{MeV},\nonumber\\
\end{eqnarray}
where $\Lambda$ is the UV momentum cutoff and $G$ is the NJL
(chiral) coupling constant. To perform the momentum integration over
${\mathbf{p}}$ and $p_{3}$, we have introduced, as in
\cite{fayazbakhsh2010}, smooth cutoff functions
\begin{eqnarray}\label{NA3}
f_{\Lambda}&=&\frac{1}{1+\exp\left(\frac{|{\mathbf{p}}|-\Lambda}{A}\right)},\nonumber\\
f_{\Lambda,B}^{p}&=&\frac{1}{1+\exp\left(\frac{\sqrt{p_{3}^{2}+2|qeB|p}-\Lambda}{A}\right)},
\end{eqnarray}
corresponding to integrals with vanishing and non-vanishing magnetic
fields, respectively. In $f_{\Lambda,B}^{p}$, $p$ labels the Landau
levels. Moreover, $A$ is a free parameter, which determines the
sharpness of the cutoff scheme. It is chosen to be $A=0.05\Lambda$,
with $\Lambda$ given in (\ref{NA2}). Using the above smooth cutoff
procedure, the above choice of parameters leads for vanishing
magnetic field and at $T=\mu=0$ to the constituent mass $m\simeq
308$ MeV.\footnote{For sharp UV-cutoff, $m$ turns out to be $m\simeq
300$ MeV, as expected.} Let us notice that the solutions of
(\ref{NA1}) are in general ``local'' minima of the theory. Keeping
$\sigma_{0}\neq 0$ and looking for ``global'' minima of the system
described by $\Omega_{\mbox{\tiny{eff}}}(m;T,\mu,eB)$ from
(\ref{NE14b}), it turns out that only in the regime $\mu\in[0,350]$
MeV, $T\in [0,390]$ MeV and $\tilde{e}B\in [0,0.8]$ GeV$^{2}$, the
global minima of $\Omega_{\mbox{\tiny{eff}}}$ are described by
nonzero $\sigma_{0}$. In these regimes, the chiral symmetry is
spontaneously broken by non-vanishing $\sigma_{0}$.\footnote{For
$m_{0}\neq 0$ the chiral symmetry of the original Lagrangian is
explicitly broken.} All our numerical computations in the present
section are therefore limited to these regimes. Note that because of
non-vanishing quark mass $m_{0}$, the transition from the chiral
symmetry broken phase to the normal phase is a smooth crossover (see
the descriptions below).
\par
In Fig. \ref{fig1}, the $T,\mu$ and $eB$ dependence of $m$ are
presented. In Fig. \ref{fig1}(a), the $T$-dependence of $m$ is
demonstrated for fixed $\mu=0$ and $eB=0,0.2,0.5$ GeV$^{2}$.
Although the transition from the chiral symmetry broken phase, with
$m\neq 0$ to the normal phase, with $m\simeq m_{0}\approx 0$, is a
smooth crossover, but as it turns out, for stronger magnetic fields
the transition to the normal phase occurs for larger values of $T$,
whereas for $eB=0$, this transition temperature into the crossover
region is smaller. Moreover, at $T\in [0,100]$ MeV, where $m$ is
almost constant, the value of $m$ increases with increasing $eB$.
All these effects are related with the phenomena of magnetic
catalysis \cite{klimenko1992, miransky1995}, according to which,
magnetic fields enhance the production of
$\sigma_{0}\sim\langle\bar{\psi}\psi\rangle$ condensate, even for
very small coupling between the fermions, and therefore catalyze the
dynamical chiral symmetry breaking. Similar effects occur also in
Fig. \ref{fig1}(b), where $m$ is plotted as a function of $\mu$, at
fixed $T=120$ MeV and for various $eB=0,0.2,0.5$ GeV$^{2}$. At
$\mu=0$, for instance, the value of $m$ increases with increasing
$eB$. In Fig. \ref{fig1}(c), the $eB$-dependence of $m$ is
demonstrated for fixed $\mu=0$ and $T=60,180$ and $220$ MeV.  As it
turns out, for fixed value of $eB$, $m$ decreases with increasing
$T$, and as it turns out, this ``melting'' effect persists in the
whole range of $eB\in [0, 0.8]$ GeV$^{2}$, although it is partly
compensated by the magnetic field in the regime $eB>0.6$ GeV$^{2}$.
Let us notice that, according to our results in
\cite{fayazbakhsh2010-1,fayazbakhsh2010}, for a certain threshold
magnetic field $eB_{t}\simeq 0.45$ GeV$^{2}$, the magnetic field is
strong enough and forces the dynamics of the system to be mainly
described by the LLL. In this regime, $m$ increases linearly with
increasing $eB$ [see Fig. \ref{fig1}(c)]. Later, in
\cite{rebhan2011}, the threshold magnetic field is estimated to be
in the order of $B\simeq 10^{19}$ Gau\ss. In the present paper,
however, the threshold magnetic field turns out to be $eB_{t}\geq
0.7$ GeV$^{2}$ [or equivalently $B\simeq 1.2\times 10^{20}$
Gau\ss].\footnote{The exact value of threshold magnetic field
$eB_{t}$ is determined from $\lfloor
\frac{\Lambda^{2}}{|qeB|}\rfloor=0$, where $\lfloor a\rfloor$ is the
greatest integer less than or equal to $a$. For up quark
$eB_{t}\simeq 0.67$ GeV$^{2}$ and for down quark $eB_{t}\simeq 1.33$
GeV$^{2}$.} The $T$ and $eB$ dependence of $m$ at fixed chemical
potential $\mu$ and various $eB$ and $T$ are discussed recently in
\cite{condensate-lattice} using lattice gauge theory methods in the
presence of constant (electro)magnetic fields. Our original results
from \cite{fayazbakhsh2010} as well as the results presented in
Figs. \ref{fig1}(a) and \ref{fig1}(c) are consistent with the
results arising from lattice simulations \cite{condensate-lattice}.
\begin{figure*}[hbt]
\includegraphics[width=5.5cm,height=4cm]{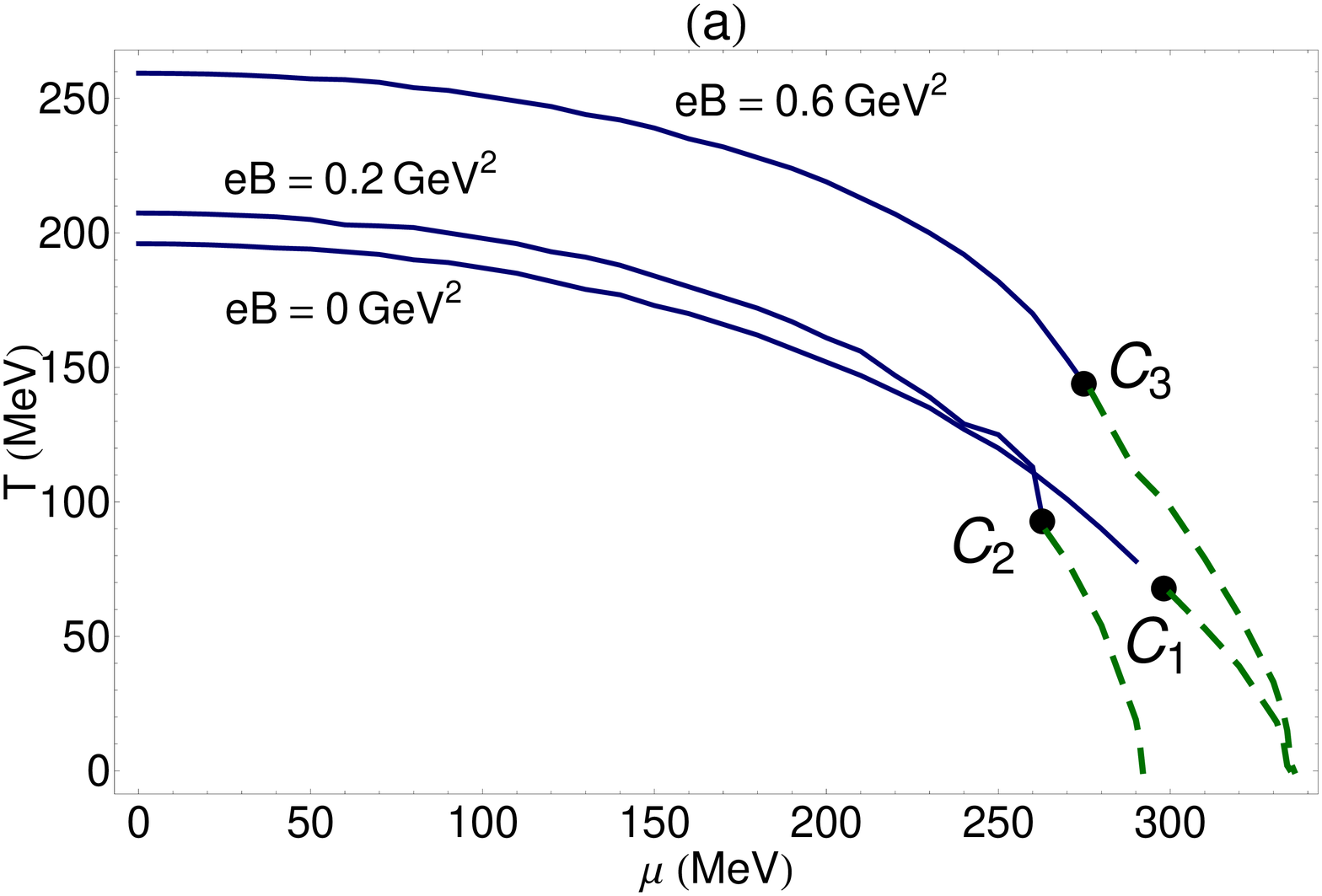}
\includegraphics[width=5.5cm,height=4cm]{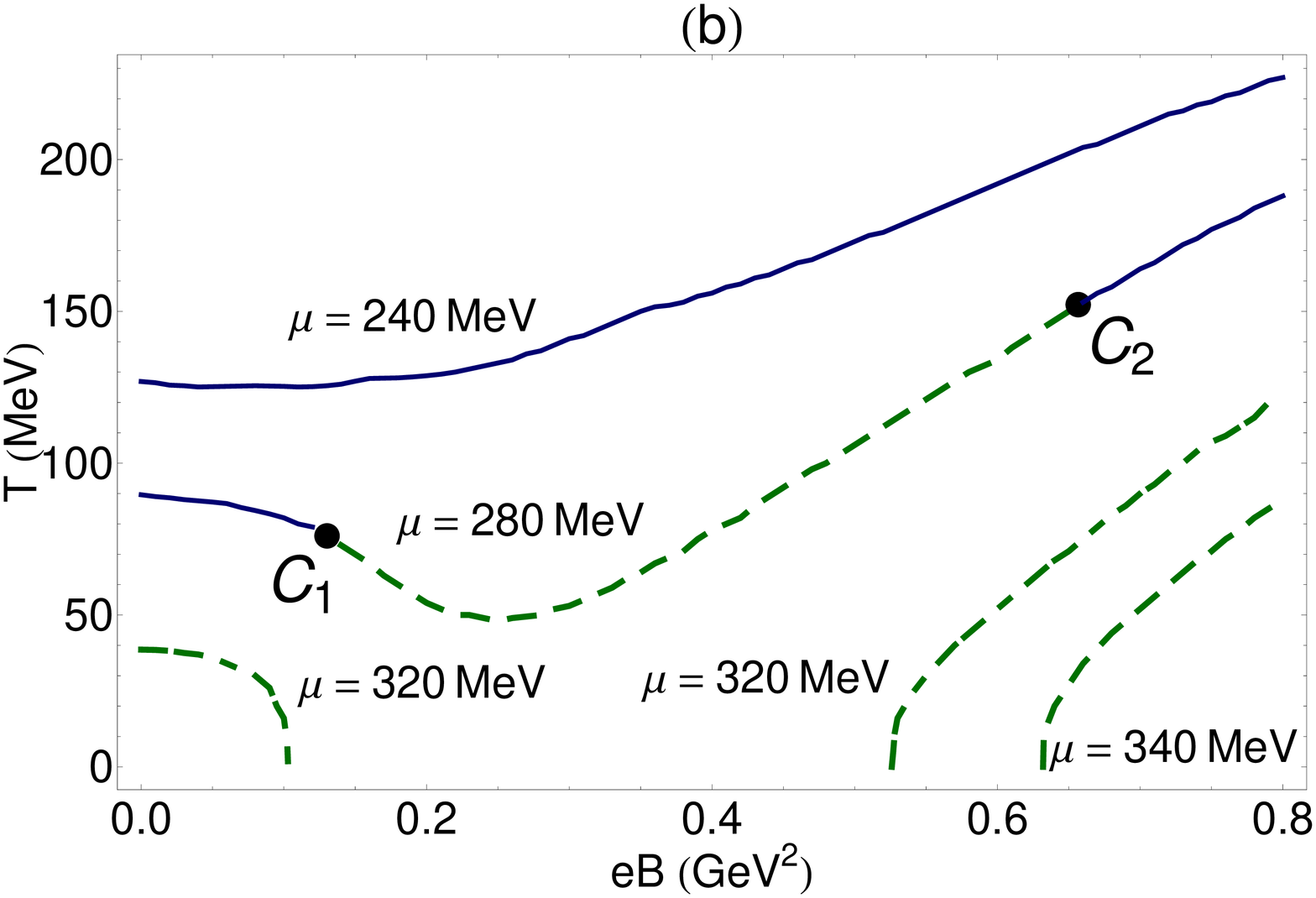}
\includegraphics[width=5.5cm,height=4cm]{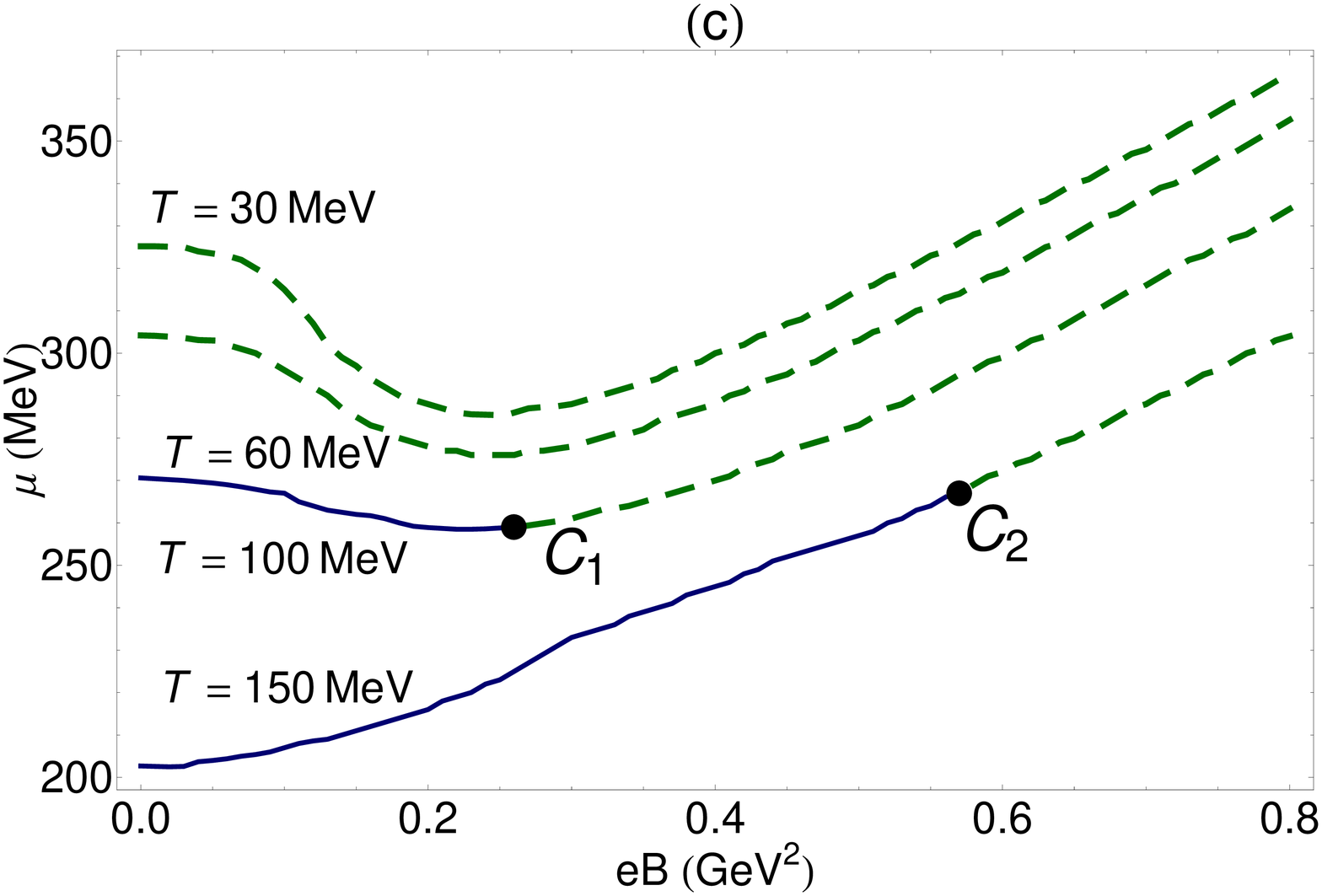}
\caption{Complete phase portrait of a two-flavor magnetized NJL
model at finite $T$, $\mu$ and $eB$ in the chiral limit of vanishing
quark mass $m_{0}$. Blue solid (green dashed) lines
 denote the second (first) phase order phase transition between the chiral symmetry broken and normal phases. Two branches of the first order
 critical line for $\mu=320$ MeV are denoted by green double-dashed lines. }\label{fig2}
\end{figure*}
\par
According to our results in \cite{fayazbakhsh2010}, in the chiral
limit $m_{0}\to 0$ and for vanishing magnetic field, at high
temperature and small chemical potential, the transition from the
chiral symmetry broken to the normal phase is of second order. In
contrast, at low temperatures and higher densities, the second order
phase transition goes over to a first order one. In the presence of
a uniform magnetic field, this picture remains essentially the same.
The only difference is that for $\mu=0$, the transition temperature
increases with increasing $eB$. Moreover, for $eB\neq 0$, the second
order phase transition occurs at higher temperatures and lower
densities comparing to the case of vanishing magnetic fields. These
two effects of the uniform magnetic field on the $T-\mu$ phase
diagram of a two-flavor NJL model in the chiral limit are
demonstrated in Fig. \ref{fig2}(a). Both effects are manifestations
of the phenomenon of magnetic catalysis in the presence of constant
magnetic fields \cite{klimenko1992, miransky1995}. In all the plots
of Fig. \ref{fig2}, the green dashed (blue solid) lines denote first
(second) order phase transitions. To determine the first and second
order phase transitions, the method described in \cite{sato1997,
inagaki2003, fayazbakhsh2010,fayazbakhsh2010-1} is used. The first
order critical lines between the chiral symmetry breaking and the
normal phase is determined by solving
\begin{eqnarray}\label{NA4}
&&\hspace{-0.8cm}\frac{\partial
\Omega_{\mbox{\tiny{eff}}}(\bar{m};T,\mu,eB)}{\partial
\bar{m}}\bigg|_{m}=0, \nonumber\\
&&\hspace{-0.8cm}\Omega_{\mbox{\tiny{eff}}}(m\neq
0;T,\mu,eB)=\Omega_{\mbox{\tiny{eff}}}(m=0;T,\mu,eB),
\end{eqnarray}
simultaneously.\footnote{In the chiral limit $m_{0}\to 0$, $m\equiv
\sigma_{0}$.} The second order critical line between these two
phases is determined using
\begin{eqnarray}\label{NA5}
\lim_{m^{2}\rightarrow 0}\frac{\partial
\Omega_{\mbox{\tiny{eff}}}(m; T,\mu,eB)}{\partial m^{2}}=0.
\end{eqnarray}
To make sure that after the second order phase transition the global
minima of the effective potential are shifted to $m=0$ in
(\ref{NA5}), and in order to avoid instabilities, an analysis
similar to \cite{berges1998} is also performed.
\par
In Fig. \ref{fig2}(b), the $T-eB$ phase diagram of our model is
plotted for various $\mu=240, 280, 320, 340$ MeV. Let us notice that
for $\mu=320$ MeV [dashed-dotted lines in Fig. \ref{fig2}(b)], the
first order critical line has two branches -- the first one for
$eB<0.1$ GeV$^{2}$ and the second one for $eB>0.5$ GeV$^{2}$, at
relatively low temperature. In the intermediate region $0.1<eB<0.5$
GeV$^{2}$, the chiral symmetry breaking phase is disfavored. In
\cite{fayazbakhsh2010}, we have studied the $T-eB$ phase diagram of
a two-flavor NJL model including meson \textit{and} diquark
condensates. We have shown that in the above mentioned intermediate
regime $0.1<eB<0.5$ GeV$^{2}$ at low temperature and for $\mu=320$
MeV, the two-flavor color superconducting (2SC) phase is favored.
For $\mu>320$ MeV, the first branch appearing for $\mu=320$ MeV and
$eB<0.1$ GeV$^{2}$ disappears, and the whole region of $eB<0.6$
GeV$^{2}$ is favored by either the normal phase, when no diquarks
exist in the model, or by the 2SC superconducting phase, when the
model includes both meson and diquark condensates (see Fig. 14 of
\cite{fayazbakhsh2010}).
\par
In Fig. \ref{fig2}(c), the $\mu-eB$ phase diagram of our two-flavor
NJL model including chiral condensates $(\sigma,\vec{\pi})$ is
plotted for various $T=30,60,100,150$ MeV. At very low temperature,
$T<100$ MeV, the transition between the chiral symmetry breaking and
normal phase is of first order (green dashed lines). Whereas at
these temperatures and for $eB<0.1$ GeV$^{2}$, the critical $\mu$ is
almost constant, it decreases by increasing the strength of the
magnetic field in the regime $0.1<eB<0.4$ GeV$^{2}$. This effect,
which is for the first time observed in
\cite{fayazbakhsh2010-1,fayazbakhsh2010}, and later also in
\cite{rebhan2011}, is called the ``inverse magnetic catalysis'',
according to which at low temperature, the addition of the magnetic
field decreases the critical chemical potential for chiral symmetry
restoration \cite{fayazbakhsh2010, rebhan2011}. However, by
increasing the magnetic field up to $eB>0.5$ GeV$^{2}$, i.e. by
entering the regime of LLL dominance, this effect is disfavored, so
that $\mu_{c}$ again increases with increasing the strength of the
magnetic field. Let us also note that similar phenomenon of inverse
magnetic catalysis appears also in Fig. \ref{fig2}(b), where for
fixed $\mu=280$ MeV, the first order critical line $T_{c}$ (green
dashed line between $C_{1}$ and $C_{2}$) decreases with
\textit{increasing} $eB$ from $0.1<eB<0.3$ GeV$^{2}$ and continues
to grow up with increasing the strength of the magnetic field up to
regime of LLL dominance, i.e $eB>0.5$ GeV$^{2}$. The inverse
magnetic catalysis effect may be related to the well-known
van-alphen--de Haas oscillations, which occur whenever Landau levels
pass the quark Fermi level \cite{alfven}. Similar effects are also
observed in \cite{inagaki2003, fayazbakhsh2010,fayazbakhsh2010-1}.
At higher temperature $T>100$ MeV and for $eB$ smaller than a
certain critical $eB_{c}$, there is a second order phase transition
between the chiral symmetry broken and normal phases (see the blue
solid lines in Fig. \ref{fig2}(c) for $T=100, 150$ MeV, that replace
the green dashed lines for $T<100$ MeV). The critical magnetic field
$eB_{c}$, for which the second order phase ends and goes over into a
first order phase transition is larger for higher temperature
[compare $eB_{c}$ for two critical points (black bullets) $C_{1}$
and $C_{2}$ in Fig. \ref{fig2}(c)]. This demonstrates the
destructive effect of the temperature, which is partly compensated
in the regime of strong magnetic fields, $eB>0.7$ GeV$^{2}$. More
details on the interplay between three parameters $T,\mu$ and $eB$
on the formation of chiral condensates $\sigma_{0}$ in the chiral
limit $m_{0}\to 0$ are discussed in \cite{fayazbakhsh2010}.
%%%%%%%%%%%%%%%%%%%%%%%%%%%%%%%%%%%%%%%%%%%%
\subsection{$(M_{\sigma}^{2},
M_{\pi^{0}}^{2})$ and $({\cal{G}}^{\mu\nu}, {\cal{F}}^{\mu\nu})$ at
$(T\neq 0, \mu=0,eB\neq 0)$}\label{sec5p2}
\par\noindent
%%%%%%%%%%%%%%%%%%%%%%%%%%%%%%%%%%%%%%%%%%%%
\begin{figure*}[hbt]
\includegraphics[width=5.5cm,height=4cm]{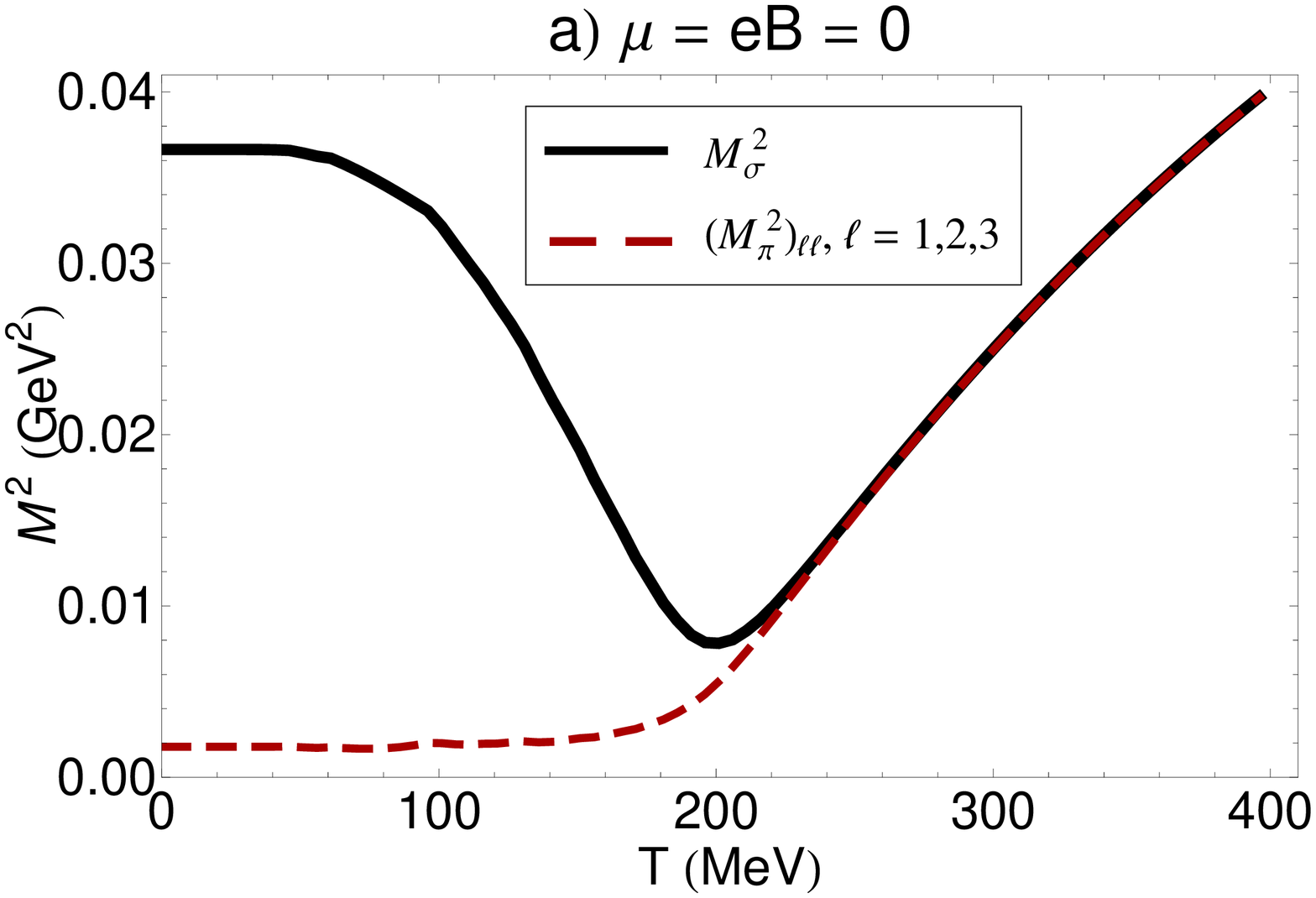}
\includegraphics[width=5.5cm,height=4cm]{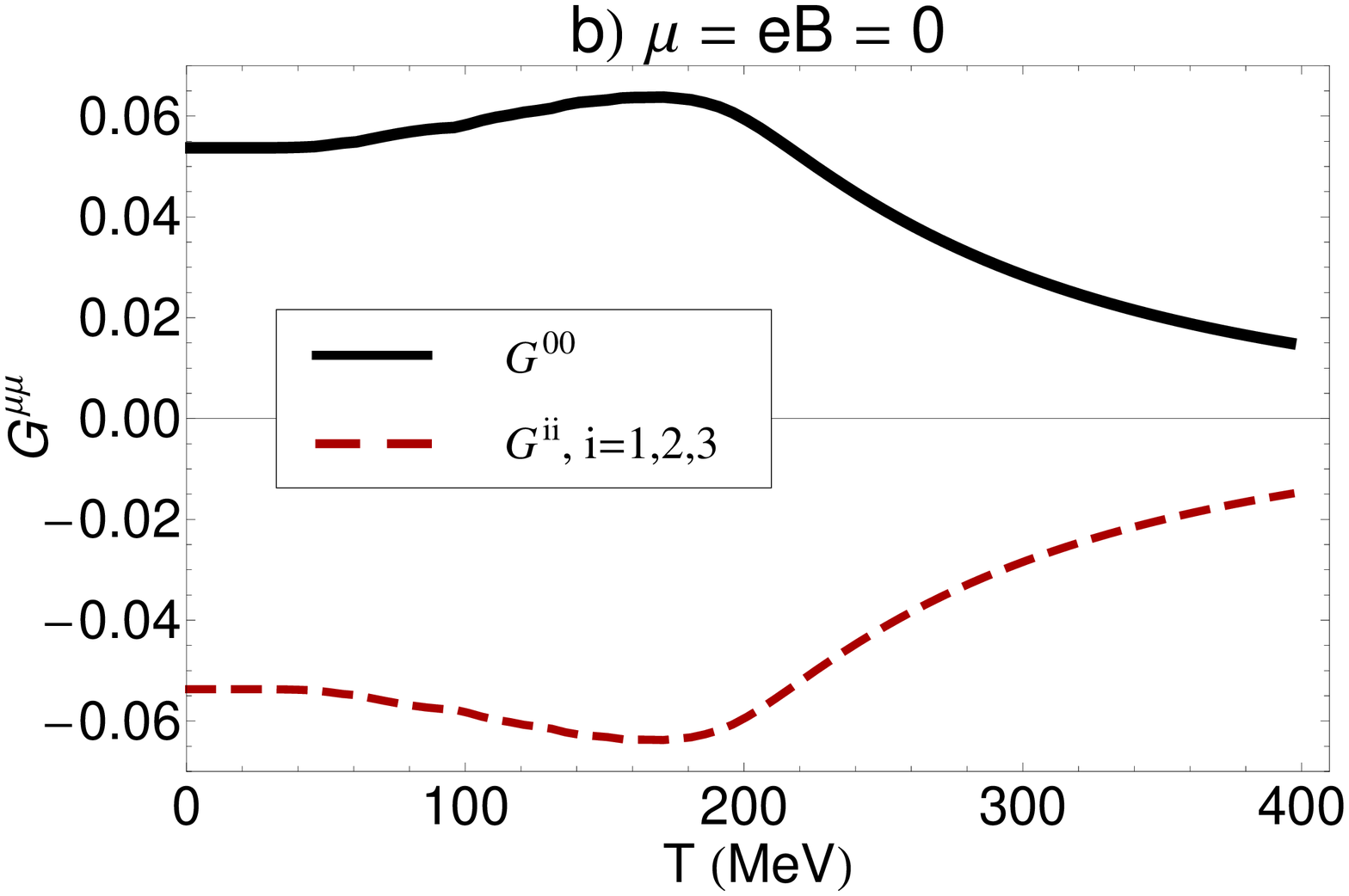}
\includegraphics[width=5.5cm,height=4cm]{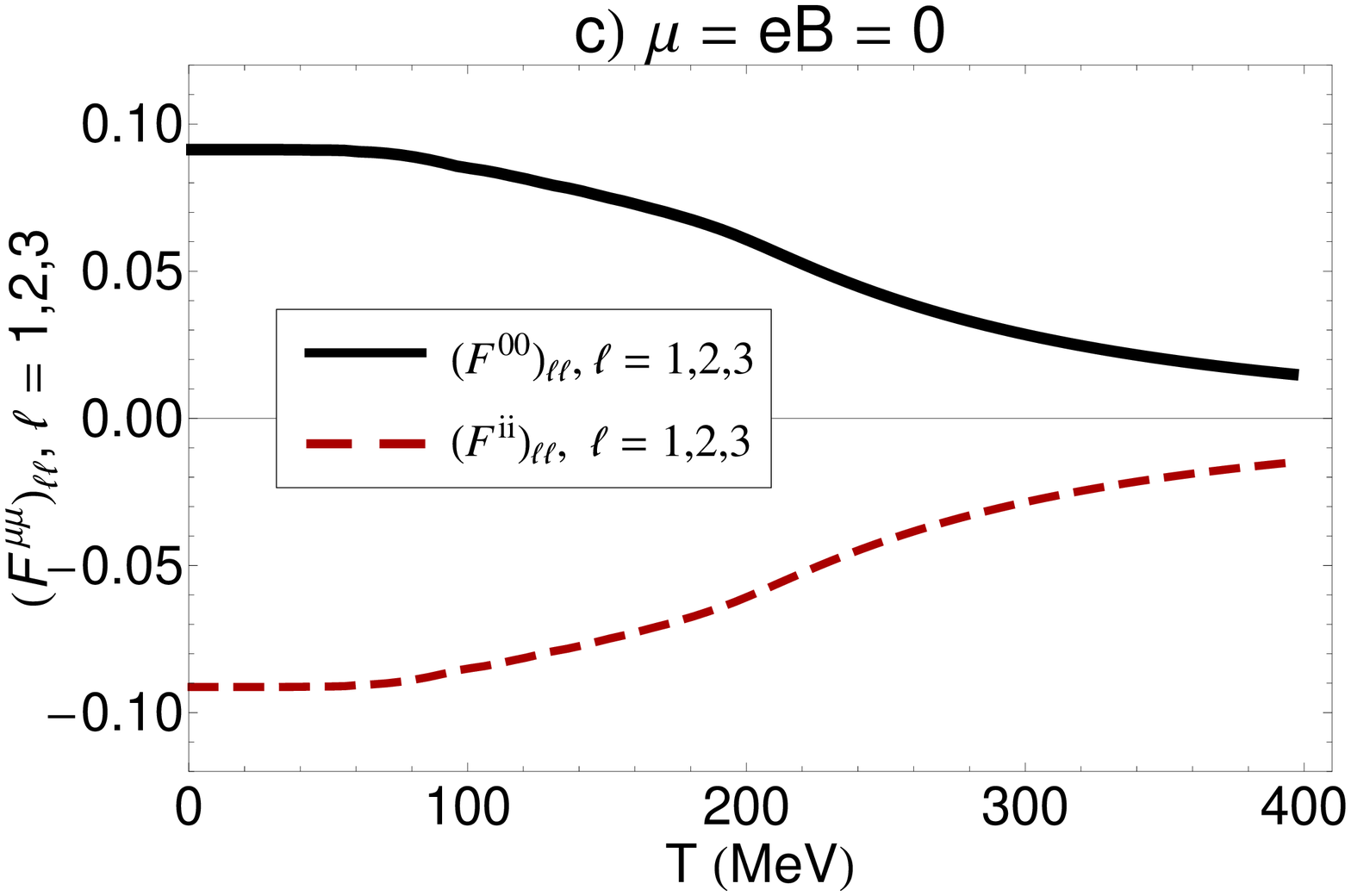}
\caption{The $T$-dependence of $(M_{\sigma}^{2},
(M_{\vec{\pi}}^{2})_{\ell\ell})$ as well as $({\cal{G}}^{00},
{\cal{G}}^{ii})$, and $(({\cal{F}}^{00})_{\ell\ell},
({\cal{F}}^{ii})_{\ell\ell})$ with $\ell,i=1,2,3$ for vanishing
magnetic field $eB$ and at $\mu=0$.}\label{fig3}
\end{figure*}
As we have described in the first part of this section, the results
presented in (\ref{ND28b}) and (\ref{ND36b}) for $M_{\sigma}^{2}$
and $M_{\pi^{0}}^{2}$, and in (\ref{ND42b}) and (\ref{ND57b}) for
${\cal{G}}^{\mu\nu}$ as well as in (\ref{ND61b}) and (\ref{ND62b})
for ${\cal{F}}^{\mu\nu}$ are given up to an integration over
$p_{3}$-momentum and a summation over Landau levels $p$. We have
performed the $p_{3}$-integration for the set of parameters
$(\Lambda,G,m_{0})$ from (\ref{NA2}) and the smooth cutoff function
(\ref{NA3}) numerically, and will present the results in what
follows. In particular, we will present the $T$-dependence of
$(M_{\sigma}^{2}, M_{\pi^{0}}^{2})$ and
$({\cal{G}}^{\mu\nu},{\cal{F}}^{\mu\nu})$ for fixed $\mu=0$ and
various $eB=0,0.03,0.2,0.3$ GeV$^{2}$.
\par
Let us start by giving the numerical values of $(M_{\sigma}^{2},
M_{\vec{\pi}}^{2})$ and $({\cal{G}}^{\mu\nu},{\cal{F}}^{\mu\nu})$ at
$T=\mu=eB=0$. According to their definitions in
(\ref{ND4b})-(\ref{ND7b}), where, for $eB=0$, $S_{Q}$ is to be
replaced by the ordinary fermion propagator
$S(z,0)=\int\frac{d^{4}p}{(2\pi)^{4}}\frac{ie^{-ip\cdot
z}}{\gamma\cdot p-m}$ at zero $(T,\mu)$, we have the following
identities and numerical values
\begin{eqnarray}\label{NA6}
M_{\sigma}^{2}&=&3.656\times 10^{-2}\mbox{GeV}^{2},\nonumber\\
M_{\pi_{\ell}}^{2}&=&1.734\times
10^{-3}\mbox{GeV}^{2},~~~\forall\ell=1,2,3,
\end{eqnarray}
as well as
\begin{eqnarray}\label{NA7}
{\cal{G}}^{00}&=&-{\cal{G}}^{ii}~~~~~\qquad \forall i=1,2,3,\nonumber\\
({\cal{F}}^{00})_{\ell\ell}&=&-({\cal{F}}^{ii})_{\ell\ell}\qquad\forall
i=1,2,3,
\end{eqnarray}
where
\begin{eqnarray}\label{NA8c}
{\cal{G}}^{00}&=&5.381\times 10^{-2},\nonumber\\
({\cal{F}}^{00})_{\ell\ell}&=&9.143\times 10^{-2},\qquad \forall
\ell=1,2,3.
\end{eqnarray}
Moreover, we have
$({\cal{F}}^{ii})_{11}=({\cal{F}}^{ii})_{22}=({\cal{F}}^{ii})_{33}$
for all $i=0,\cdots,3$.
\par
At finite temperature and vanishing $\mu$ and $eB$, although the
above relations (\ref{NA6}) and (\ref{NA7}) between different
components of $M_{\vec{\pi}}^{2}$ as well as ${\cal{G}}^{\mu\nu}$
and ${\cal{F}}^{\mu\nu}$ are still valid, i.e. we have
\begin{eqnarray}\label{NA8}
(M_{\vec{\pi}}^{2})_{11}=(M_{\vec{\pi}}^{2})_{22}=(M_{\vec{\pi}}^{2})_{33},
\end{eqnarray}
as well as
\begin{eqnarray}\label{NA9}
{\cal{G}}^{00}&=&-{\cal{G}}^{11}=-{\cal{G}}^{22}=-{\cal{G}}^{33},\nonumber\\
({\cal{F}}^{00})_{\ell\ell}&=&-
({\cal{F}}^{ii})_{\ell\ell},\qquad\forall \ell,i=1,2,3,
\end{eqnarray}
but their values become temperature dependent. In Fig. \ref{fig3},
the $T$-dependence of $(M_{\sigma}^{2},
(M_{\vec{\pi}}^{2})_{\ell\ell})$ as well as $({\cal{G}}^{00},
{\cal{G}}^{ii})$, and $(({\cal{F}}^{00})_{\ell\ell},
({\cal{F}}^{ii})_{\ell\ell})$ with $\ell,i=1,2,3$ are plotted for
vanishing $eB$ and $\mu$. As it is demonstrated in Fig.
\ref{fig3}(a), $M^{2}_{\sigma}$ and $(M_{\vec{\pi}}^{2})_{\ell\ell}$
are degenerate at $T> 220$ MeV. This is because the difference
between these two functions are in terms proportional to the
constituent quark mass $m=m_{0}+\sigma_{0}$, that, according to Fig.
\ref{fig1}(a) almost vanishes in the crossover region $T>220$ MeV.
Later, we will show that the degeneracy of $M^{2}_{\sigma}$ and
$(M_{\vec{\pi}}^{2})_{\ell\ell}$ for $T>220$ MeV leads to the
expected degeneracy of $\sigma$ and $\pi^{0}$ meson masses
$(m_{\sigma},m_{\pi^{0}})$ for vanishing $eB$ and $\mu$ in the
crossover region $T>220$ MeV \cite{buballa2012}.
\begin{figure}[hbt]
\includegraphics[width=7.7cm,height=4.5cm]{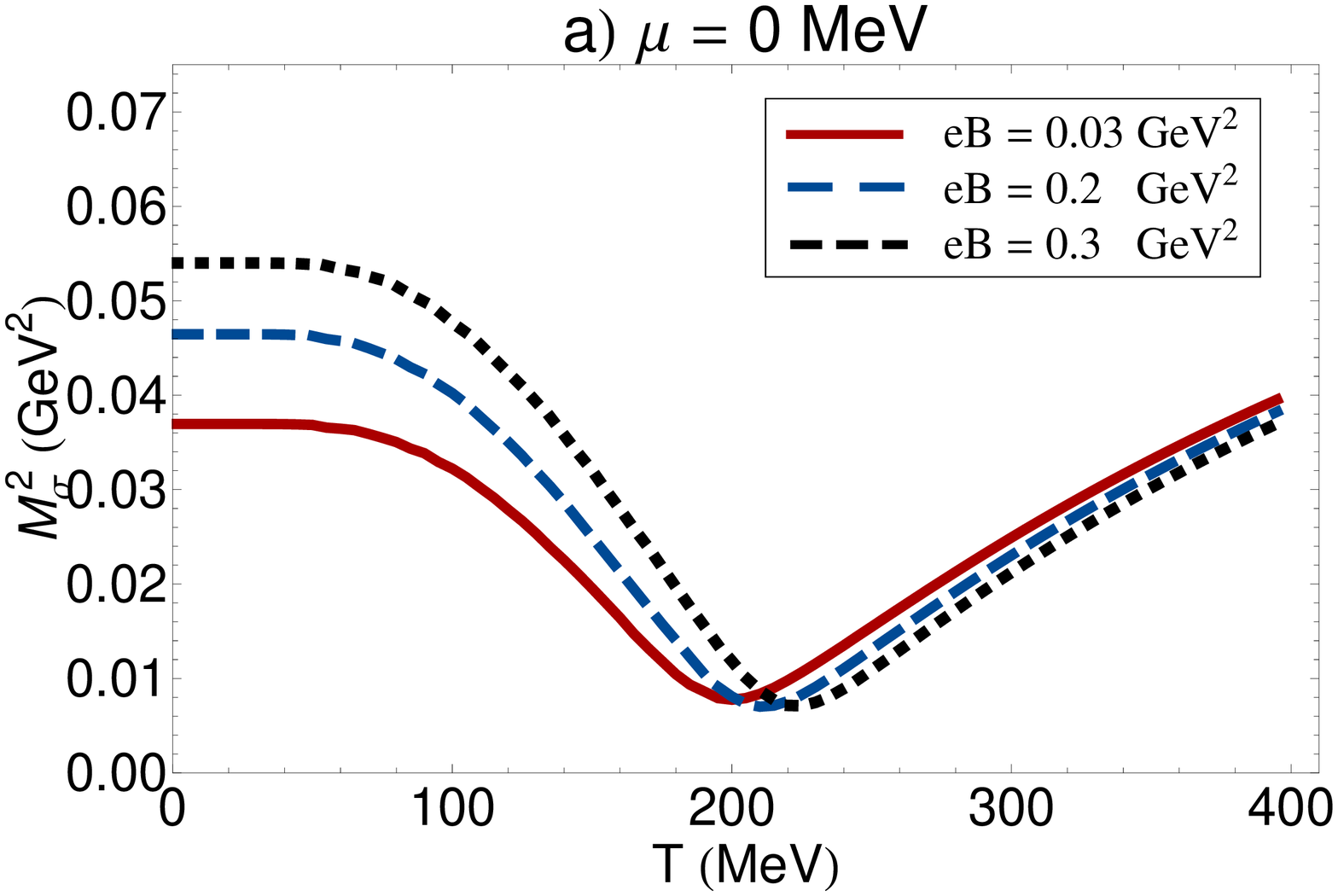}
\includegraphics[width=7.7cm,height=4.5cm]{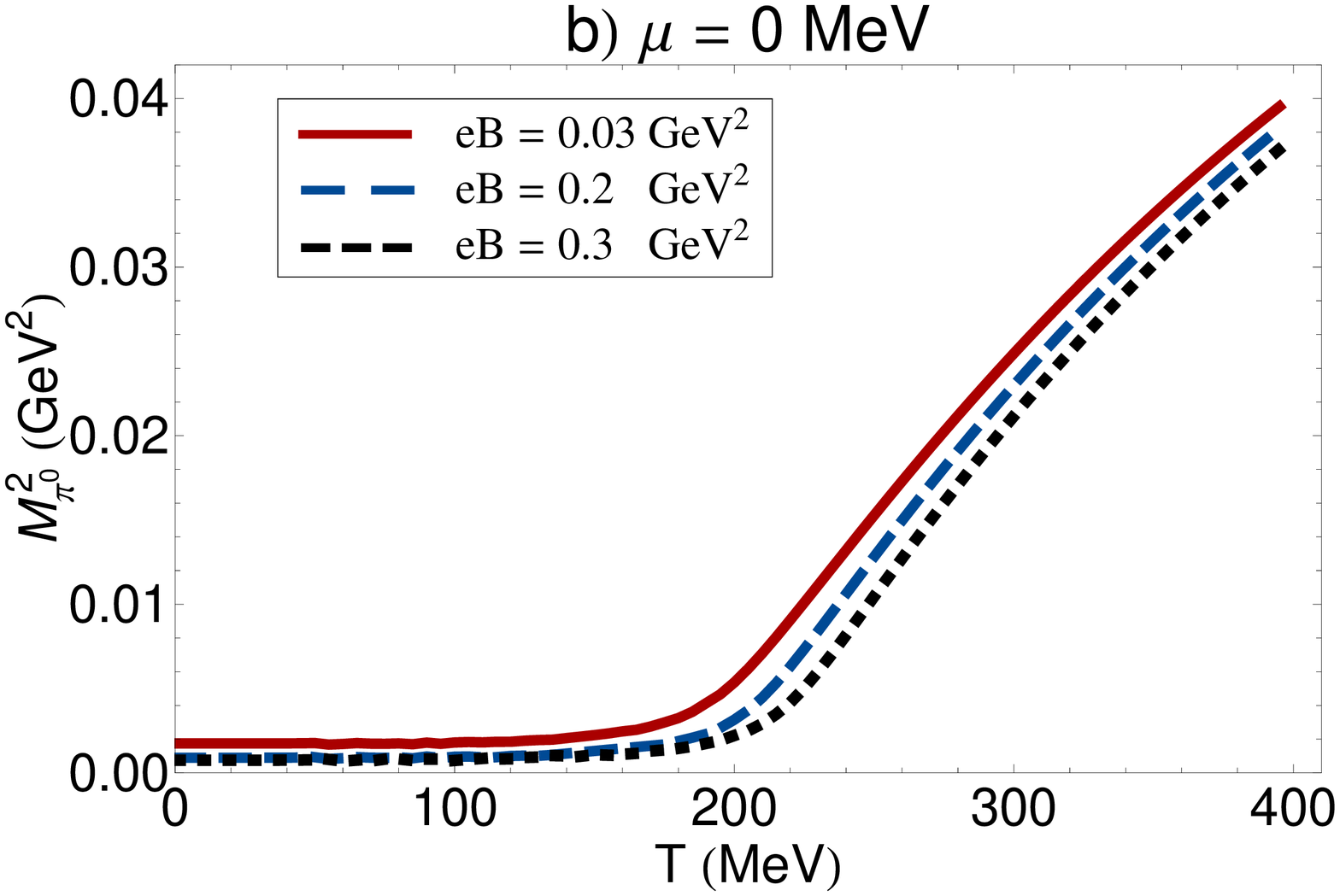}
\caption{The coefficient $M_{\sigma}^{2}$ (panel a) and
$M_{\pi^{0}}^{2}$ (panel b), are plotted as functions of
$T\in[0,400]$ MeV at $\mu=0$ and for $eB=0.03,0.2,0.3$ GeV$^{2}$
(red solid, blue dashed and black dotted lines, respectively)
.}\label{fig4}
\end{figure}
\par\noindent
Let us finally consider the case of $(T,eB\neq
0,\mu=0)$.\footnote{In this paper, we are interested on the effects
of magnetic fields on the meson masses and their refraction indices
at $T\neq 0$ and $\mu=0$. The results for $T\neq 0$ and $\mu\neq 0$
as well as the $eB$-dependence of these quantities will be presented
elsewhere \cite{sadooghi2012-3}.} As it turns out, the degeneracy in
$(M_{\vec{\pi}}^{2})_{\ell\ell}$ as well as ${\cal{G}}^{ii}$ and
$({\cal{F}}^{ii})_{\ell\ell}$, with $\ell,i=1,2,3$ at $(T,eB\neq
0,\mu=0)$ breaks down by finite magnetic fields. In other words, for
$(T,eB\neq 0, \mu=0)$, in contrast to (\ref{NA8}), we have
\begin{eqnarray}\label{NA10}
(M_{\vec{\pi}}^{2})_{11}=(M_{\vec{\pi}}^{2})_{22}\neq
(M_{\vec{\pi}}^{2})_{33}.
\end{eqnarray}
Moreover, in contrast to (\ref{NA9})
\begin{eqnarray}\label{NA11}
{\cal{G}}^{00}=-{\cal{G}}^{33}\neq {\cal{G}}^{11}={\cal{G}}^{22}.
\end{eqnarray}
Similarly, in contrast to (\ref{NA9}), although
$({\cal{F}}^{\mu\mu})_{11}=({\cal{F}}^{\mu\mu})_{22}\neq
({\cal{F}}^{\mu\mu})_{33}$, for all $\mu=0,\cdots,3$, but
\begin{eqnarray}\label{NA12}
({\cal{F}}^{00})_{\ell\ell}=-({\cal{F}}^{33})_{\ell\ell}\neq
({\cal{F}}^{11})_{\ell\ell}=({\cal{F}}^{22})_{\ell\ell},
\end{eqnarray}
$\forall~\ell=1,2,3$. In Fig. \ref{fig4}, the $T$-dependence of
$M_{\sigma}^{2}$ (panel a) and $M_{\pi^{0}}^{2}$ [or equivalently,
$(M_{\vec{\pi}}^{2})_{33}$] (panel b) are demonstrated at $\mu=0$
and for non-vanishing $eB=0.03, 0.2, 0.3$ GeV$^{2}$. The exact
$(T,\mu,eB)$ dependence of
$(M_{\vec{\pi}}^{2})_{11}=(M_{\vec{\pi}}^{2})_{22}$ will be used in
\cite{sadooghi2012-3}, to determine the $(T,\mu,eB)$-dependence of
charged pion masses.
\begin{figure}[hbt]
\includegraphics[width=7.7cm,height=4.5cm]{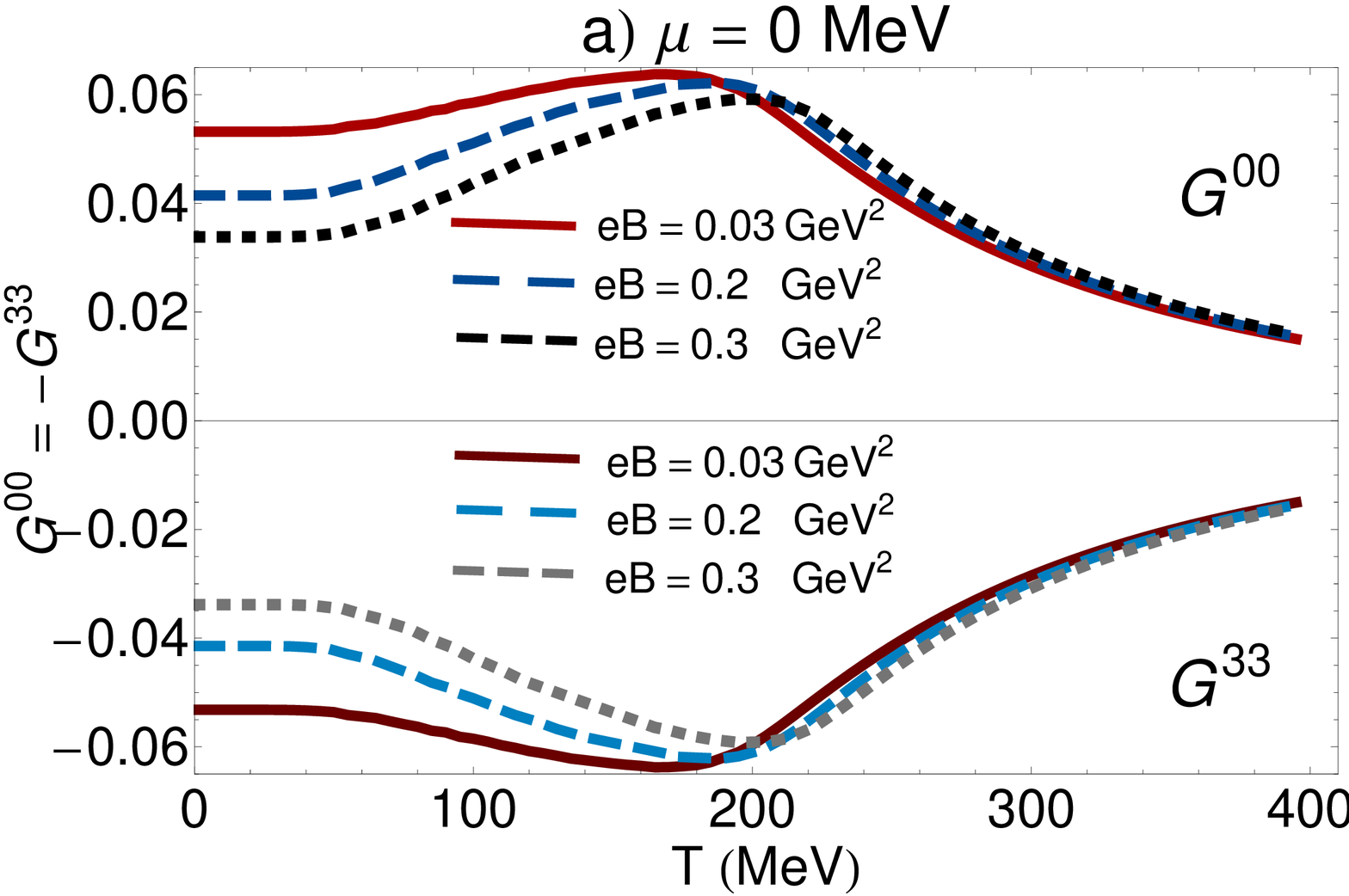}
\par\hspace{0.2cm}
\includegraphics[width=8.7cm,height=5cm]{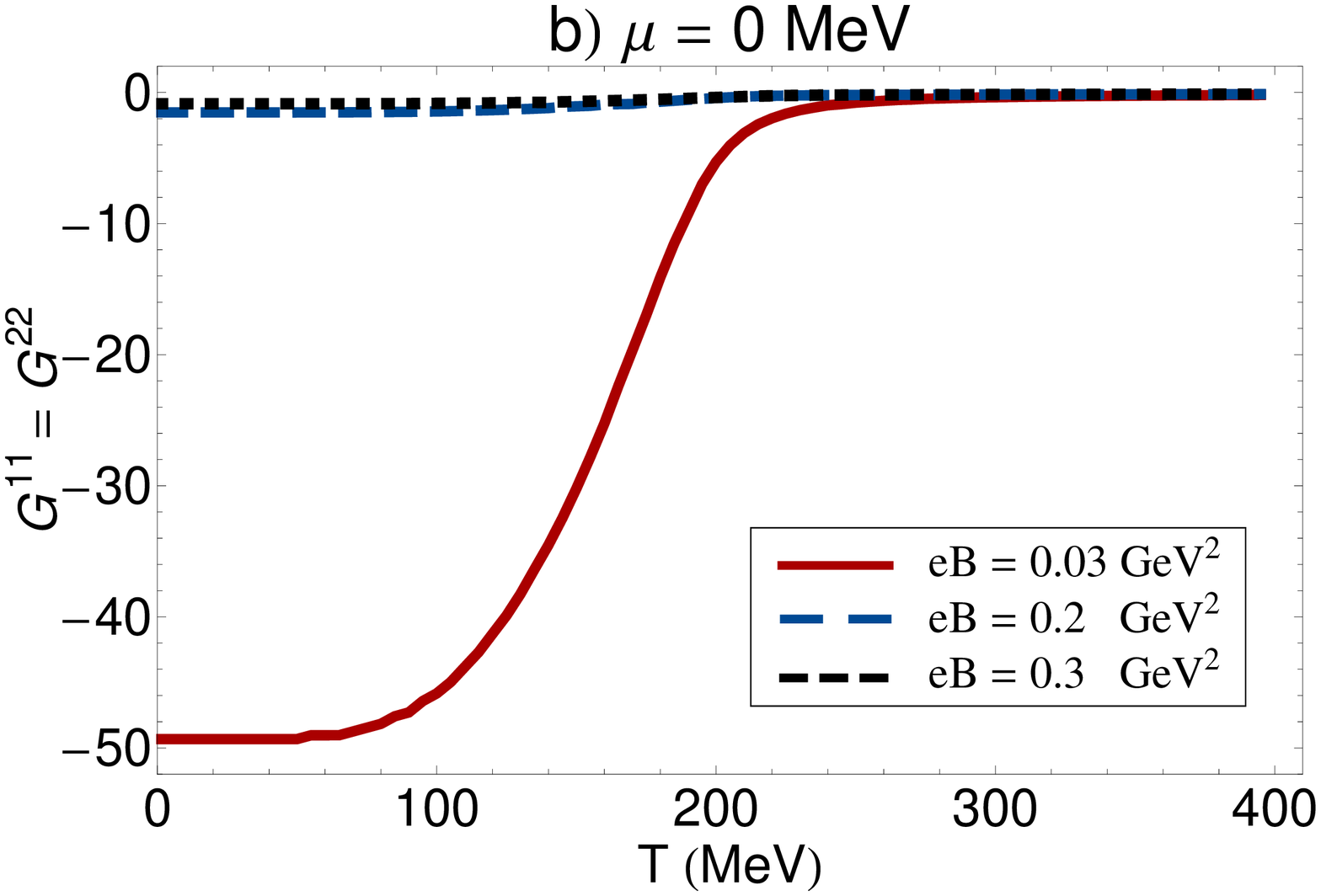}
\caption{The coefficients ${\cal{G}}^{00}=-{\cal{G}}^{33}$ (panel
a), ${\cal{G}}^{11}={\cal{G}}^{22}$ (panel b) are plotted as
functions of $T\in[0,400]$ MeV for $\mu=0$ and $eB=0.03,0.2,0.3$
GeV$^{2}$.}\label{fig5}
\end{figure}
\par
In Fig. \ref{fig5}, the $T$-dependence of ${\cal{G}}^{00}$ and
${\cal{G}}^{33}$ (${\cal{G}}^{00}=-{\cal{G}}^{33}$) (panel a) as
well as ${\cal{G}}^{11}={\cal{G}}^{22}$ (panel b) are plotted for
$\mu=0$ and $eB=0.03, 0.2, 0.3$ GeV$^{2}$. Whereas ${\cal{G}}^{00}$
is positive, ${\cal{G}}^{11}={\cal{G}}^{22}$ and ${\cal{G}}^{33}$
are negative. Later, we will use the matrix elements of
$M_{\sigma}^{2}$ from Fig. \ref{fig4} and the coefficients
${\cal{G}}^{\mu\mu}, \mu=0,\cdots,3$ from Fig. \ref{fig5}, to
determine the $T$-dependence of $m_{\sigma}$ at $\mu=0$ and for
various $eB\neq 0$.
\begin{figure*}[hbt]
\includegraphics[width=7.7cm,height=4.5cm]{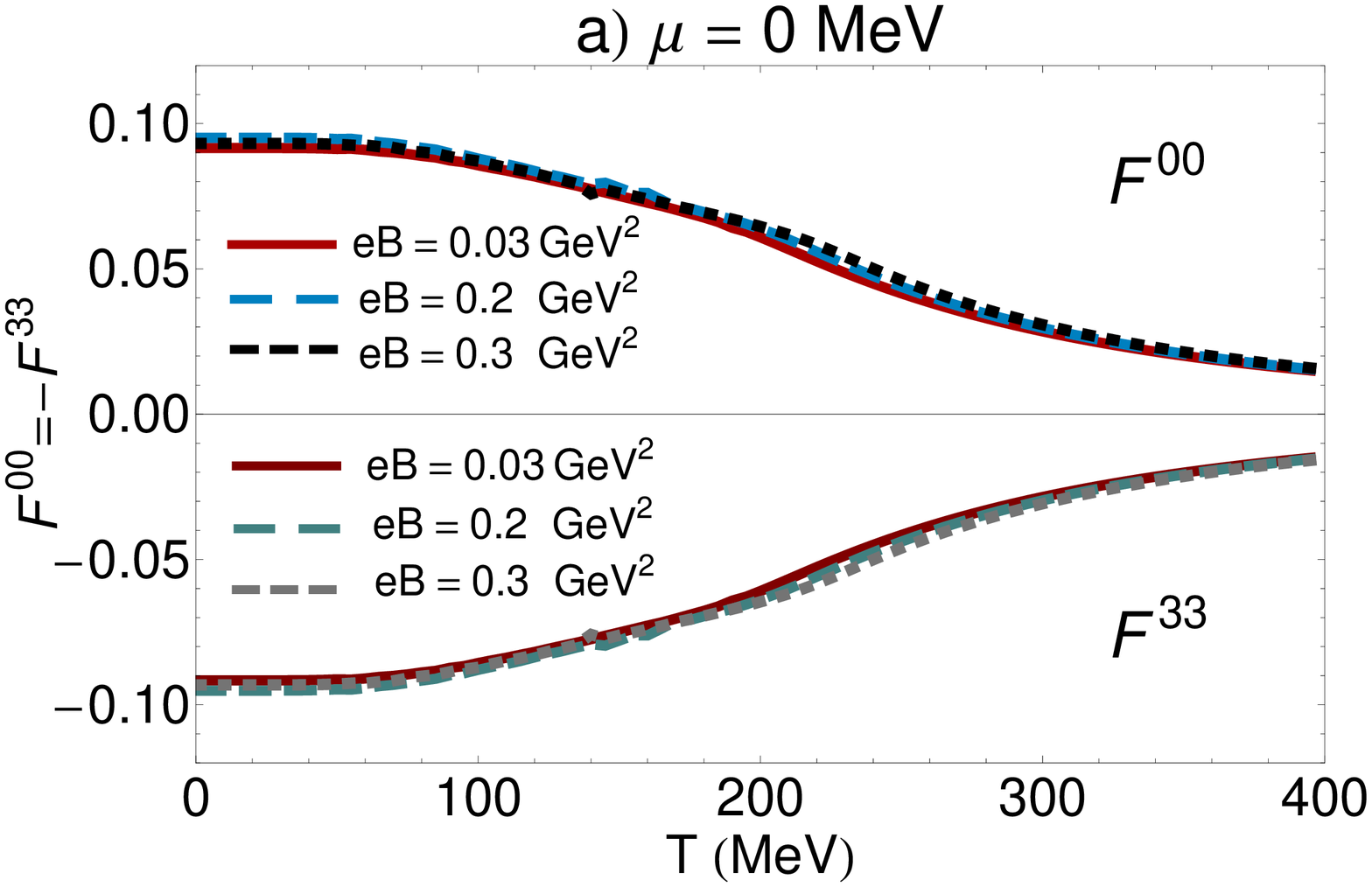}
\includegraphics[width=7.7cm,height=4.5cm]{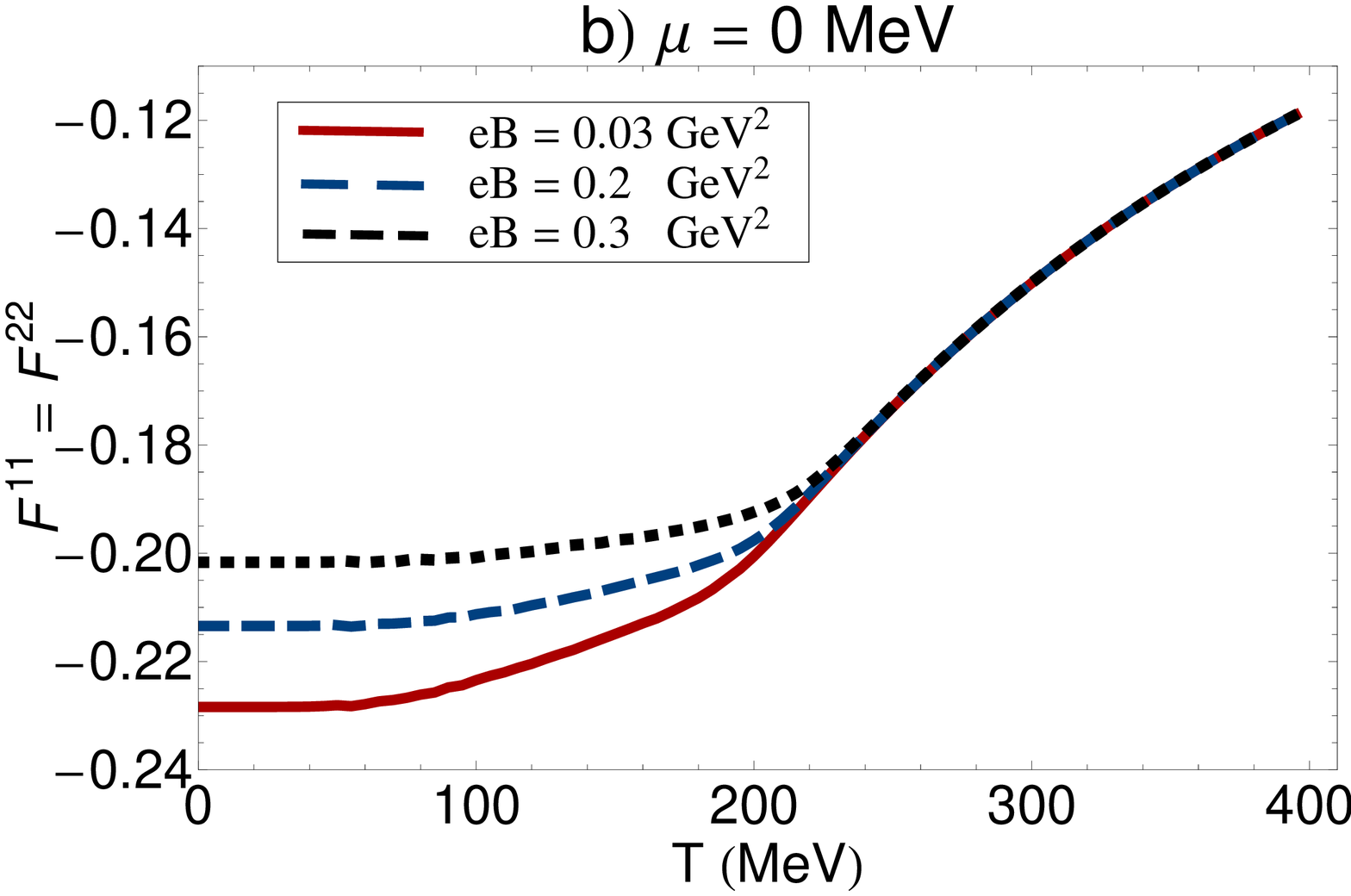}
\caption{The diagonal elements of ${\cal{F}}^{\mu\mu},
\mu=0,\cdots,3$ matrices are plotted as functions of $T\in[0,400]$
MeV at vanishing chemical potential ($\mu=0$) and for $eB=0.03,0.2,
0.3$ GeV$^{2}$. The identities from (\ref{NA12}) for $\ell=3$,
${\cal{F}}^{11}={\cal{F}}^{22}$ and
${\cal{F}}^{00}=-{\cal{F}}^{33}$, are explicitly demonstrated in
these plots.}\label{fig6}
\end{figure*}
\par
In Fig. \ref{fig6}, the $T$-dependence of ${\cal{F}}^{\mu\mu},
\mu=0,\cdots,3$ [or equivalently, $({\cal{F}}^{\mu\mu})_{33}$]
matrices are plotted for vanishing chemical potential and $eB=0.03,
0.2,0.3$ GeV$^{2}$. In the subsequent section, we will in particular
use $M_{\pi^{0}}^{2}$, ${\cal{F}}^{00}$ and ${\cal{F}}^{33}$ to
determine the $T$-dependence of $\pi^{0}$ pole and screening masses
as well as the direction-dependent refraction indices of neutral
pion in the longitudinal and transverse directions with respect to
the direction of the external magnetic field.
%%%%%%%%%%%%%%%%%%%%%%%%%%%%%%%%%%%%%%%%%%%%
\subsection{Masses and directional refraction indices of neutral mesons}\label{sec5p3}
\par\noindent
%%%%%%%%%%%%%%%%%%%%%%%%%%%%%%%%%%%%%%%%%%%%
In this section, we will use the results obtained in Sec.
\ref{sec5p2} to determine the $T$-dependence of pole and screening
masses as well as the direction-dependent refraction indices of
neutral mesons, $\sigma$ and $\pi^{0}$, in a hot and dense
magnetized quark matter. In what follows, we will first define these
quantities according to the descriptions presented in Sec.
\ref{sec2} and the corresponding energy dispersion relations for
neutral and charged mesons $\sigma$ and $\pi_{\ell}, \ell=1,2,3$
mesons [see also (\ref{NN15}) and (\ref{NN15b})],
\begin{eqnarray}\label{NA13}
E_{\sigma}^{2}&=&\frac{1}{{\cal{G}}^{00}}\left({\cal{G}}^{11}p_{1}^{2}+{\cal{G}}^{22}p_{2}^{2}+{\cal{G}}^{33}p_{3}^{2}+M_{\sigma}^{2}\right),
\nonumber\\
\lefteqn{\hspace{-0.5cm}E_{\pi_{\ell}}^{2}\hspace{0.05cm}=
}\nonumber\\
&&\hspace{-1.1cm}\frac{1}{({\cal{F}}^{00})_{\ell\ell}}\big[({\cal{F}}^{11})_{\ell\ell}p_{1}^{2}+({\cal{F}}^{22})_{\ell\ell}p_{2}^{2}+
({\cal{F}}^{33})_{\ell\ell}p_{3}^{2}+
M_{\pi_{\ell}}^{2}\big].\hspace{-0.2cm}\nonumber\\
\end{eqnarray}
The pole and screening masses of $\sigma$-mesons, $m_{\sigma}$ and
$m_{\sigma}^{(i)}, i=1,2,3$, are then defined by
\begin{eqnarray}\label{NA14}
\hspace{-0.5cm}m_{\sigma}=\bigg[\frac{\mbox{Re}~M_{\sigma}^{2}}{\mbox{Re}~{\cal{G}}^{00}}\bigg]^{1/2},~~\mbox{as
well as}~~m_{\sigma}^{(i)}=\frac{m_{\sigma}}{u_{\sigma}^{(i)}},
\end{eqnarray}
where, $u_{\sigma}^{(i)}$, is the directional refraction index of
$\sigma$-mesons in the $i$-th direction,
\begin{eqnarray}\label{NA15}
u_{\sigma}^{(i)}=\bigg|\frac{\mbox{Re}~{\cal{G}}^{ii}}{\mbox{Re}~{\cal{G}}^{00}}\bigg|^{1/2},
\qquad i=1,2,3.
\end{eqnarray}
The $(T,\mu,eB)$-dependence of $m_{\sigma}, m_{\sigma}^{(i)}$ and
$u_{\sigma}^{(i)}$ are given by plugging $M_{\sigma}^{2}$ and
${\cal{G}}^{\mu\mu}, \mu=0,\cdots,3$ from (\ref{ND28b}),
(\ref{ND42b}) and (\ref{ND57b}) in the above relations. As concerns
the pions, we choose the basis $(\pi^{\pm},\pi^{0})$ instead of the
real basis $\vec{\pi}=(\pi_{1},\pi_{2},\pi_{3})$. Here,
$\pi^{\pm}\equiv (\pi_{1}\pm i\pi_{2})/\sqrt{2}$ and $\pi^{0}\equiv
\pi_{3}$. Using this new imaginary basis, the energy dispersion
relations $E_{\pi_{\ell}}$ from (\ref{NA13}) for
$(\pi^{\pm},\pi^{0})$ turn out to be
\begin{eqnarray}\label{NA16}
\lefteqn{E_{\pi^{\pm}}^{2}\equiv
\frac{eB(2\ell+1)}{[({\cal{F}}^{00})_{11}\mp
i({\cal{F}}^{00})_{12}]}
}\nonumber\\
&&+\frac{[({\cal{F}}^{33})_{11}\mp
i({\cal{F}}^{33})_{12}]}{[({\cal{F}}^{00})_{11}\mp
i({\cal{F}}^{00})_{12}]}p_{3}^{2}+\frac{[(M_{\vec{\pi}}^{2})_{11}\mp
i(M_{\vec{\pi}}^{2})_{12}]}{[({\cal{F}}^{00})_{11}\mp
i({\cal{F}}^{00})_{12}]},\nonumber\\
\lefteqn{E_{\pi^{0}}^{2}\equiv
\frac{({\cal{F}}^{11})_{33}}{({\cal{F}}^{00})_{33}}~p_{1}^{2}
}\nonumber\\
&&+ \frac{({\cal{F}}^{22})_{33}}{({\cal{F}}^{00})_{33}}~p_{2}^{2}+
\frac{({\cal{F}}^{33})_{33}}{({\cal{F}}^{00})_{33}}~p_{3}^{2}+
\frac{(M_{\vec{\pi}}^{2})_{33}}{({\cal{F}}^{00})_{33}}.
\end{eqnarray}
Note that since $\pi^{\pm}$ are charged pseudoscalar particles,
their energy dispersion relations in the presence of constant
magnetic fields have discrete contributions. According to our
results in \cite{jafarisalim2006}, the energy levels are labeled by
$\ell$, in the form given in the first term in (\ref{NA16}). The
above dispersion relations for charged pions are comparable with the
dispersion relations presented recently in \cite{anderson2012-2}
(see Eq. (2.10) in \cite{anderson2012-2}). According to the
formalism presented originally in \cite{miransky1995} and
generalized to a multi-flavor system in the present paper, in
contrast to the relations presented in \cite{anderson2012-2} for
charged pions, the nontrivial form factors
$({\cal{F}}^{\mu\mu})_{\ell m}$, $\forall~ \ell,m\neq 3$ in
(\ref{NA16}), consider the effect of external magnetic fields on
charged quarks produced at the early stage of the heavy-ion
collisions. Moreover, in the formalism presented in
\cite{anderson2012-2}, in contrast to the dispersion relations
presented in (\ref{NA16}), the energy dispersion relation of neutral
pion is unaffected by the external magnetic field. Using
(\ref{NA16}), and in analogy to (\ref{NA14}), the pions pole masses
are defined by
\begin{eqnarray}\label{NA17}
m_{\pi^{\pm}}&=&\bigg[\frac{\mbox{Re}~[(M_{\vec{\pi}}^{2})_{11}\mp
i(M_{\vec{\pi}}^{2})_{12}]}{\mbox{Re}~[({\cal{F}}^{00})_{11}\mp
i({\cal{F}}^{00})_{12}]}\bigg]^{1/2},\nonumber\\
m_{\pi^{0}}&=&\bigg[\frac{\mbox{Re}~(M_{\pi}^{2})_{33}}{\mbox{Re}~({\cal{F}}^{00})_{33}}\bigg]^{1/2}.
\end{eqnarray}
In particular, the screening mass and the refraction index of
neutral pions in the $i$-th direction are given by
\begin{eqnarray}\label{NA18}
\hspace{-0.5cm}m_{\pi^{0}}^{(i)}=\frac{m_{\pi^{0}}}{u_{\pi^{0}}^{(i)}},~~~\mbox{and}~~~
u_{\pi^{0}}^{(i)}=\bigg|\frac{\mbox{Re}~({\cal{F}}^{ii})_{33}}{\mbox{Re}~({\cal{F}}^{00})_{33}}\bigg|^{1/2},
\end{eqnarray}
respectively. In this paper, we will focus on the $T$-dependence of
the mass and refraction index of neutral pions at fixed $\mu$ and
finite $eB$. The study of the effect of constant magnetic fields on
charged pion masses and refraction indices will be postponed to a
future publication \cite{sadooghi2012-3}.
\par
Let us start with the case $T=\mu=eB=0$. Using the numerical results
from (\ref{NA6}) and (\ref{NA7}), in this case, the $\sigma$-meson
mass and refraction index are given by
\begin{eqnarray}\label{NA19}
m_{\sigma}\simeq 824.3~\mbox{MeV},\qquad\mbox{and}\qquad
u_{\sigma}^{(i)}=1,
\end{eqnarray}
and therefore
\begin{eqnarray}\label{NA21b}
m_{\sigma}^{(i)}=m_{\sigma},~\forall i=1,2,3.
\end{eqnarray}
Similarly, the $\vec{\pi}$-meson mass and refraction index at
$T=\mu=eB=0$ read
\begin{eqnarray}\label{NA20}
m_{\pi_{\ell}}\simeq 137.7~\mbox{MeV},\qquad\mbox{and}\qquad
u_{\pi_{\ell}}^{(i)}=1,
\end{eqnarray}
$\forall~\ell,i=1,2,3$, and therefore
\begin{eqnarray}\label{NA22b}
m_{\pi_{\ell}}^{(i)}=m_{\pi_{\ell}},~\forall \ell,i=1,2,3.
\end{eqnarray}
At $(T\neq 0,\mu=eB=0)$, $m_{\sigma}$ is given by (\ref{NA14}).
Similarly, according to (\ref{NA13}), the pion masses
$m_{\vec{\pi}}$ are defined by
\begin{eqnarray}\label{NA21}
m_{\pi_{\ell}}=\bigg[\frac{\mbox{Re}~(M_{\vec{\pi}}^{2})_{\ell\ell}}{\mbox{Re}~
({\cal{F}}^{00})_{\ell\ell}}\bigg]^{1/2},\qquad \forall\ell=1,2,3.
\end{eqnarray}
\begin{figure}[hbt]
\includegraphics[width=7.7cm,height=4.5cm]{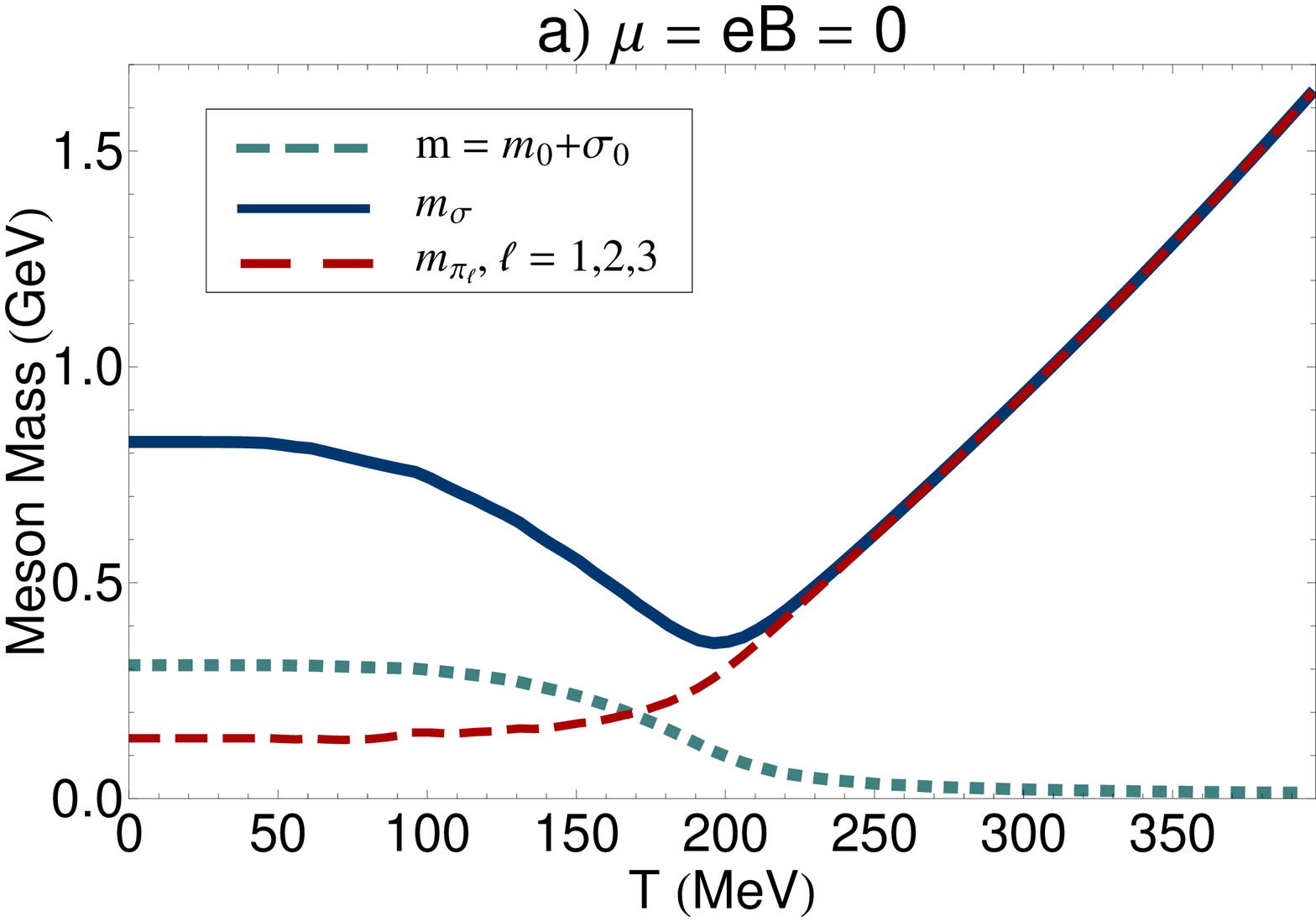}
\includegraphics[width=7.7cm,height=4.5cm]{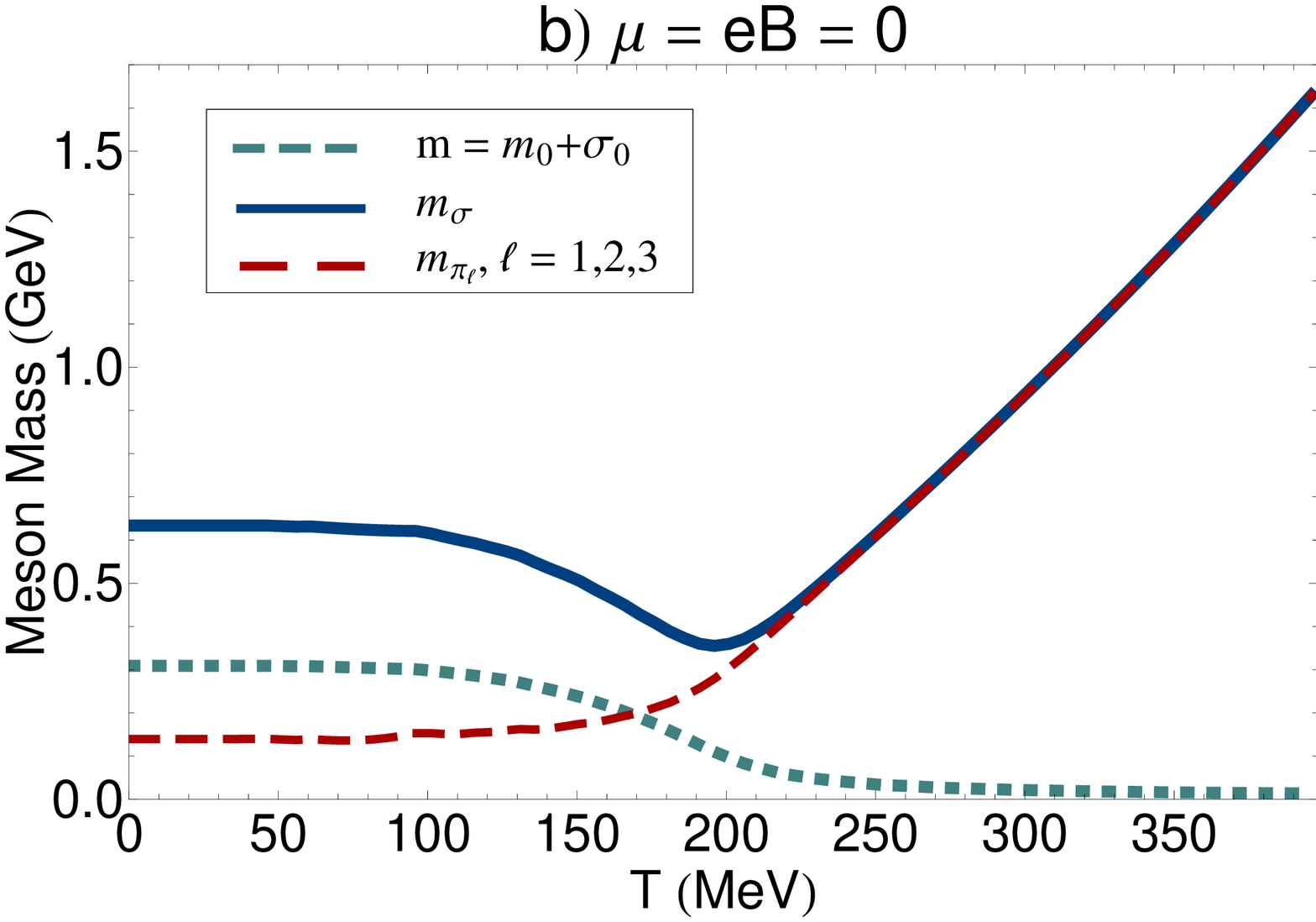}
\caption{(a) The $T$-dependence of $\sigma$ and $\vec{\pi}$ mesons
masses is demonstrated at $\mu=0$ and for vanishing magnetic field
(solid line for $m_{\sigma}$ and dashed line for $m_{\vec{\pi}}$).
Comparing these curves with the $T$-dependence of the constituent
quark mass $m=m_{0}+\sigma_{0}$ (dotted line), shows that a mass
degeneracy between $m_{\sigma}$ and $m_{\vec{\pi}}$ occurs in the
crossover region at $T>220$ MeV. (b) The $T$-dependence of the pole
masses of neutral mesons is plotted for the case when
$F_{2}^{\mu\nu}$ in (\ref{NN10}) vanishes. The numerical results for
$m_{\sigma}$ and $m_{\vec{\pi}}$ are in good agreement with the
results recently presented in \cite{buballa2012}.}\label{fig7}
\end{figure}
\begin{figure}[hbt]
\includegraphics[width=7.7cm,height=4.5cm]{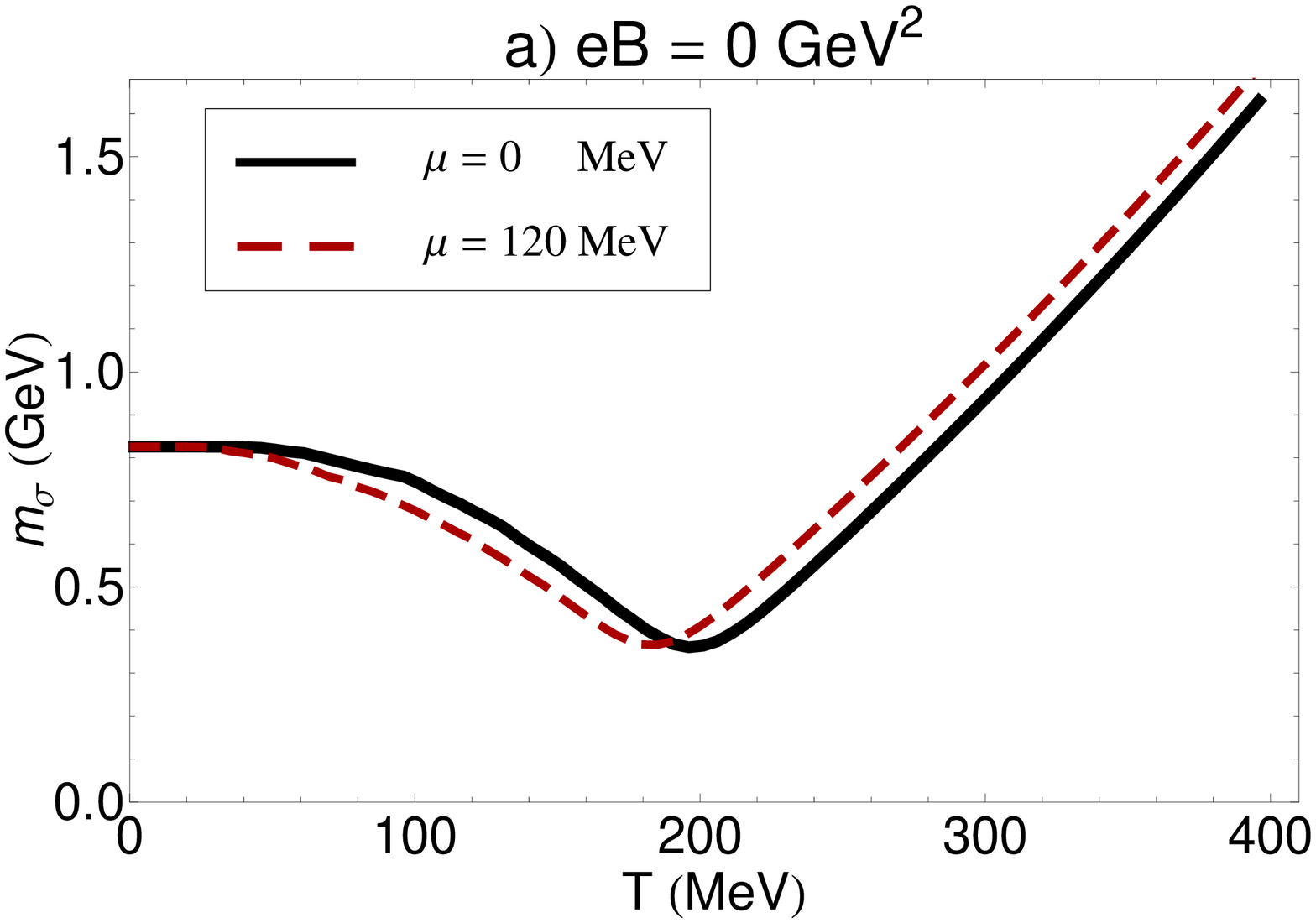}
\includegraphics[width=7.7cm,height=4.5cm]{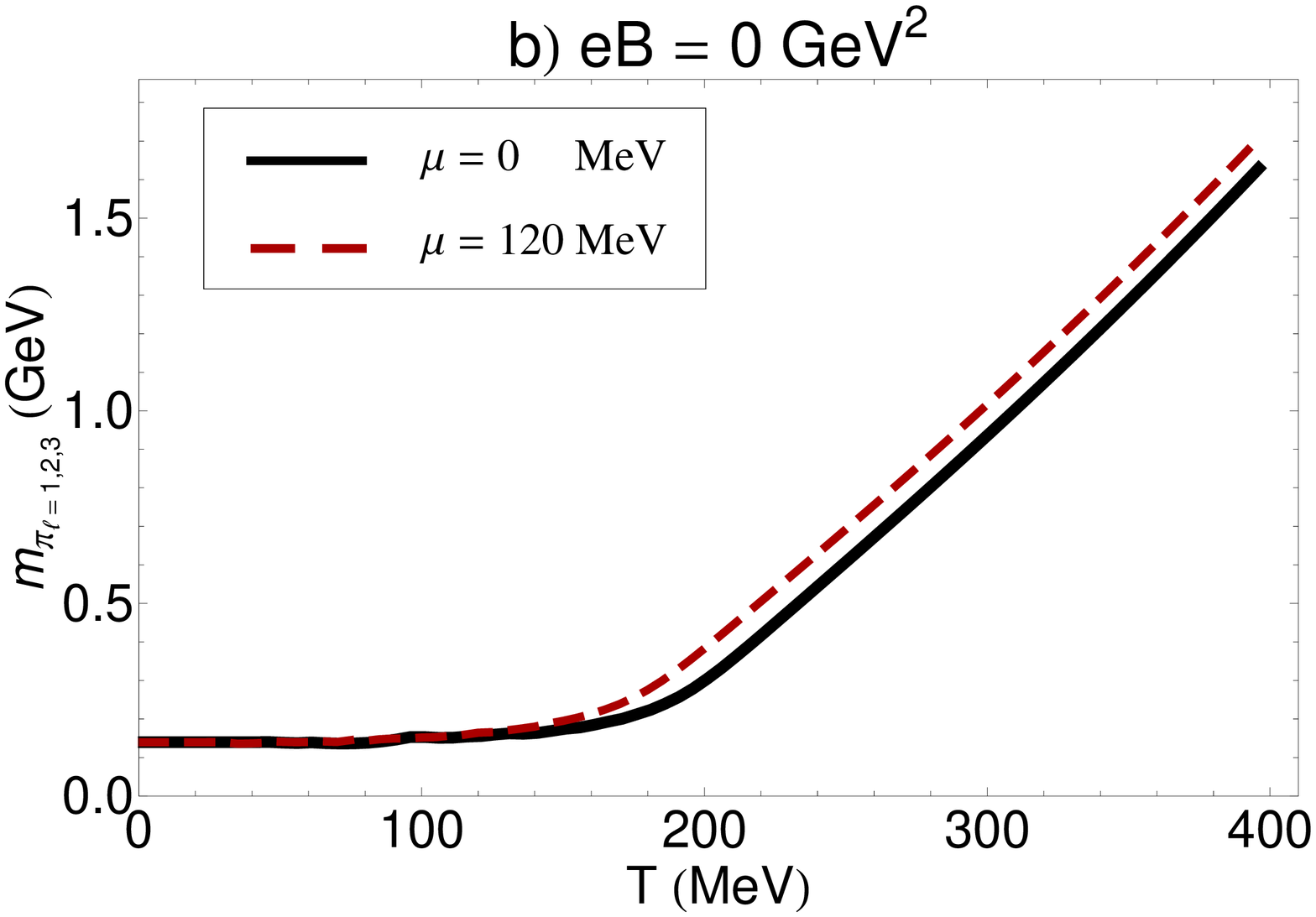}
\caption{The $T$-dependence of $m_{\sigma}$ (panel a) and
$m_{\vec{\pi}}$ (panel b) is plotted for $eB=0$ and at $\mu=0,120$
MeV. As in the case of $\mu=0$, at $\mu=120$ MeV, the neutral meson
masses are still degenerate in the crossover region $T>220$
MeV.}\label{fig8}
\end{figure}
\par\noindent Because of the identity (\ref{NA8}), which is still valid at
$(T\neq 0,\mu=eB=0)$, the masses of $\vec{\pi}=(\pi_{1}, \pi_{2},
\pi_{3})$ are degenerate, as in $T=0$ case [see (\ref{NA20})]. In
Fig. \ref{fig7}(a), the $T$-dependence of $m_{\sigma}$ and
$m_{\vec{\pi}}$ is plotted for $\mu=eB=0$ (black solid line for
$m_{\sigma}$ and red dashed line for $m_{\vec{\pi}}$). Here, the
$T$-dependence of the coefficients ${\cal{G}}^{\mu\mu}$ and
${\cal{F}}^{\mu\mu}$ from Fig. \ref{fig3} is used. We have also
plotted the $T$-dependence of the constituent mass
$m=m_{0}+\sigma_{0}$ in Fig. \ref{fig7}(a) (dotted line). Comparing
these curves, it turns out that, as expected, the mass degeneracy of
$\sigma$ and $\vec{\pi}$ meson masses occurs in the crossover region
$T>220$ MeV. To compare the result presented in Fig. \ref{fig7}(a),
with the recent results for $m_{\vec{\pi}}$ and $m_\sigma$,
presented e.g. in \cite{buballa2012}, we have set $F_{2}^{\mu\nu}=0$
in (\ref{NN10}), and determined the pole masses of neutral mesons
and the chiral condensate using the same method as presented in this
paper. The numerical results for neutral meson masses and chiral
condensate for $T\neq 0$, $\mu=eB=0$ and vanishing $F_{2}^{\mu\nu}$
are plotted in Fig. \ref{fig7}(b). As it turns out, only
$m_{\sigma}$ changes relative to the case where $F_{2}^{\mu\nu}\neq
0$ [see Fig. \ref{fig7}(b)]. The numerical results are in good
agreement with the results presented in \cite{buballa2012}.
\par
As concerns the screening mass and refraction index of $\vec{\pi}$
mesons at $(T\neq 0, \mu=eB=0)$, we use the results of Fig.
\ref{fig3}, and in analogy to the definitions (\ref{NA14}) and
(\ref{NA15}) define the screening mass and refraction index of
$\vec{\pi}$ mesons by
\begin{eqnarray}\label{NA22}
m_{\pi_{\ell}}^{(i)}&=&\frac{m_{\pi_{\ell}}}{u_{\pi_{\ell}}^{(i)}},\qquad\mbox{where}\nonumber\\
u_{\pi_{\ell}}^{(i)}&=&\bigg|\frac{\mbox{Re}~({\cal{F}}^{ii})_{\ell\ell}}{\mbox{Re}~({\cal{F}}^{00})_{\ell\ell}}\bigg|^{1/2},
~~\forall \ell,i=1,2,3.
\end{eqnarray}
Using the definitions (\ref{NA15}) and (\ref{NA22}), and the
numerical results of ${\cal{G}}^{00}$ as well as
$({\cal{F}}^{00})_{\ell\ell}$ from Fig. \ref{fig3} at $(T\neq
0,\mu=eB=0)$, the $T$-dependence of the screening mass and
refraction index of $(\sigma,\vec{\pi})$ mesons can be determined
for all directions $i=1,2,3$. As it turns out, as in $T=0$ case, we
have
\begin{eqnarray}\label{NA23}
\begin{array}{rclcrcl}
u_{\sigma}^{(i)}&=&1,&\qquad&\forall i&=&1,2,3,\\
u_{\pi_{\ell}}^{(i)}&=&1,&\qquad&\forall \ell,i&=&1,2,3,
\end{array}
\end{eqnarray}
and therefore
\begin{eqnarray}\label{NA27b}
\begin{array}{rclcrcl}
m_{\sigma}^{(i)}&=&m_{\sigma},&\qquad&\forall i&=&1,2,3,\\
m_{\pi_{\ell}}^{(i)}&=&m_{\pi_{\ell}},&\qquad&\forall
\ell,i&=&1,2,3,
\end{array}
\end{eqnarray}
for the whole interval $T\in[0,400]$. These results are compatible
with the identities (\ref{NA9}).  The fact that $u_{\pi_{\ell}}=1$
seems to be in contradiction with the results from \cite{son2000,
ayala2002}, where it is shown that at finite temperature because of
different pion decay constants in the spatial and temporal
directions, $f_{s}$ and $f_{t}$, at finite temperature, the
refraction index $u=\frac{f_{s}}{f_{t}}$ appearing in the energy
dispersion relation $\omega^{2}=u^{2}(\mathbf{p}^{2}+m^{2})$ is
smaller than one. Note, however, that in \cite{son2000, ayala2002},
the pions are self-interacting and $f_{s}$ and $f_{t}$ receive
$T$-dependent contributions from one-loop pion-self energy diagram,
that includes a $(\vec{\pi}^{2})^{2}$ vertex. In contrast, the pions
considered in the present paper are free.
\begin{figure*}[hbt]
\includegraphics[width=5.5cm,height=4cm]{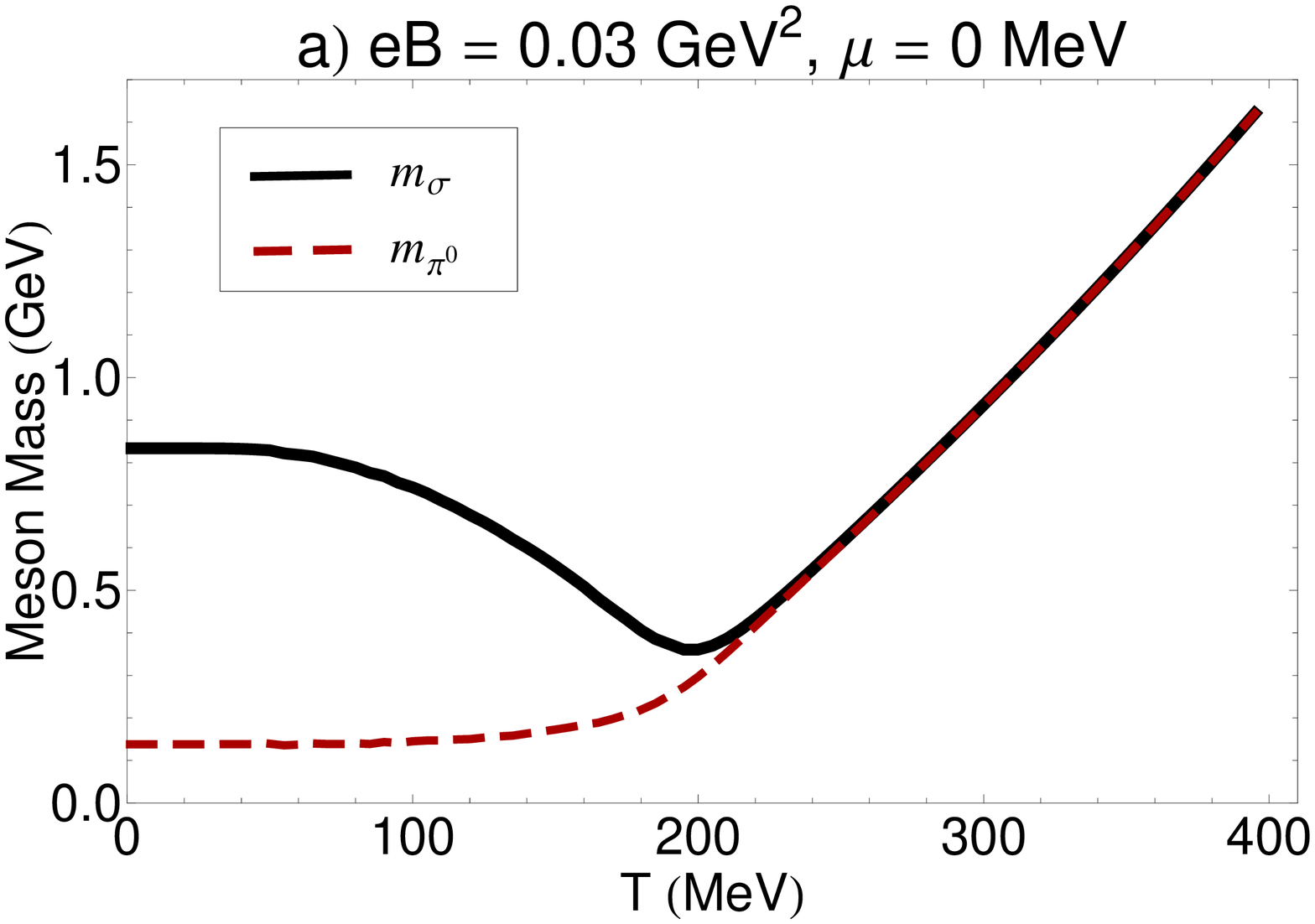}
\includegraphics[width=5.5cm,height=4cm]{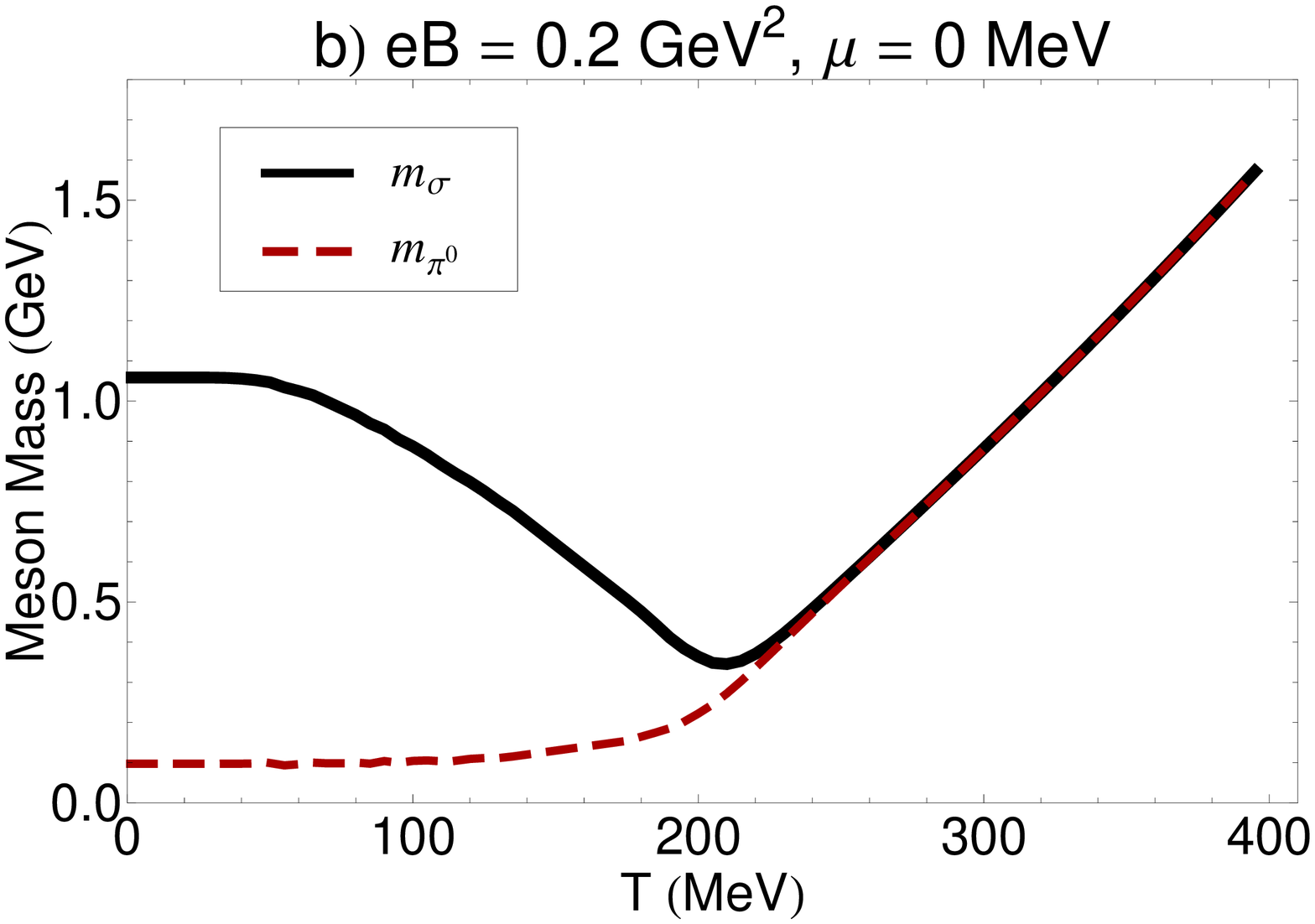}
\includegraphics[width=5.5cm,height=4cm]{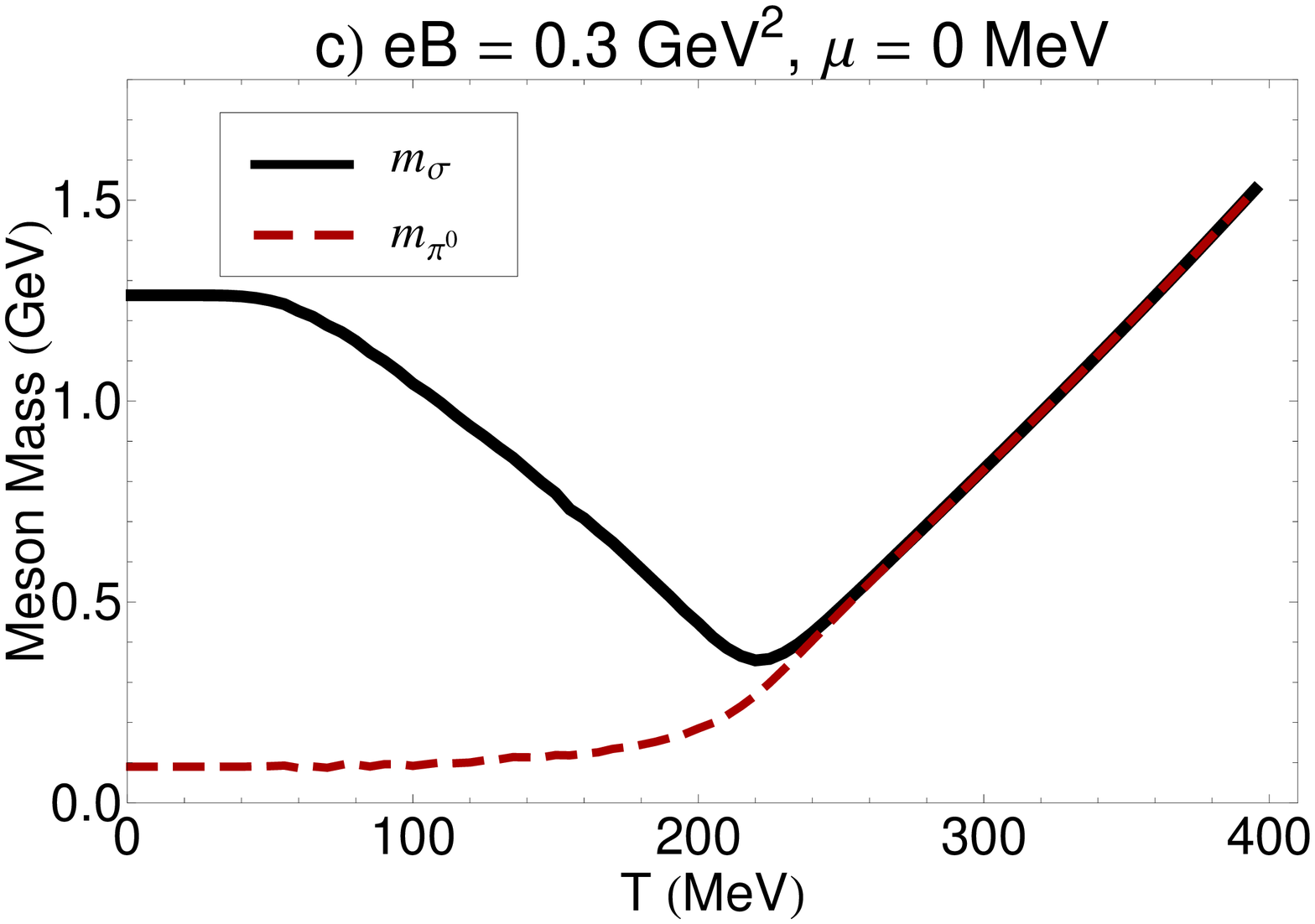}
\caption{The $T$-dependence of $m_{\sigma}$ and $m_{\pi^{0}}$ is
plotted for $eB=0.03,0.2,0.3$ GeV$^{2}$ and at $\mu=0$ MeV.
}\label{fig9}
\end{figure*}
\par
Let us also notice that the above results are still valid at
non-vanishing $\mu$ and for vanishing $eB$. In Fig. \ref{fig8}, we
have compared $m_{\sigma}$ and $m_{\pi_{\ell}}, \ell=1,2,3$ at
$\mu=0$ with their values at $\mu=120$ MeV for vanishing $eB$. Small
deviations from their value at $\mu=0$ appears for $m_{\sigma}$
(panel a). For $m_{\pi_{\ell}}, \ell=1,2,3$ (panel b) the difference
between $m_{\pi_{\ell}}$ at $\mu=0$ and $\mu=120$ MeV becomes larger
with increasing temperature. As it turns out, at non-vanishing
chemical potential, the degeneracy of the pion masses
$m_{\pi_{\ell}}$ is still valid for all $\ell=1,2,3$. Moreover,
$m_{\sigma}$ and $m_{\vec{\pi}}$ are also degenerate in the
crossover region $T>220$ MeV for $\mu=120$ MeV, as in the $\mu=0$
case.
\par
At finite $T$ and for non-vanishing magnetic fields, the pion masses
are not degenerate, i.e. we have $m_{\pi^{+}}\neq m_{\pi^{-}}\neq
m_{\pi^{0}}$ [see (\ref{NA10}) and (\ref{NA11}) and the definitions
of $m_{\pi^{\pm}}$ and $m_{\pi^{0}}$ from (\ref{NA17})]. In this
paper, we will focus on the $T$-dependence of $m_{\sigma}$ and
$m_{\pi^{0}}$. In Figs. \ref{fig9}(a)-\ref{fig9}(c), the
$T$-dependence of $(m_{\sigma},m_{\pi^{0}})$ masses are plotted for
$eB=0.03,0.2,0.3$ GeV$^{2}$ and at $\mu=0$. The expected degeneracy
of $m_{\sigma}$ and $m_{\pi^{0}}$ mesons in the crossover region can
be observed in all the plots of Fig. \ref{fig9}. However, as it
turns out, the overlap interval depends on $eB$ for fixed $\mu$.
Denoting the minimum temperature for which the overlap interval
starts with $T_{o}$, then for $eB=0.03$ GeV$^{2}$ we have
$T_{o}\simeq 210$ MeV, whereas for $eB=0.2,0.3$ GeV$^{2}$, $T_{o}$
are given by $T_{o}\simeq 220$ MeV and $T_{o}\simeq 240$ MeV,
respectively.
\begin{figure}[hbt]
\includegraphics[width=7.7cm,height=4.5cm]{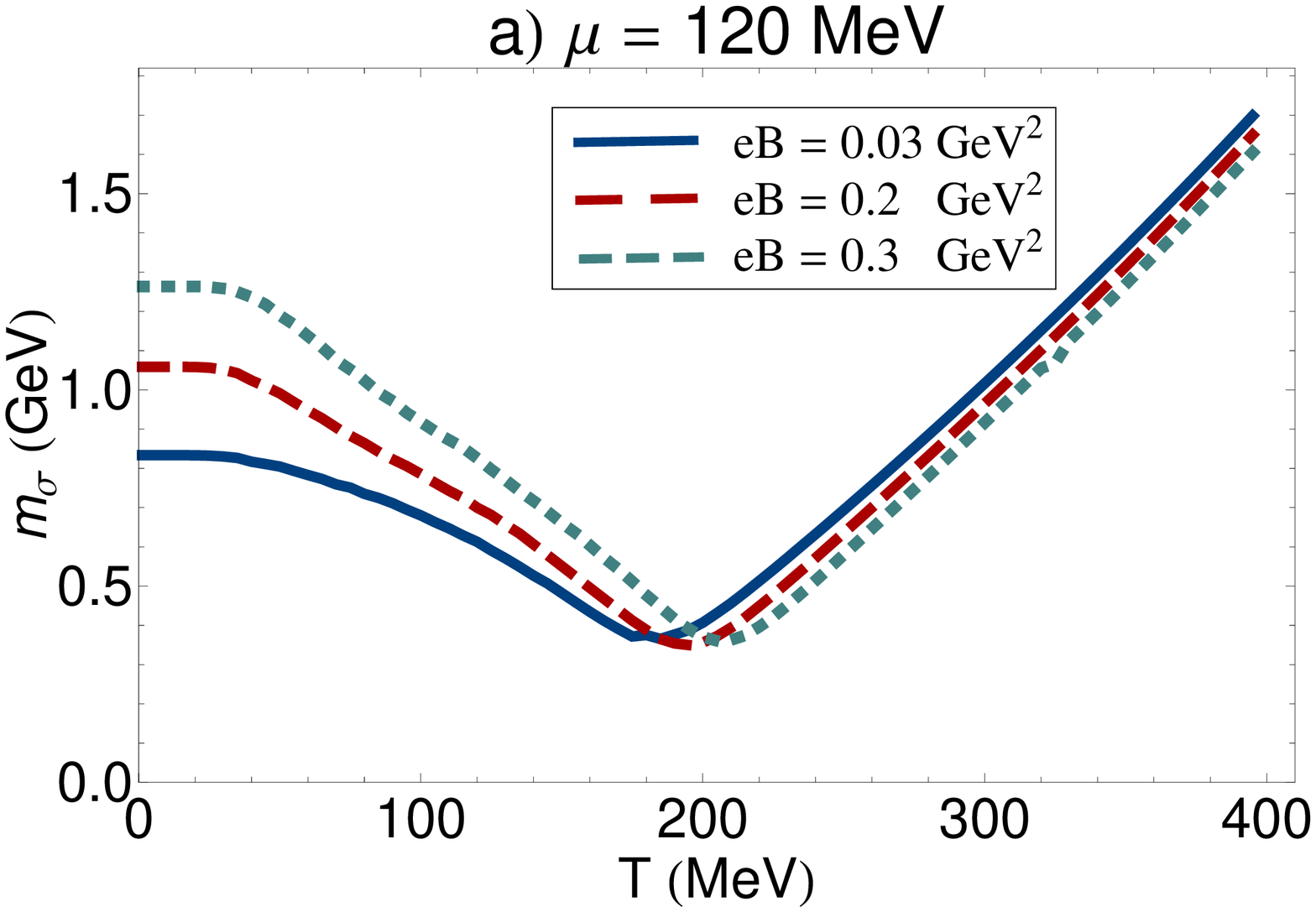}
\includegraphics[width=7.7cm,height=4.5cm]{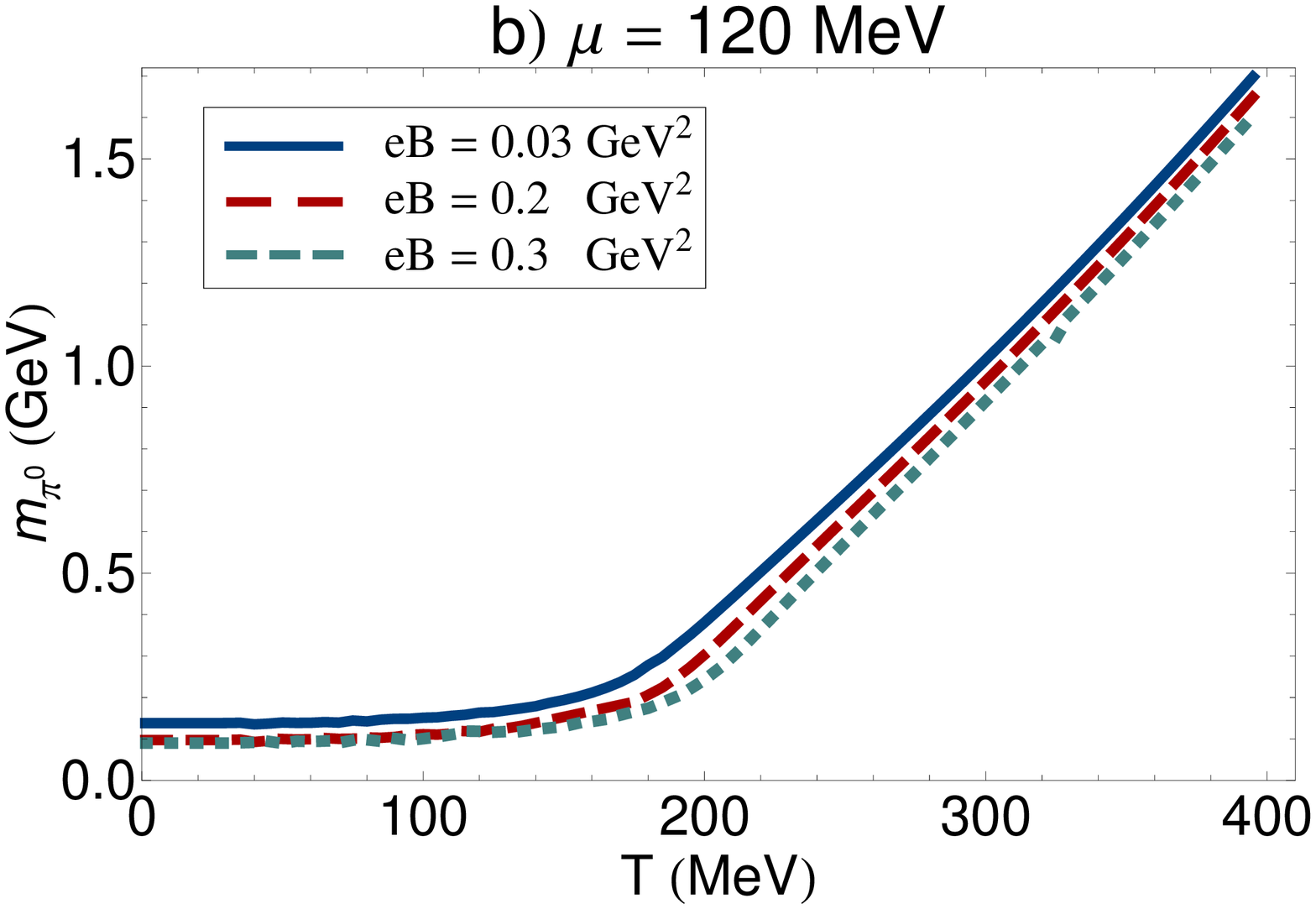}
\caption{The $T$-dependence of $m_{\sigma}$ (panel a) and
$m_{\pi^{0}}$ (panel b) is plotted for $eB=0.03,0.2,0.3$ GeV$^{2}$
at $\mu=120$ MeV.}\label{fig10}
\end{figure}
\par
In Fig. \ref{fig10}, we have compared the $T$-dependence of the
masses of $\sigma$ and $\pi^{0}$ mesons for fixed $\mu=120$ MeV and
various $eB=0.03,0.2,0.3$ GeV$^{2}$. As it turns out, at temperature
below (above) the crossover region, the $\sigma$-meson masses
increase (decrease) with increasing the magnetic field strength.
This qualitative behavior of the $T$-dependence of $m_{\sigma}$ for
various $eB\neq 0$ is comparable with the results presented in
\cite{skokov2011} (see Fig. 3 in \cite{skokov2011}). The difference
arises from the fact that, in contrast to the present paper, the
quantum fluctuations of $\sigma$-mesons is considered in
\cite{skokov2011}. And, in contrast to the present paper, the
contribution of $F_{2}^{\mu\nu}$ appearing in (\ref{NN10}) is not
considered in \cite{skokov2011}.
\begin{figure}[hbt]
\includegraphics[width=7.7cm,height=4.5cm]{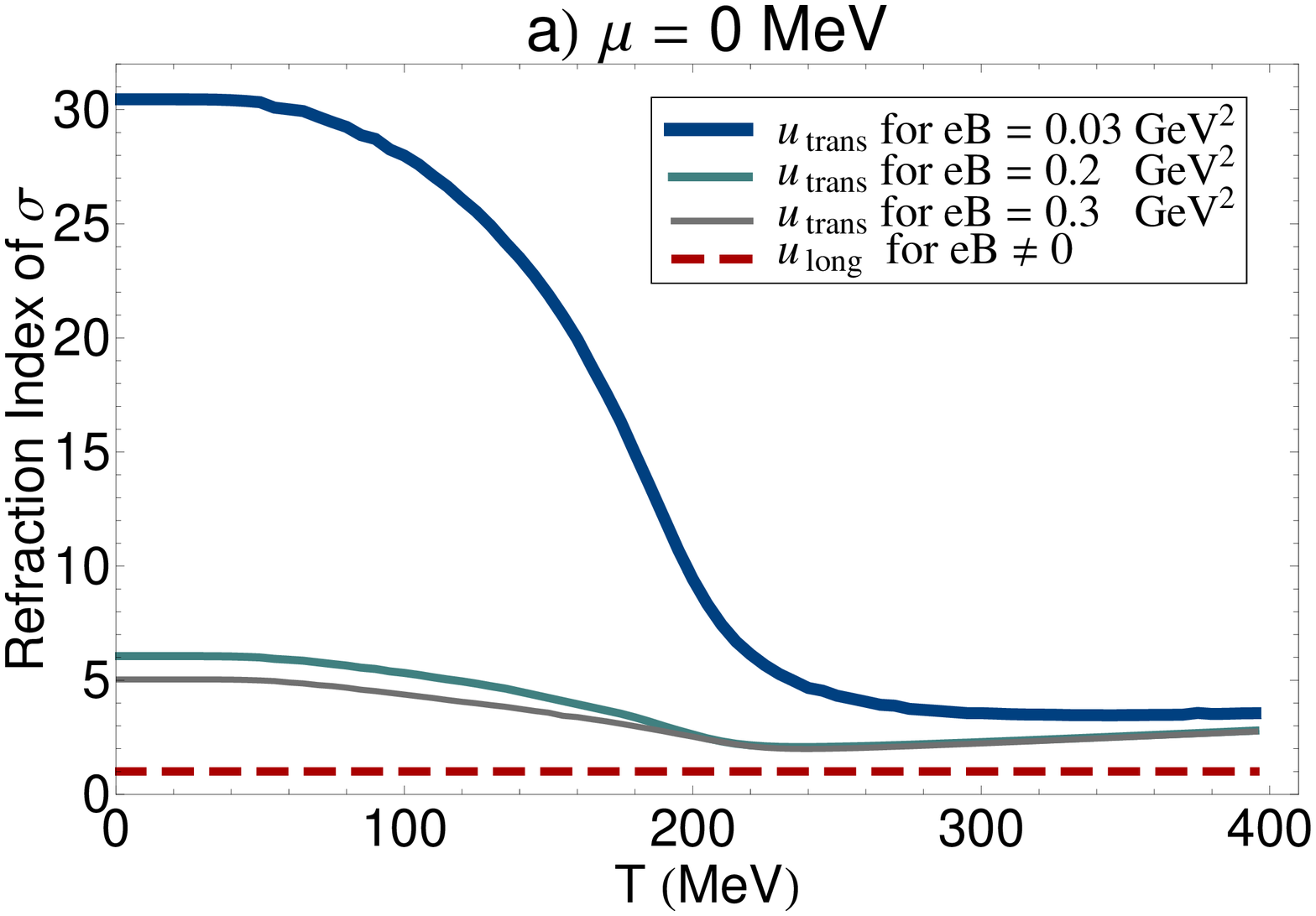}
\includegraphics[width=7.7cm,height=4.5cm]{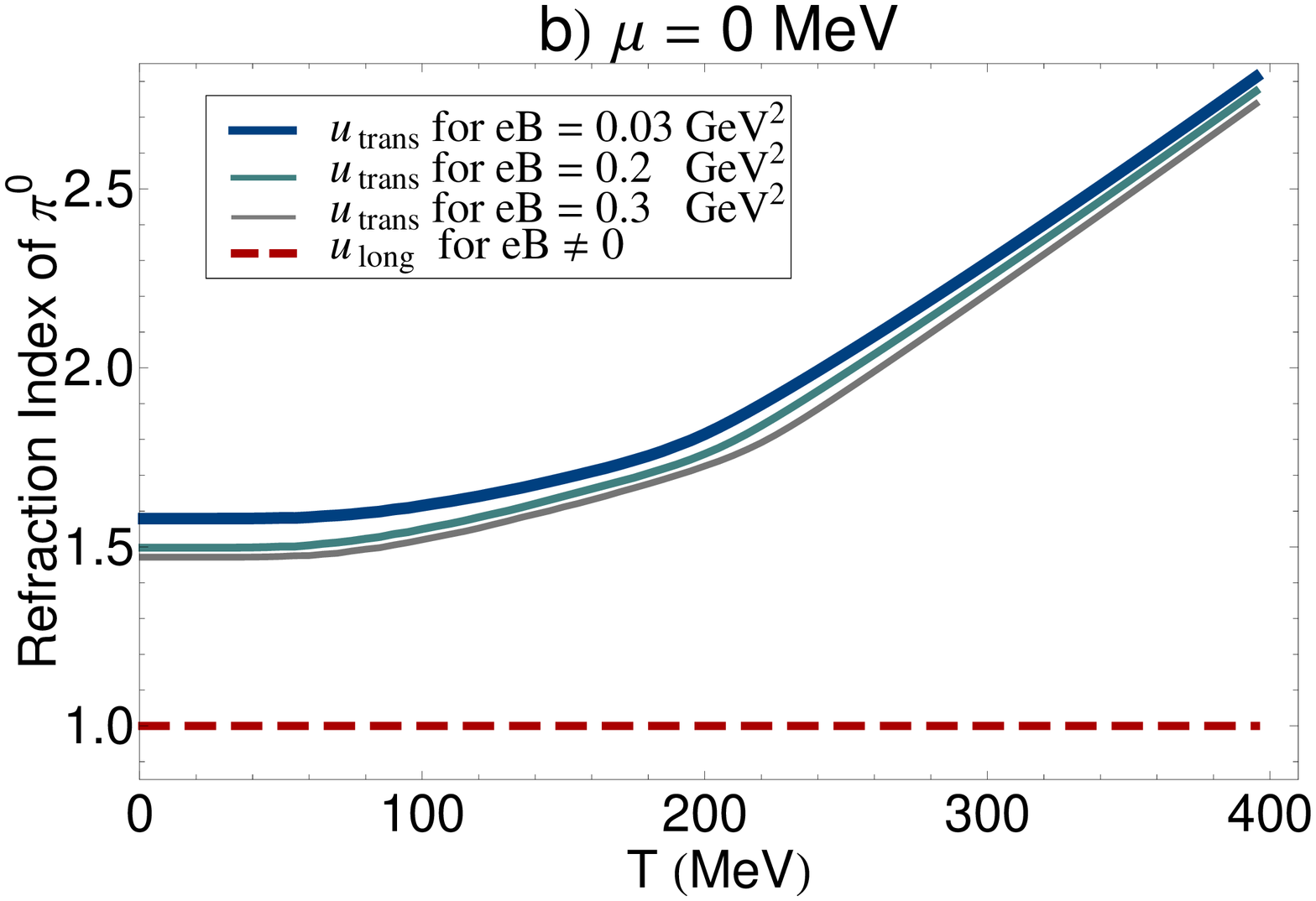}
\caption{The $T$-dependence of the transverse and longitudinal
refraction indices of $\sigma$ (panel a) and $\pi^{0}$ mesons (panel
b) is plotted for various $eB$. The longitudinal refraction index of
neutral mesons is equal to unity and independent of $T$ (red dashed
lines). The transverse refraction index of neutral mesons decreases
with increasing the strength of $eB$. }\label{fig11}
\end{figure}
\begin{figure*}[hbt]
\includegraphics[width=5.5cm,height=4cm]{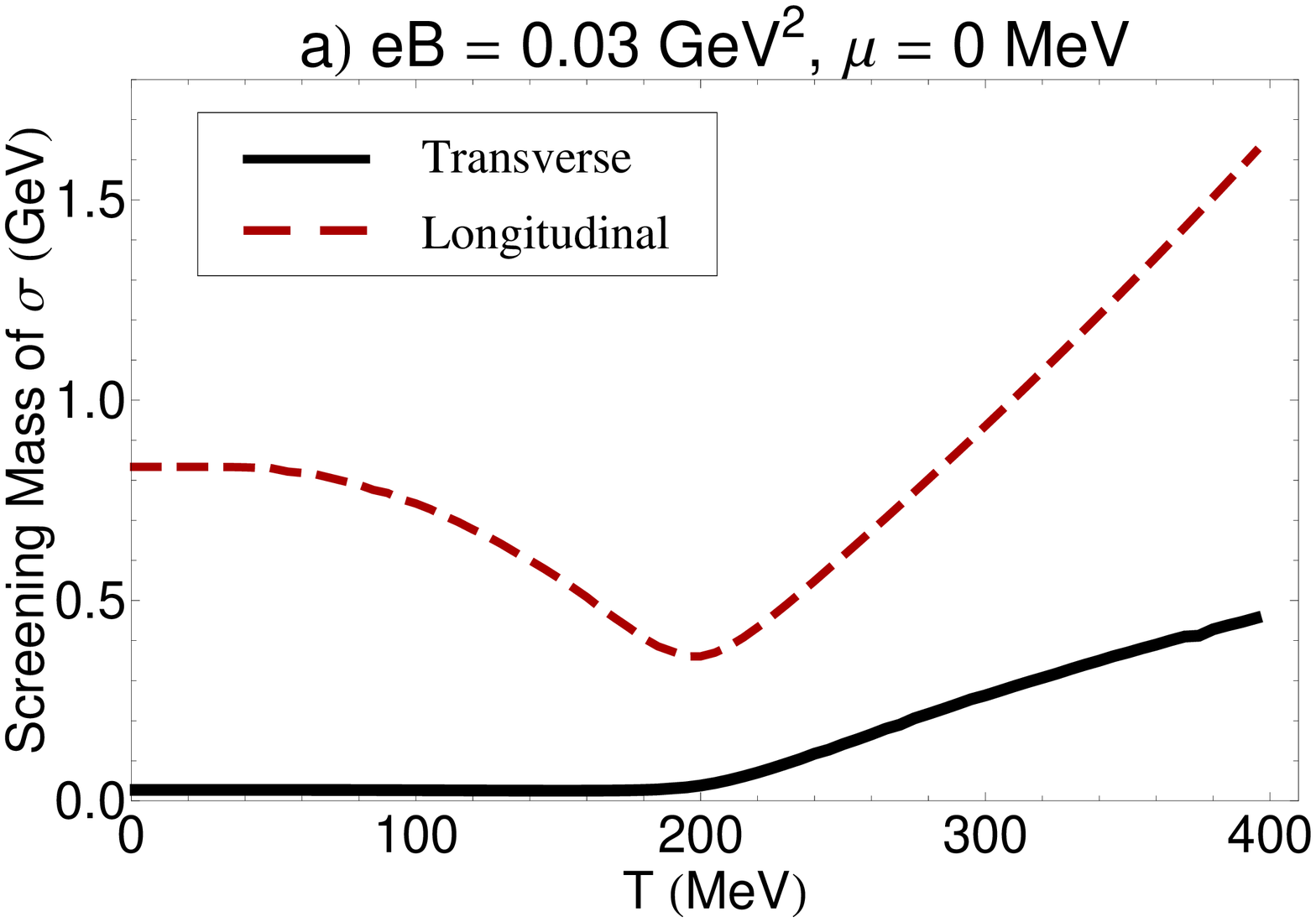}
\includegraphics[width=5.5cm,height=4cm]{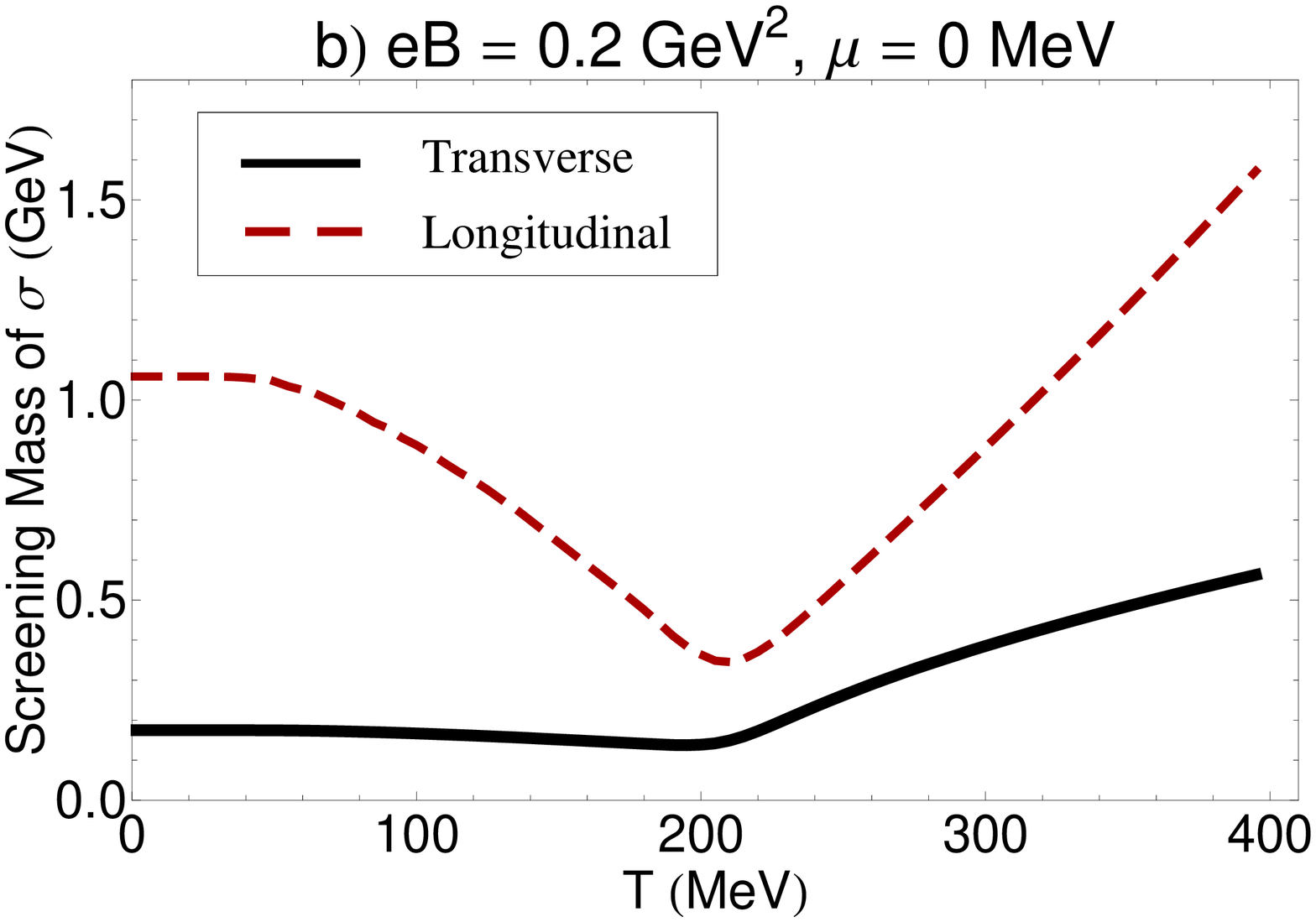}
\includegraphics[width=5.5cm,height=4cm]{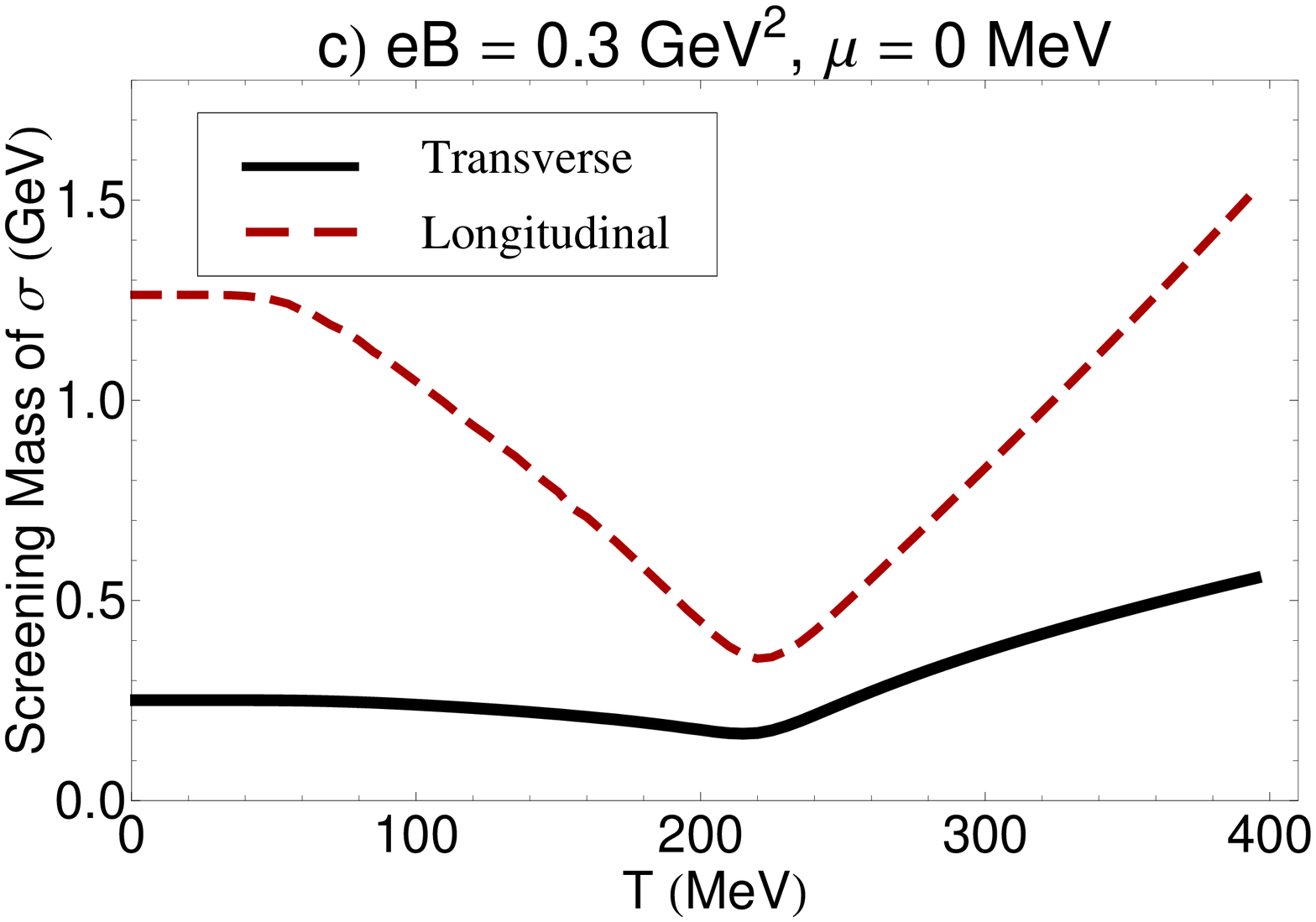}
\caption{The $T$-dependence of the screening mass of $\sigma$
mesons, $m_{\sigma}^{(i)}$, from (\ref{NA14}) in the transverse
$(i=1,2)$ and longitudinal $(i=3)$ directions is plotted for various
$eB$. As it turns out, the screening mass of the $\sigma$-meson in
the longitudinal direction $m_{\sigma}^{(i)}, i=3$ is the same as
its pole mass $m_{\sigma}$ (see $m_{\sigma}$ in Fig. \ref{fig9}).
}\label{fig12}
\end{figure*}
\begin{figure*}[hbt]
\includegraphics[width=5.5cm,height=4cm]{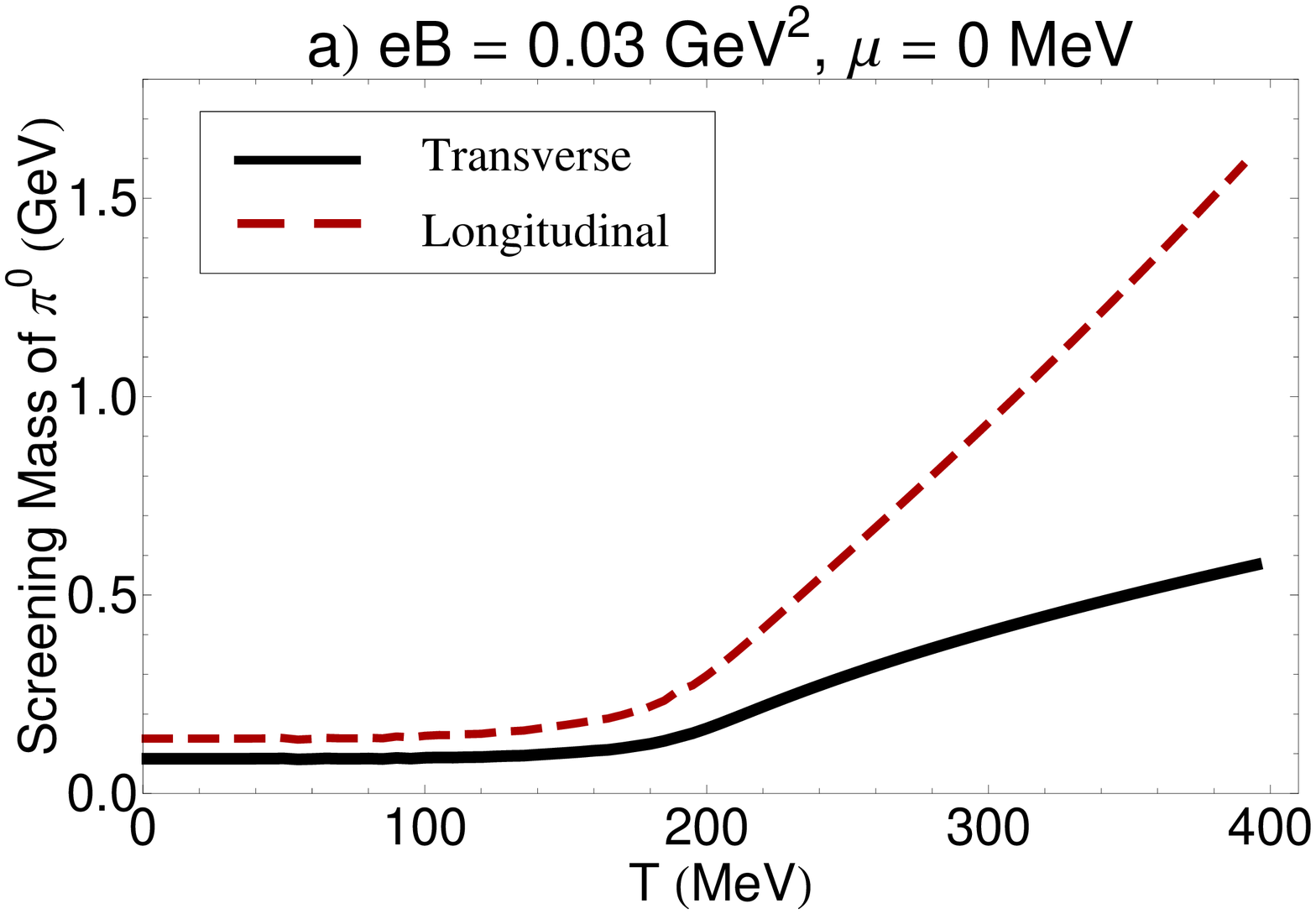}
\includegraphics[width=5.5cm,height=4cm]{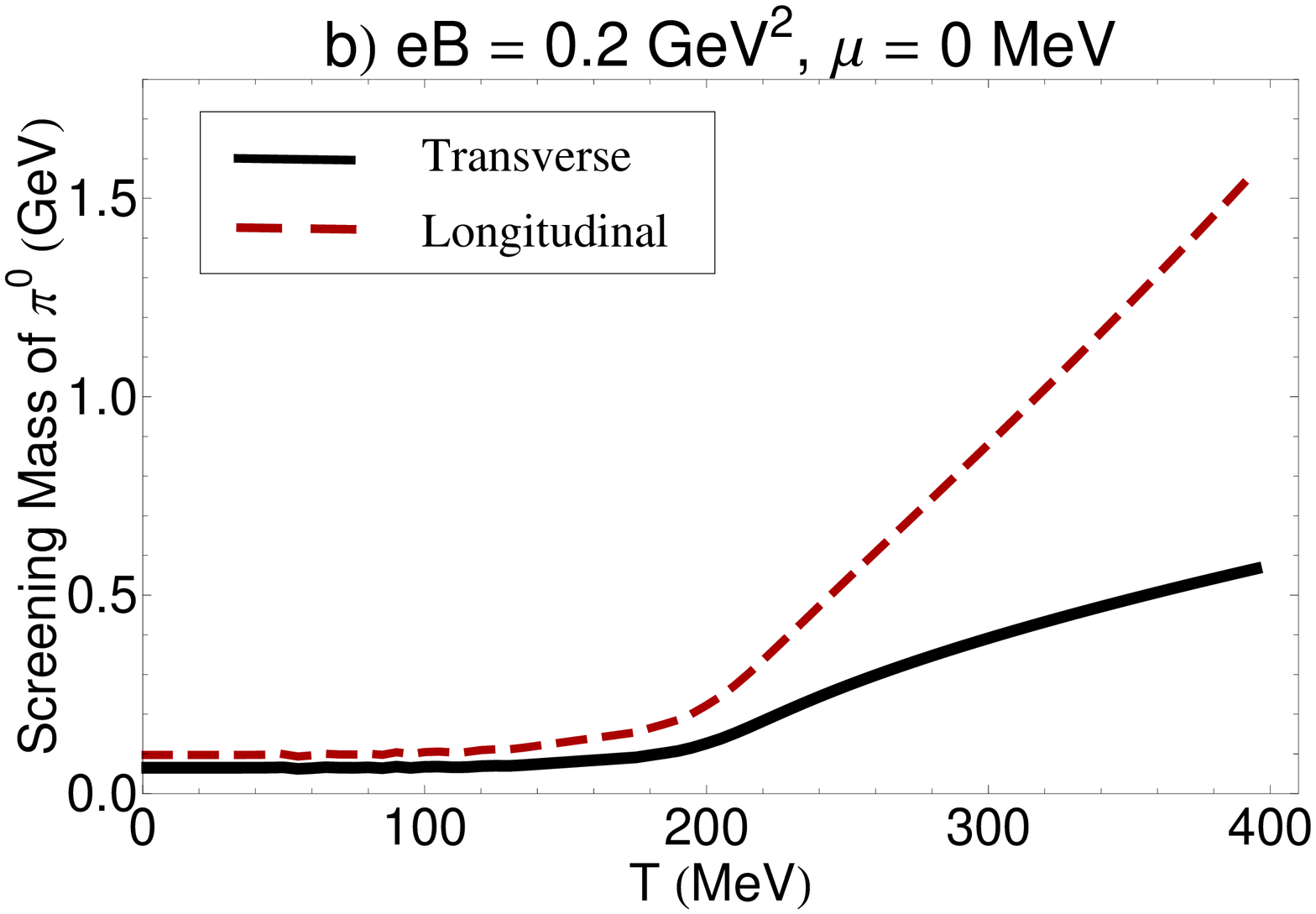}
\includegraphics[width=5.5cm,height=4cm]{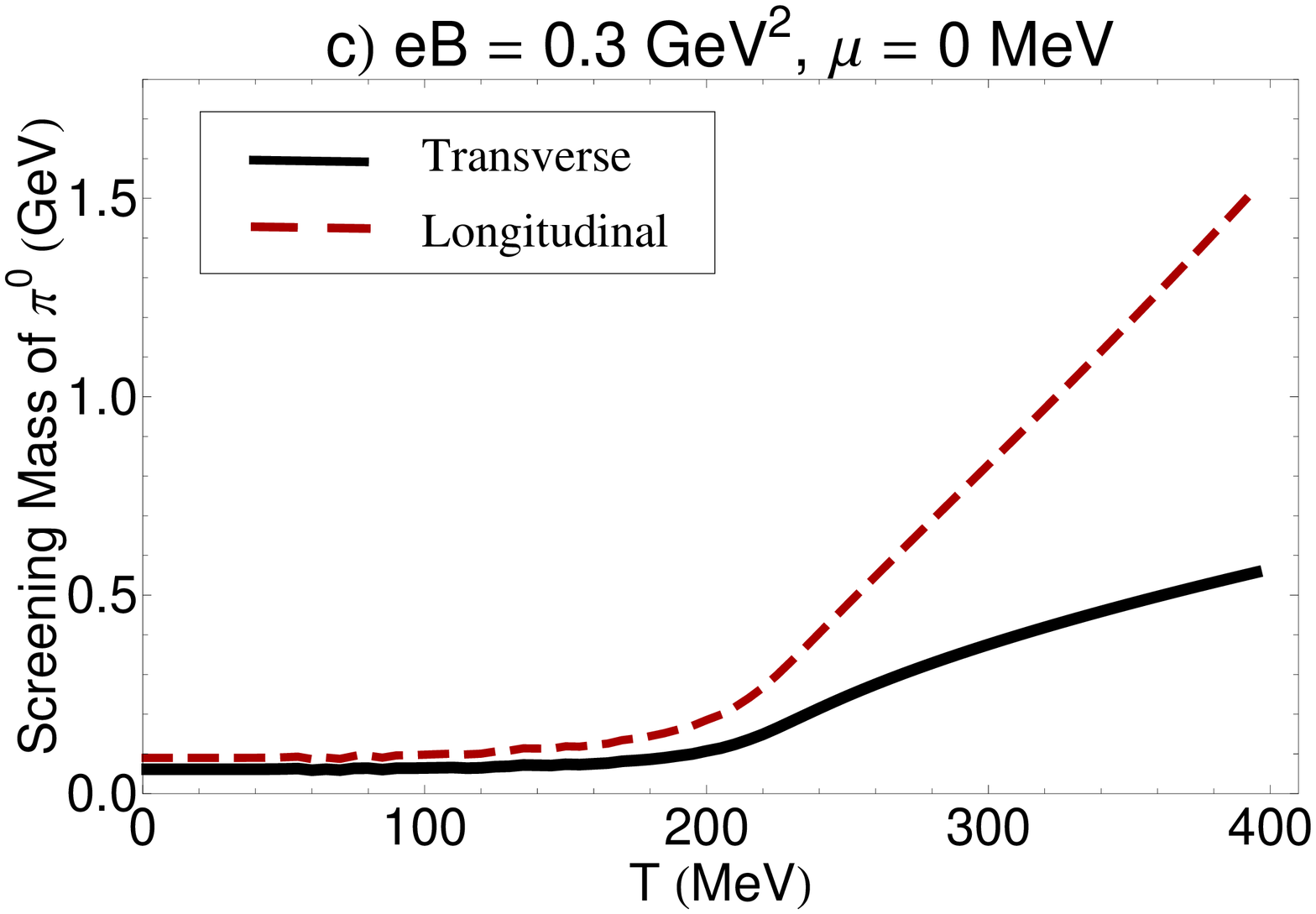}
\caption{The $T$-dependence of the screening mass of $\pi^{0}$
mesons, $m_{\pi^{0}}^{(i)}$, from (\ref{NA18}) in the transverse
$(i=1,2)$ and longitudinal $(i=3)$ directions is plotted for various
$eB$. As it turns out, the screening mass of the $\pi^{0}$-meson in
the longitudinal direction $m_{\pi^{0}}^{(i)}, i=3$ is the same as
its pole mass $m_{\pi^{0}}$ (see $m_{\pi^{0}}$ in Fig.
\ref{fig9}).}\label{fig13}
\end{figure*}
\begin{figure}[hbt]
\includegraphics[width=7.7cm,height=5cm]{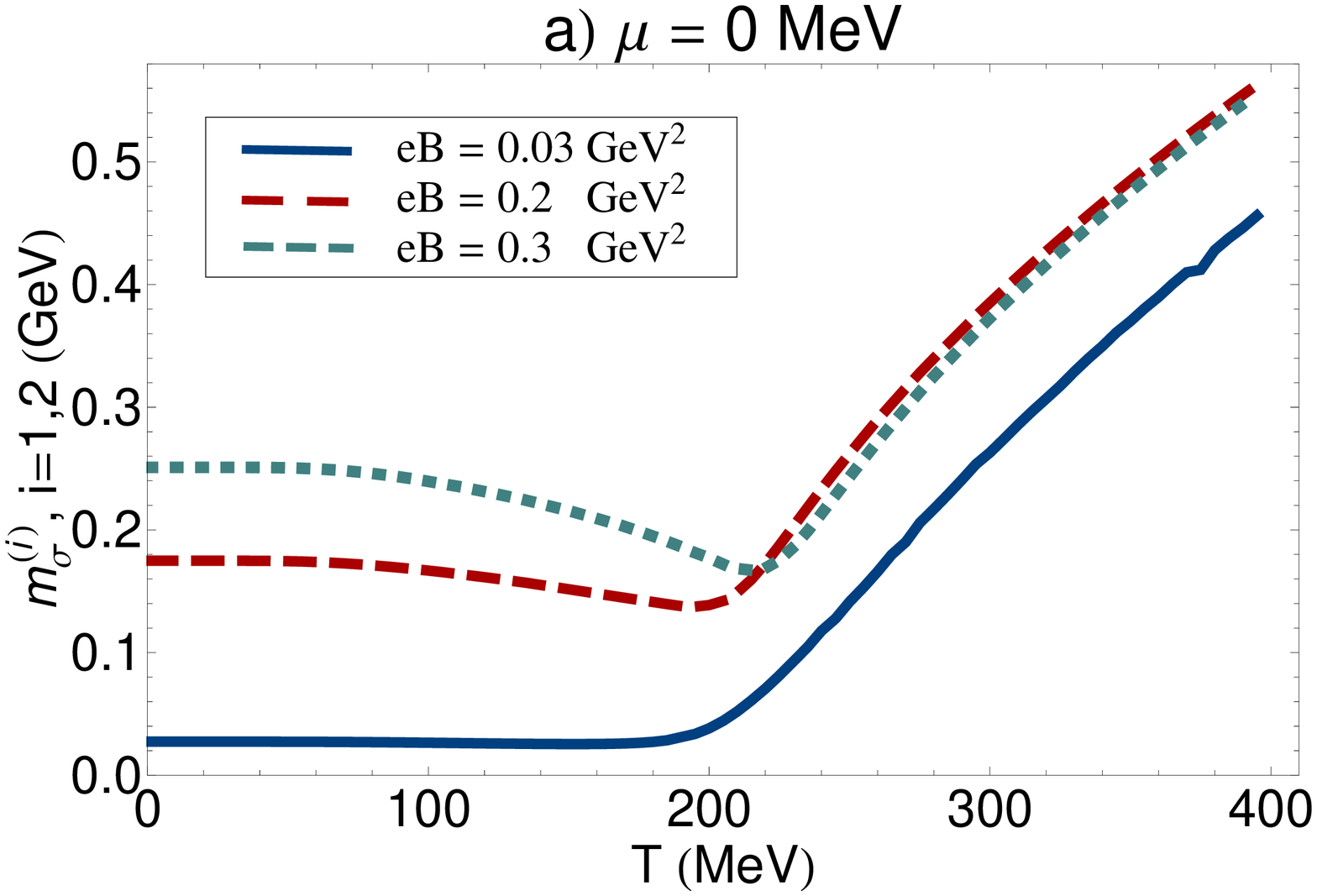}
\includegraphics[width=7.7cm,height=5cm]{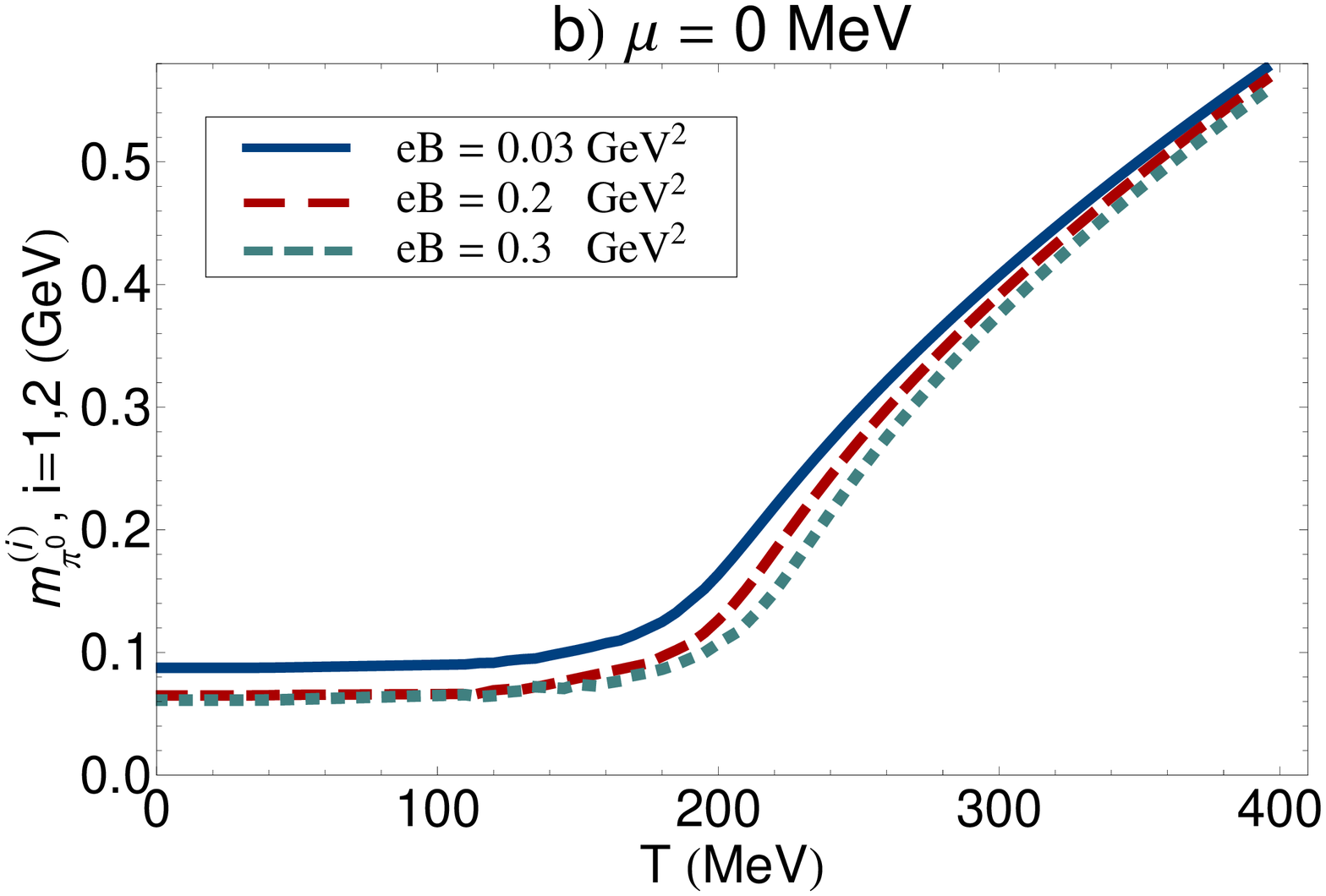}
\caption{The $T$-dependence of the screening mass of $\sigma$ (panel
a) and $\pi^{0}$ (panel b) mesons in the transverse direction is
plotted for $eB=0.03,0.2,0.3$ GeV$^{2}$ at $\mu=0$
MeV.}\label{fig14}
\end{figure}
\begin{figure}[hbt]
\includegraphics[width=7.7cm,height=5cm]{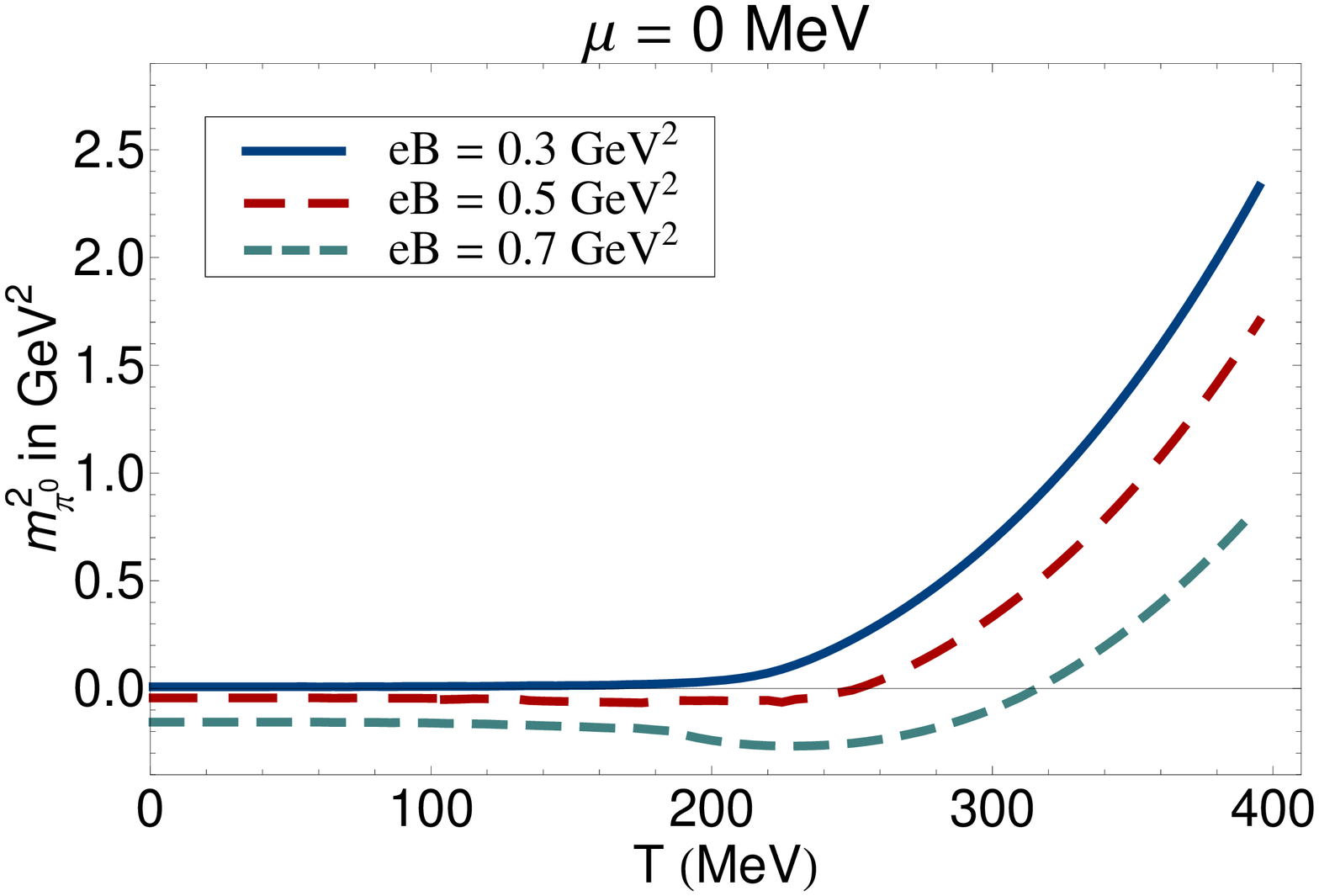}
\caption{The $T$-dependence of the squared mass of neutral pion is
plotted for $eB=0.3, 0.5, 0.7$ GeV$^{2}$.}\label{fig15}
\end{figure}
\par
Using the definitions of the directional refraction index of neutral
mesons, $u_{\sigma}^{(i)}$ and $u_{\pi^{0}}^{(i)}$, from
(\ref{NA15}) and (\ref{NA18}), as well as the $T$-dependence of
${\cal{G}}^{\mu\mu}, \mu=0,\cdots,3$ and ${\cal{F}}^{\mu\mu},
\mu=0,\cdots,3$ from Figs. \ref{fig5} and \ref{fig6}, the
$T$-dependence of the transverse ($i=1,2$) and longitudinal ($i=3$)
refraction indices of $\sigma$ and ${\pi}^{0}$ mesons, are plotted
in Fig. \ref{fig11}. From ${\cal{G}}^{00}=-{\cal{G}}^{33}$ as well
as $({\cal{F}}^{00})_{33}=-({\cal{F}}^{33})_{33}$ in (\ref{NA11}) as
well as (\ref{NA12}), the refraction index of neutral mesons in the
longitudinal direction turn out to be equal to unity, independent of
$T$ and $\mu$ (see the horizontal red dashed line in Fig.
\ref{fig11}). In contrast, the relations ${\cal{G}}^{11}\neq
{\cal{G}}^{00}$ as well as $({\cal{F}}^{11})_{33}\neq
({\cal{F}}^{00})_{33}$ from (\ref{NA11}) as well as (\ref{NA12}),
lead to $u_{\sigma}^{(i)}\neq 1$ as well as $u_{\pi^{0}}^{(i)}\neq
1$ for $i=1,2$. In Fig. \ref{fig11}, the $T$-dependence of the
transverse and longitudinal refraction indices of \textit{free} and
\textit{neutral} mesons are plotted for $eB=0.03, 0.2, 0.3$
GeV$^{2}$ and $\mu=0$ MeV. As it turns out, in the presence of
constant magnetic fields, the transverse refraction indices of
neutral mesons are always larger than unity. Moreover, the
transverse refraction index of $\sigma$ ($\pi^{0}$) meson decreases
(increases) with increasing temperature. Note that transverse
refraction indices of neutral mesons decrease with increasing the
strength of the background magnetic fields.  It is interesting to
add the effect of meson fluctuations to the above results and
recalculate the $T$-dependence of longitudinal and transverse
refraction indices of neutral mesons for non-vanishing $eB$ and
$\mu$.
\par
Using the definition of the screening masses $m_{\sigma}^{(i)}$ from
(\ref{NA14}) and $m_{\pi^{0}}^{(i)}$ from (\ref{NA18}), we arrive at
the $T$-dependence of the screening masses of neutral mesons at
$\mu=0$ and for fixed $eB=0.03,0.2,0.3$ MeV. Since in the
longitudinal direction ($i=3$), the directional refraction index of
$\sigma$ and $\pi^{0}$ mesons is equal to unity, the screening
masses of the neutral mesons in this direction are the same as their
pole masses $m_{\sigma}$ and $m_{\pi^{0}}$. In Figs. \ref{fig12} and
\ref{fig13}, the $T$-dependence of the screening masses of $\sigma$
and $\pi^{0}$ mesons in the transverse ($i=1,2$) and longitudinal
($i=3$) directions with respect to the direction of the external
magnetic field are demonstrated. As it turns out, the screening mass
of $\sigma$ and $\pi^{0}$ mesons in the transverse directions are
for all fixed $eB$ always smaller than the screening masses in the
longitudinal direction. In Figs. \ref{fig14}(a) and (b), we have
compared the screening masses of $\sigma$ and $\pi^{0}$ mesons in
the transverse directions ($i=1,2$), respectively. Comparing with
the plots of Figs. \ref{fig10}(a) and (b), it turns out, that, in
contrast to $m_{\pi^{0}}^{(i)}, i=1,2$, the behavior of
$m_{\sigma}^{(i)}, i=1,2,$ by increasing the strength of the
magnetic field is different from that of $m_{\sigma}$. And, whereas
$m_{\sigma}^{(i)}, i=1,2$, increases, in general with $eB$,
$m_{\pi^{0}}^{(i)}, i=1,2$, decreases with increasing the strength
of the background magnetic field.
\par
At this stage a remark concerning the effects of stronger magnetic
fields, $eB>0.4$ GeV$^{2}$, is in order. In Fig. \ref{fig15}, the
squared mass of neutral pion, $m_{\pi^{0}}^{2}$, is plotted for
$eB=0.3,0.5,0.7$ GeV$^{2}$ (or equivalently $eB\sim 26~m_{\pi}^{2}$,
$37~m_{\pi}^{2}$ with $m_{\pi}=138$ MeV). As it turns out, for
$eB=0.5$ GeV$^{2}$ ($eB=0.7$ GeV$^{2}$), in the regime of $T<250$
MeV ($T<320$ MeV), the squared mass of neutral pion is negative.
This the pions are tachyonic. As we have mentioned before, for
$eB>0.5$ GeV$^{2}$, only lower Landau levels contribute to
$m_{\pi^{0}}^{2}$ [see Footnote 10]. The fact that in the regime of
LLL dominance and at relatively low temperature tachyonic modes
appear, is in consistency with the recent results presented in
\cite{gorbar2012}. Here, it is shown, that at sufficiently low
temperature and in the LLL approximation tachyonic instabilities
appears in the NJL model in $2+1$ dimensions. The tachyonic
instabilities appearing in $m_{\pi^{0}}^{2}$ from Fig. \ref{fig14}
is another example of the appearance of these instabilities at low
temperature and strong magnetic field in $3+1$ dimensional NJL
model.
%%%%%%%%%%%%%%%%%%%%%%%%%%%%%%%%%%%%%%%%%
\section{Summary and conclusions}\label{sec6}
%%%%%%%%%%%%%%%%%%%%%%%%%%%%%%%%%%%%%%%%%
\par\noindent
In this paper, we studied the effects of uniform magnetic fields on
the properties of free neutral mesons, $\sigma$ and $\pi^{0}$,  in a
hot and dense quark matter. The aim was, in particular, to explore
possible effects of a background (constant) magnetic field on the
temperature dependence of the pole and screening masses as well as
the directional refraction indices of these mesons. To do this,
first, using an appropriate derivative expansion up to second order,
the one-loop effective action of a two-flavor NJL model at finite
$(T,\mu,eB)$ including $\sigma$ and $\vec{\pi}$ mesons is
determined. Then, using the formalism, presented in Sec. \ref{sec2},
the masses and refraction indices of these composite fields are
computed from their energy dispersion relations.
\par
As it turns out, the one-loop effective action of this model
consists of two parts, the effective kinetic part, including
non-trivial form factors, and the effective potential part, from
which we explored in Sec. \ref{sec5p1}, the complete phase portrait
of the model in $T-\mu$, $T-eB$ and $\mu-eB$ planes for various
fixed $eB, \mu$ and $T$, respectively. Here, we have mainly reviewed
the results previously presented in \cite{fayazbakhsh2010} for a
two-flavor NJL model including mesons and diquarks. We have shown
that the magnetic catalysis of dynamical chiral symmetry breaking
affects the phase portrait of this model in two different ways: i)
The type of the chiral phase transition changes from second to first
order in the presence of constant magnetic fields, and  ii) the
transition temperatures and chemical potentials from the chiral
symmetry broken to chirally symmetric phase increase, in general,
with increasing the strength of the external magnetic fields. Only
at low temperatures $T<50$ MeV and high chemical potentials
$280<\mu<340$ MeV and for weak magnetic fields $eB<0.2$ GeV$^{2}$,
the transition temperature decreases with increasing the strength of
$eB$. This is related to the phenomenon of inverse magnetic
catalysis, discussed in \cite{fayazbakhsh2010, rebhan2011}.
\par
In the rest of the paper, we mainly focused on the kinetic part of
the one-loop effective action. Using the formalism originally
presented in \cite{miransky1995} for a single flavor NJL model, and
generalizing it to a multi-flavor system, in Sec. \ref{sec2}, we
have determined, in Sec. \ref{sec4}, the nontrivial form factors and
squared mass matrices corresponding to neutral mesons at finite
$(T,\mu,eB)$, up to an integration over $p_{3}$-momentum and a
summation over Landau levels. They are then performed numerically in
Sec. \ref{sec5}, where, in particular, the $T$-dependence of the
form factors and squared mass matrices of the neutral mesons are
presented for several fixed magnetic fields and zero chemical
potential. Using these quantities, we have eventually determined,
the $T$-dependence of the pole and screening masses as well as the
directional refraction index of $\sigma$ and $\pi^{0}$ mesons for
fixed magnetic fields and at vanishing as well as finite chemical
potential.
\par
Because of the assumed isospin symmetry, implying $m_{u}=m_{d}$,
charged and neutral meson masses are expected to be degenerate for
vanishing magnetic fields and at zero temperature and chemical
potential. However, as it turns out, this degeneracy breaks down in
the presence of constant magnetic fields, so that we have
$m_{\pi^{0}}\neq m_{\pi^{+}}\neq m_{\pi^{-}}$, even at zero $(T,
\mu)$. This effect is mainly because of the dimensional reduction
from $D$ to $D-2$ dimensions in the presence of constant magnetic
fields, which affects the dynamics of a fermionic system in the
longitudinal and transverse directions with respect to the direction
of the external magnetic field. As a consequence, directional
anisotropy in various quantities corresponding to the particles in
the presence of a uniform magnetic field is implied.
\par
In the present paper, we have only studied the $T$-dependence of the
masses of \textit{neutral} mesons for fixed magnetic fields and
chemical potentials. The $T$-dependence of \textit{charged} meson
masses at finite $eB$ and $\mu$, will be presented elsewhere
\cite{sadooghi2012-3}. As concerns the $\sigma$-meson mass,
$m_{\sigma}$, the expected mass degeneracy with the mass of neutral
pions, $m_{\pi^{0}}$, in the crossover region, $T>220$ MeV, is
observed for various fixed $eB$ and $\mu$. Moreover, as it turns
out, $m_{\pi^{0}}$ decreases with increasing the strength of the
magnetic field. In contrast, $m_{\sigma}$ increases with increasing
$eB$ only at low temperature $T<220$ MeV, while it decreases with
increasing $eB$ in the crossover region, $T>220$ MeV. This
qualitative behavior is consistent with the result previously
presented in \cite{skokov2011} in the framework of a
Polyakov-Quark-Meson model in $3+1$ dimensions.
\par
As concerns the refraction indices of neutral mesons, it turns out
that in the presence of constant magnetic fields, the longitudinal
and transverse refraction indices with respect to the direction of
the external magnetic field are different. Moreover, whereas the
longitudinal refraction index of neutral mesons is equal to unity,
their transverse refraction index is larger than unity. The observed
anisotropy in the refraction indices of neutral mesons is because of
the explicit breaking of Lorentz symmetry in the presence of
constant and uniform magnetic fields. The anisotropy observed in the
directional refraction index of neutral mesons is also reflected in
their screening masses, which are different in the longitudinal and
transverse directions with respect to the direction of $eB$.
According to their definitions, and because of the above mentioned
results for directional refraction indices in finite $eB$, the
screening masses of the neutral mesons in the longitudinal direction
are the same as their pole masses, while in the transverse
direction, independent of $T$ and $\mu$, their screening masses are
always smaller than their pole masses. They increase with increasing
temperature at a fixed $eB$ and $\mu$. Moreover, whereas the
screening mass of $\sigma$ in the transverse direction increases in
general with the strength of the background magnetic field, the
screening mass of $\pi^{0}$, in the same direction, decreases with
$eB$.
\par
It is worth to note that the results obtained in this paper, showing
qualitatively the effect of strong magnetic fields on the properties
of neutral mesons in a hot and magnetized quark matter, can, apart
from the physics of magnetars, be also relevant for the physics of
heavy ion collisions at RHIC and LHC. As it is known from
\cite{mclerran2007, skokov2010}, magnetic fields are supposed to be
produced in the early stage of non-central heavy-ion collisions,
and, depending on the initial conditions, e.g. the energies of
colliding nucleons and the corresponding impact parameters, they are
estimated to be in the order $eB\sim 1.5~m_{\pi}^{2}$ ($eB\sim 0.03$
GeV$^{2}$) at RHIC and $eB\sim 15~m_{\pi}^{2}$ ($eB\sim 0.3$
GeV$^{2}$) at LHC energies. Although the created magnetic field is
extremely short-living and decays very fast, it can affect the
properties of charged quarks produced in the earliest stage of
heavy-ion collisions. The way we have introduced the magnetic fields
in, e.g., (\ref{ND1b}), where the external magnetic field interacts
essentially with charged quarks, opens the possibility to describe
qualitatively the effect of external magnetic fields on
\textit{neutral} mesons built from these magnetized and charged
quarks. Note that neutral mesons, by themselves, have, because of
the lack of electric charge no interaction with the external
magnetic fields. Thus, the method used in the present paper, is in
contrast to the method used in \cite{andersen2011-pions,
anderson2012-2}, where the external magnetic field interacts only
with charged pions appearing in a magnetized chiral perturbative
Lagrangian.
\par
Being motivated by these facts, we mainly focused, in this paper, on
the effects of weak and intermediate magnetic fields, $eB\leq 0.3$
GeV$^{2}$. In Fig. \ref{fig14}, however, we have plotted the squared
mass of neutral pion as a function of temperature for $eB=0.5, 0.7$
GeV$^{2}$. Here, we have shown that at low temperature and for
strong magnetic fields, where LLL approximation is reliable,
$m_{\pi^{0}}^{2}$ becomes negative. The appearance of these kind of
tachyonic instabilities at low temperature and in the presence of
strong magnetic fields is recently observed in \cite{gorbar2012} in
the framework of an NJL model in $2+1$ dimensions, which has
application in condensed matter physics. Our results are consistent
with the main conclusions presented in \cite{gorbar2012}.
\par
Let us also notice that the model used in the present paper can be
extended in many ways, e.g. by improving the method leading to the
kinetic coefficients and mass matrices of the mesons using
functional renormalization group (RG) method, which is recently used
in \cite{skokov2011, pawlowski2012, andersen2012}.
%%%%%%%%%%%%%%%%%%%%%%%%%%%%%%%%%%%%
\section{Acknowledgments}
The authors thank F. Ardalan and H. Arfaei for valuable discussions.
N. S. is grateful to R. D. Pisarski for useful comments on pion
velocity. S. S. thanks the supports of the Physics Department of
SUT, where the analytical computation of the kinetic coefficients is
performed in the framework of her master thesis. N. S. thanks the
hospitality of the Institute for Theoretical Physics of the Goethe
University of Frankfurt, Germany, where the final stage of this work
is performed. Her visit is supported by the Helmholz International
Center for FAIR within the framework of the LOEWE program launched
by the state of Hesse.
%%%%%%%%%%%%%%%%%%%%%%%%%%%%%%%%%%%%%%%%%%%%%%%%%%%%%%%%%%%%%%%%%%%%%%%%
\begin{appendix}
%%%%%%%%%%%%%%%%%%%%%%%%%%%%%%%%%%%%%%%%%%%%
\section*{Appendix: Dimensional Regularization of (\ref{NE13b})}\label{appA}
\setcounter{section}{1} \setcounter{equation}{0} \par\noindent
%%%%%%%%%%%%%%%%%%%%%%%%%%%%%%%%%%%%%%%%%%%%
In this appendix, we will use an appropriate dimensional
regularization to regularize the $(T,\mu)$-independent part of the
effective potential
\begin{eqnarray}\label{appB1}
\lefteqn{\hspace{-0.5cm}\Omega_{\mbox{\tiny{eff}}}^{(1)}(m;eB,T=\mu=0)}\nonumber\\
&&\hspace{-0.5cm}\equiv
-3\sum\limits_{q\in\{\frac{2}{3},-\frac{1}{3}\}}|qeB|\sum^{\infty}_{p=0}\alpha_{p}\int_{-\infty}^{+\infty}
\frac{dp_{3}}{4\pi^{2}}E_q,
\end{eqnarray}
appearing in (\ref{NE13b}). Here, $E_{q}$ is given in (\ref{NE10b}).
Using the definition of $\alpha_{p}=2-\delta_{p0}$, we get
\begin{eqnarray}\label{appB2}
\lefteqn{\hspace{-0.5cm}\Omega_{\mbox{\tiny{eff}}}^{(1)}(m;eB,T=\mu=0)=-3\sum\limits_{q\in\{\frac{2}{3},-\frac{1}{3}\}}|qeB|
}\nonumber\\
&&\times
\int_{-\infty}^{+\infty}\frac{dp_{3}}{4\pi^{2}}\left(\sum\limits_{p=0}^{+\infty}2E_q-E_{q}(p=0)\right).
\end{eqnarray}
The above integral can be dimensionally regularized using
\begin{eqnarray}\label{appB3}
\int_{-\infty}^{+\infty}\frac{d^{d}p}{(2\pi)^{d}}{(\phi^2+p^2)}^{-\alpha}=
\frac{\Gamma(\alpha-\frac{d}{2})}{(4\pi)^{\frac{d}{2}}\Gamma(\alpha)~\phi^{2\alpha-d}}.
\end{eqnarray}
Setting $\alpha=-1/2$, $d=1-\epsilon$, with $\epsilon$ a small and
positive number, we arrive first at
\begin{eqnarray}\label{appB4}
\lefteqn{\hspace{-0cm}\Omega_{\mbox{\tiny{eff}}}^{(1)}(m;eB,T=\mu=0)=\frac{3\Gamma(-1+\frac{\varepsilon}{2})}{4\pi^{2}}
}\nonumber\\
&&\times\sum\limits_{q\in\{\frac{2}{3},-\frac{1}{3}\}}|qeB|^{2}\left\{\sum^{\infty}_{p=0}\frac{2}{(x_q+p)^{-1+\frac{\varepsilon}{2}}}
-\frac{1}{x_q^{-1+\frac{\varepsilon}{2}}}\right\},\nonumber\\
\end{eqnarray}
where $x_q\equiv\frac{m^{2}}{2|qeB|}$. Replacing the sum over the
Landau levels $p$ with the generalized Riemann-Hurwitz
$\zeta$-function \cite{gradshteyn}, $\zeta\left(s,a\right)\equiv
\sum_{p=0}^{\infty}\left(a+p\right)^{-s}$, we get
\begin{eqnarray}\label{appB6}
\lefteqn{\hspace{-0.8cm}\Omega_{\mbox{\tiny{eff}}}^{(1)}(m;eB,T=\mu=0)=\frac{3}{8\pi^{2}}\sum\limits_{q\in\{\frac{2}{3},-\frac{1}{3}\}}(2|qeB|)^{2-\frac{\epsilon}{2}}
}\nonumber\\
&&\hspace{-0.5cm}\times\Gamma(-1+\frac{\epsilon}{2})
\left\{\zeta\left(-1+\frac{\varepsilon}{2},x_q\right)-\frac{1}{2x_q^{-1+\frac{\varepsilon}{2}}}\right\}.
\end{eqnarray}
Expanding the above expression in the orders of $\epsilon$ up to
${\cal{O}}(\epsilon)$ and eventually taking the limit $\epsilon\to
0$, we arrive at
\begin{eqnarray}\label{appB7}
\lefteqn{\Omega_{\mbox{\tiny{eff}}}^{(1)}(m;eB,T=\mu=0)}\nonumber\\
&&=\lim\limits_{\varepsilon\to
0}\frac{3}{4\pi^{2}}\sum\limits_{q\in\{\frac{2}{3},-\frac{1}{3}\}}|qeB|^{2}\left\{
\frac{(1+6x_q^2)}{3\varepsilon}\right. \nonumber\\
&&\left.+\frac{(1-\gamma_E)(1+6x_{q}^{2})}{6}-x_q\ln
x_q-2\zeta'(-1,x_q)\right.\nonumber\\
&&\hspace{0cm}\left.-\frac{1}{6}\ln\left(2|qeB|\right)-x_{q}^{2}\ln\left(2|qeB|\right)\right\}.
\end{eqnarray}
Here, we have used the polynomial expansion of
$\zeta(-1,x_q)=-\frac{1}{2}\left(\frac{1}{6}-x_q+x_q^{2}\right)$ and
the notation
\begin{eqnarray}\label{appB8}
\zeta'(-1,x_q)\equiv\frac{d\zeta(s,x_q)}{ds}\bigg|_{s=-1}.
\end{eqnarray}
In (\ref{appB7}), $\gamma_{E}\simeq 0.577$ is the Euler-Mascheroni
constant. To eliminate the divergent term, proportional to
$\epsilon^{-1}$ in (\ref{appB7}), we use the method introduced in
\cite{providencia2008}, and add/subtract to
$\Omega_{\mbox{\tiny{eff}}}^{(1)}(m;eB,T=\mu=0)$ the contribution of
the vacuum pressure
\begin{eqnarray}\label{appB9}
P_{0}=2N_{c}N_{f}\int
\frac{d^{3}\mathbf{p}}{(2\pi)^{3}}\left(\mathbf{p}^{2}+m^{2}\right)^{1/2},
\end{eqnarray}
where $N_{c}$ and $N_{f}$ are the number of colors and flavors,
respectively. But before doing this, we will first bring $P_{0}$ in
an appropriate form. Using (\ref{appB3}) with $\alpha=-1/2$, setting
$d=3-\epsilon$, and eventually expanding the resulting expression in
the orders of $\epsilon$ up to ${\cal{O}}(\epsilon)$, the vacuum
pressure, $P_{0}$, can be brought in the form
\begin{eqnarray}\label{appB10}
P_{0}=\lim\limits_{\epsilon\to
0}\left\{\frac{N_{c}N_{f}m^{4}}{8\pi^{2}}\left(\frac{(-3+2\gamma_{E})}{4}-\frac{1}{\epsilon}+\frac{\ln
m^{2}}{2}\right)\right\}.\nonumber\\
\end{eqnarray}
Replacing, according to the definition of $x_{q}$, $m^{2}$ with
$m^{2}=2|qeB|x_{q}$, and $N_{f}$ with a summation over $q$, we
arrive at
\begin{eqnarray}\label{appB11}
P_{0}&=&\lim\limits_{\epsilon\to
0}\bigg[-\frac{3}{4\pi^{2}}\sum\limits_{q\in\{\frac{2}{3},-\frac{1}{3}\}}|qeB|^{2}\left(\frac{x_{q}^{2}(3-2\gamma_{E})}{2}\right.\nonumber\\
&&\left.+\frac{2x_{q}^{2}}{\epsilon}-x_{q}^{2}\ln x_{q}-x_{q}^{2}\ln
(2|qeB|)\right)\bigg],
\end{eqnarray}
where $N_{c}=3$ is chosen. Equivalently, $P_{0}$ can be evaluated
using a sharp cutoff $\Lambda$ \cite{providencia2008},
\begin{eqnarray}\label{appB12}
P_{0}&=&-\frac{3}{4\pi^{2}}\bigg[m^{4}\ln\left(\frac{\Lambda+\sqrt{\Lambda^{2}+m^{2}}}{m}\right)
\nonumber\\
&&-\Lambda(2\Lambda^{2}+m^{2})\sqrt{\Lambda^{2}+m^{2}}\bigg].
\end{eqnarray}
Adding and subtracting $P_{0}$ to $\Omega_{\mbox{\tiny{eff}}}^{(1)}$
from (\ref{appB7}), we finally get
\begin{widetext}
\begin{eqnarray}\label{appB13}
\lefteqn{\Omega_{\mbox{\tiny{eff}}}^{(1)}(m;eB,T=\mu=0)
=-\frac{3}{2\pi^{2}}\sum\limits_{q\in\{\frac{2}{3},-\frac{1}{3}\}}|qeB|^{2}\left\{\zeta'\left(-1,x_{q}\right)+\frac{x_{q}^{2}}{4}+\frac{x_{q}}{2}(1-x_{q})\ln
x_{q}\right\}
}\nonumber\\
&&
+\frac{3}{4\pi^{2}}\left\{m^{4}\ln\left(\frac{\Lambda+\sqrt{\Lambda^{2}+m^{2}}}{m}\right)-\Lambda(2\Lambda^{2}+m^{2})\sqrt{\Lambda^{2}+m^{2}}\right\}+\mbox{$x_{q}$
independent terms}.
\end{eqnarray}
\end{widetext}
Since $\Omega_{\mbox{\tiny{eff}}}^{(1)}(m;eB,T=\mu=0)$ is a part of
the effective potential in the gap equation with respect to $m$, and
we are only interested on the minima of this potential, we have
neglected the $x_{q}$ (or equivalently $m$) independent terms in
(\ref{appB13}). Adding the tree level and the temperature dependent
parts of the effective action, the full effective action of a
two-flavor magnetized NJL model is given by (\ref{NE14b}).
\end{appendix}
%%%%%%%%%%%%%%%%%%%%%%%%%%%%%%%%%%%%%%%%%%%

\end{document}